\documentclass[epjH]{svjour}
\usepackage{graphicx}
\begin{document}
\title{Towards High-Energy Neutrino Astronomy} %\hspace{0.5cm}
%A Historical Review}
\subtitle{A Historical Review}
\author{Christian Spiering%\inst{1}
%\fnmsep
\thanks{\email{christian.spiering@desy.de}} }
%\and Second author\inst{2} \and ... }
%
\institute{DESY, Platanenallee, D-15738 Zeuthen} % \and the second here \and ...}
\abstract{
The search for the sources of cosmic rays is a three-fold assault, using charged cosmic rays,
gamma rays and neutrinos. The first conceptual ideas to detect high energy neutrinos date back to the late fifties.
The long evolution towards detectors with a realistic discovery potential
started in the seventies and eighties, with the pioneering works 
in the Pacific Ocean close to Hawaii  and in Lake Baikal in Siberia.
But only now, half a century after the first concepts, such a detector 
is in operation: IceCube at the South Pole. 
We do not yet know whether with  IceCube we will indeed 
detect extraterrestrial high energy neutrinos
or whether this will  remain the privilege of next generation telescopes. 
But whatever the answer will be: the path to the present detectors was a 
remarkable journey. This review sketches its main
milestones. 
} %end of abstract
\maketitle

\section{Introduction}

The year 2012 marks the hundredth anniversary of the detection of
cosmic rays by Viktor Hess \cite{Hess-1912}. As we know today,
cosmic rays consist of protons and nuclei of heavier elements;
electrons contribute only on the percent level.  Since cosmic rays are
electrically charged, they are deflected by cosmic magnetic fields
on their way to Earth. Precise pointing -- i.e. astronomy -- is
only possible with electrically neutral, stable particles: electromagnetic
waves (i.e. gamma rays at the energies under consideration) and
neutrinos. High energy neutrinos,
with energies much beyond a GeV, must be emitted as a by-product
of collisions of charged cosmic rays with matter. Actually, only
neutrinos provide incontrovertible evidence for 
acceleration of hadrons since gamma rays may also evolve from
inverse Compton scattering of accelerated electrons and other electromagnetic
processes. 

Since neutrinos can escape much denser celestial environments
than light, they can be tracers of processes which stay
hidden to traditional and gamma ray astronomy. At the same time, however, their
extremely low reaction cross section  makes their detection a challenge
($\sigma_{\nu\,p} \sim E_{\nu} \times 10^{-38}$\,cm$^2$, with $E_{\nu}$ in GeV).

\begin{figure}[ht]
\begin{center}
%\sidecaption
\includegraphics[width=10cm]{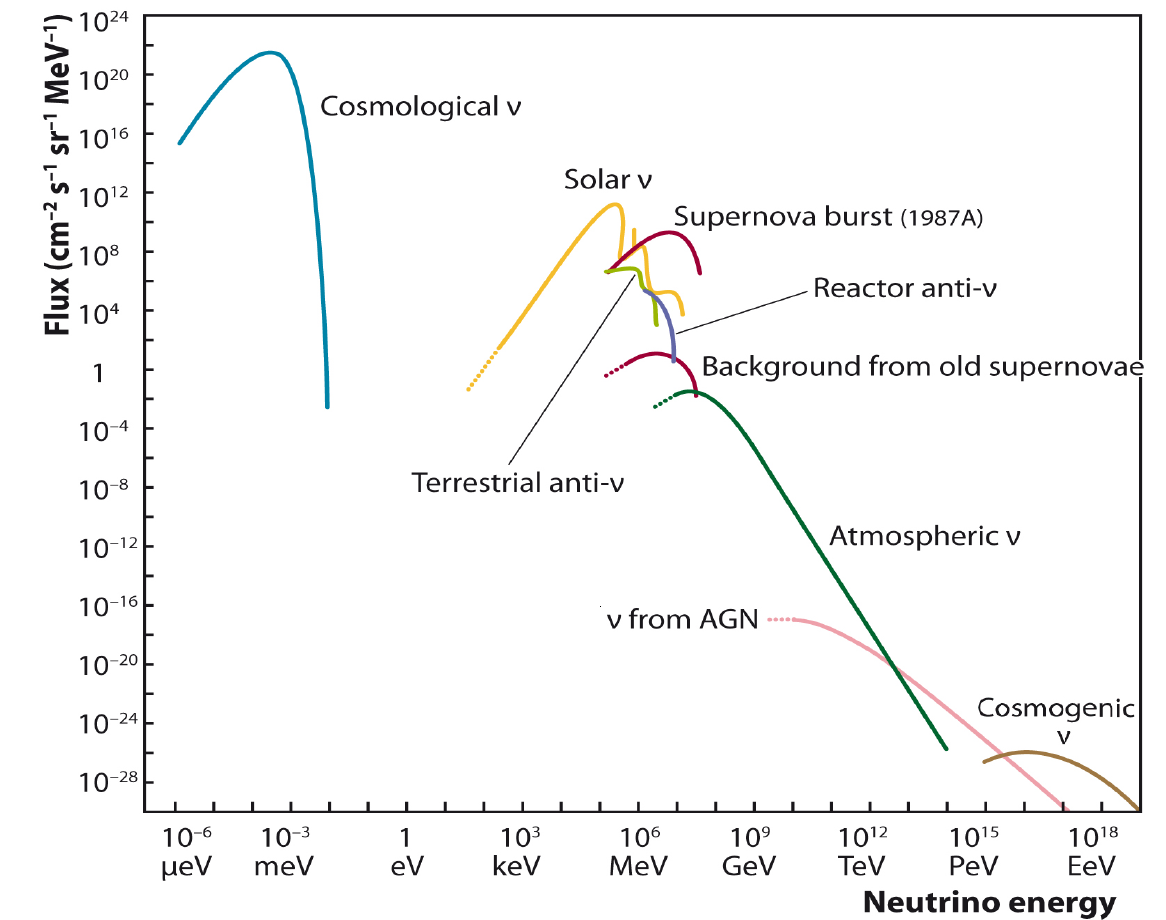}
\caption
{Measured and expected fluxes of natural and reactor neutrinos (see text for explanations). The energy range from keV to  several GeV is the domain of
underground detectors. The region from tens of GeV to about
100 PeV, with its much smaller
fluxes, is addressed by Cherenkov light detectors underwater and in ice.
The highest energies are only accessible with huge detector volumes
and methods described in section \ref{new-methods}.}
\label{all-nu}
\end{center}
\end{figure}

Neutrino astronomy is reality already now in the {\it low-energy} sector, where
the detection of neutrinos from the Sun and the Supernova SN\,1987A 
has been accomplished and was honored
by the 2002 Nobel Prize for physics. Figure~\ref{all-nu} shows a compilation of
the spectra of dominant natural and artificial neutrino fluxes.

The range of $\mu$eV and meV is that of cosmological (or "relic") neutrinos, i.e. the 1.9 Kelvin neutrino
counterpart to the 2.7 Kelvin cosmic microwave background.  No practicable idea exists on how 
to detect these neutrinos, since their reaction cross section 
as well as the energy of the recoil products from their interactions are frustratingly
small. 

The keV-MeV range is populated by neutrinos
from the Sun, from supernovae, from nuclear reactors and the from the interior of the
Earth. Neutrinos from a nuclear reactor first recorded in
1956 by Clyde Cowan and Frederick Reines  \cite{Reines-1956} mark the discovery of neutrinos.
It was acknowledged with the 1995 Nobel Prize for physics. 
Solar neutrinos have been measured first by Ray Davis in 1968 \cite{Davis-1968} 
in the Homestake mine in USA. 
The apparent deficit of neutrinos
observed by Davis  --  the long-standing "solar neutrino puzzle" -- could eventually
be explained by neutrino oscillations which transform a large part of the original
solar electron neutrinos to muon and tau neutrinos; with respect to these, the detector of Davis
and many of its successors were blind. Supernova neutrinos 
from the supernova 1987A in the Large Magellanic Cloud have been recorded
at February 23, 1987 by three detectors: Kamiokande in Japan, 
IMB in the USA (both water Cherenkov detectors)
and the Baksan scintillation  detector in Russia.  The Nobel Prize for physics 2002
was awarded to Masatoshi Koshiba (spokesman of the
Kamioka collaboration) and Ray Davis, for "pioneering contributions to astrophysics,
in particular the detection of neutrinos from the Sun and
a supernova". Neutrinos from radioactive decay processes in the interior of the Earth
("geo-" or "terrestrial" neutrinos) have been
identified only recently \cite{Kamland-2005,Borexino-2010}.

Next on the energy scale come "atmospheric neutrinos''
created in cosmic ray interactions in the Earth's atmosphere. They have been
detected in 1965 and will be in the focus of section \ref{sec-concepts}. 

The highest energies are the domain
%This article mainly describes the attempts to detect high-energy extraterrestrial
%neutrinos. Such neutrinos are expected 
of neutrinos from sources like supernova remnants, Gamma Ray Bursts or 
Active Galactic Nuclei (marked AGN in the figure) or from interactions of 
ultra-energetic protons with the 2.7\,K cosmic microwave background 
(marked "cosmogenic") \cite{Berezinsky-Zatsepin}. These cosmic
neutrinos will hopefully be detected by neutrino telescopes in this decade,
even though predictions for their fluxes are uncertain by orders of magnitude in
many cases. 

This review is about neutrinos related to cosmic rays. 
First ideas to detect extraterrestrial high energy neutrinos data back to
the end of the fifties, i.e. we look back to a journey of more than fifty
years. I will focus to the first four decades and keep
the developments of the last decade comparatively short.
I refer the reader to the 2011 review of the field \cite{Katz-Spiering-2011} for 
more detailed information on actual results and plans for future
detectors.

\section{From first concepts to the detection of atmospheric neutrinos}
\label{sec-concepts}

The initial idea of neutrino astronomy beyond the solar system  
rested on two arguments: The first was the
expectation that a supernova stellar collapse in our galaxy would be accompanied
by an enormous burst of neutrinos in the 5-10 MeV range. The second was the expectation 
that fast rotating pulsars must accelerate charged particles in their Tera-Gauss magnetic
fields. Either in the source or on their way to Earth they must hit matter, generate
pions and neutrinos as decay products of the pions. 

\begin{equation}
 p + \mbox{nucleus}  \rightarrow \pi + X  \hspace{1cm} \mbox{and} \hspace{1cm}
\pi \rightarrow \mu + \nu
\end{equation}
%with the subsequent decays $\pi^0\to\gamma\gamma$, $\pi^\pm\to\mu^\pm\nuan_\mu$
%and $\mu^+\to e^+\nubar_\mu\nu_e$, $\mu^-\to e^-\nu_\mu\nubar_e$. The resulting
%neutrino flavour ratio is approximately $\nu_e:\nu_\mu:\nu_\tau=1:2:0$ at the
%sources; neutrino oscillation turns this into a ratio of $\nu_e:\nu_\mu:\nu_\tau
%=1:1:1$ upon arrival at Earth (see Sect.~\ref{sec-sci-par-osc}).

%Actually, before the discovery of the muon neutrino in 1962 this sequence was
%not proven. It was clear, however, that in neutrino reactions
%charged leptons, i.e. muons or electrons, would be generated and that these
%charged particles must be detectable via ionization or Cherenkov radiation.

The first ideas to detect cosmic high energy neutrinos underground or underwater
date back to the late fifties.  In the 1960 Annual Review of Nuclear Science,
Kenneth Greisen and Frederick Reines discussed the motivations and prospects for such detectors.
In his paper entitled {\it Cosmic Ray Showers} \cite{Greisen-1960}, Greisen writes:
\begin{quotation}
Let us now consider the feasibility of detecting the neutrino flux. 
As a detector,
we propose a large Cherenkov counter, about 15 m in diameter, located
in a mine far underground. The counter should be surrounded with photomultipliers
to detect the events, and enclosed in a shell of scintillating material to distinguish
neutrino events from those caused by $\mu$ mesons. Such a detector would be
rather expensive, but not as much as modern accelerators and large
radio telescopes. The mass of the sensitive detector could be about 3000 tons of 
inexpensive liquid.
\end{quotation} 
Later he estimates the rate of neutrino events from the Crab Nebula
as one count per three years and
%and assumes that one could distinguish
%that from any background (astonishingly, since he must have
%been aware of the existence of the large background from atmospheric neutrinos). 
optimistically concludes:
\begin{quotation}
Fanciful though this proposal seems, we suspect that within the next decade
cosmic ray neutrino detection will become one of the tools of both
physics and astronomy.
\end{quotation}

F.\,Reines in his article {\it Neutrino Interactions} \cite{Reines-1960} is more conservative
with respect to extraterrestrial neutrinos: 
\begin{quotation}
At present no acceptable theory of the origin and extraterrestrial diffusion
exists so that the cosmic neutrino flux can not be usefully predicted.
\end{quotation} 
At this time, he could not be aware of the physics potential of atmospheric
neutrinos and continues:  
\begin{quotation}
The situation
is somewhat simpler in the case of cosmic-ray neutrinos ("atmospheric neutrinos"
in present language. C.S.)  -- they are both more predictable and of less
intrinsic interest.
\end{quotation}

In the same year, on the 1960 Rochester Conference, Moisei Markov published his
groundbreaking idea   \cite{Markov-1960}
\begin{quotation}
...to install detectors deep in a lake or a sea and  to determine the 
direction of  charged particles with the help of Cherenkov radiation.
\end{quotation}
This appeared to be the only way to reach detector volumes beyond the scale of
$10^4$ tons.

During the sixties, no predictions or serious estimates for neutrino fluxes from 
cosmic accelerators were published. Actually, many of the objects nowadays considered
as top candidates for neutrino emission were discovered only in the sixties and
seventies (the first quasar 1963, pulsars 1967, X-ray binaries with a black hole
1972, gamma ray bursts 1973).  The situation changed dramatically in the seventies,
when these objects were identified as possible neutrino emitters, triggering
an enormous amount of theoretical activity.

Different to extraterrestrial neutrino fluxes, the calculation of the flux of atmospheric 
neutrinos became more reliable. 
The idea that neutrinos might be generated
in air showers from cosmic rays, astonishingly, dates back to 1936.
This was a only two years after Enrico
Fermi had published his first paper on the theory of $\beta$ decay \cite{Fermi-1934}
and six years after Wolfgang Pauli had postulated the neutrino \cite{Pauli-1930}.
In \cite{Heisenberg-1936} Werner Heisenberg writes:
\begin{quotation}
Instead (of protons and neutrons, C.S.) Pauli's  hypothetical 'neutrinos' should
contribute substantially to the penetrating radiation. This is because in
each shower ... neutrinos should be generated which then
would lead to the generation of small secondary showers. The cross section
for the generation of these secondary showers would likely not be much smaller
than $10^{-26}$\,cm$^2$.
Contrary to the low-energy
neutrinos from $\beta$ decay one should be able to detect the energetic neutrinos
from cosmic rays via their interactions.
\end{quotation} 
Heisenberg's idea that neutrinos must be generated in showers was
correct, but he overestimated the cross section by about
ten orders of magnitude and therefore
came to the wrong conclusion that the neutrinos would
generate secondary showers and contribute to the multiplication process.
The final prediction that neutrinos generated in atmospheric showers
would be detectable is certainly true, but in another way than
Heisenberg assumed: namely, by consequently filtering away all other
particles from the showers, i.e. by going deep underground.

First serious estimates for the fluxes of atmospheric neutrinos
were published in the early %sixties
1960s 
\cite{Zatsepin-1961,Cowsik-1963,Osborne-1965,Volkova-1965}. 
Actually first calculations had been made in 1958 by Igor
Zheleznykh, at that time a diploma student of Markov. 
In his diploma work he estimated the flux of atmospheric neutrinos
and the flux of neutrinos from the Crab Nebula.
The pioneering attempts of the two Moscow groups are described in
detail in the recollections of Zheleznykh \cite{Zheleznykh-2008}.
His diploma work laid the basis of his common paper with Markov 
\cite{Markov-1961} (Actually this paper had appeared already
in 1960 in a collection of the Joint Institute of Nuclear Research, Dubna
\cite{Markov-1960a}).
It is interesting to read in this paper a footnote acknowledging 
an communication from Georgi Zatsepin who was working on
the same subject:
\begin{quotation}
We have been kindly informed by Professor G. Zatsepin that the contribution
of neutrinos originating in $\mu$-decay may be essential for
neutrinos of the order of several GeV.
\end{quotation}

\noindent
which refers to the decays
\begin{equation}
\mu^+ \rightarrow e^+ + {\bar{\nu}}_{\mu} + \nu_e   \hspace{1cm} \mbox{and} \hspace{1cm}
\mu^- \rightarrow e^-  + \nu_{\mu} +  {\bar{\nu}}_e
\end{equation}

Actually this decay doubles the number of muon neutrinos and adds electron neutrinos.
Only at higher energies muons have no time to decay before
reaching the Earth where they are slowed down and stopped. In this
case energetic neutrinos from muon decay can indeed be neglected.
Reading the equation today one must keep in mind that in 1961, although the
formula already appears in the papers of  Markov and Zhelesnykh and 
of Zatsepin and Kuzmin, the
existence of separate muon and electron neutrinos was not yet
proven. The existence of the muon neutrino was confirmed only in 1962.
 
The real explosion of papers on atmospheric neutrinos, however, happened between 
1980 and 1990 when the large underground detectors 
became operational and the field turned into a
precision science (see section \ref{subsec-Underground}).
Still, from the perspective of the sixties and early seventies, 
the study of atmospheric neutrinos  appeared  equally interesting as the search for 
extraterrestrial neutrinos \cite{Markov-1961}. 
Neutrino oscillations did not yet play a role in the discussions of the late sixties and appeared
only in the seventies on the shopping list. 
However, atmospheric neutrinos offered a possibility to study
neutrino cross sections in an energy region which was
not accessible to accelerator experiments at that time.
Using the language of the seventies, these studies would have given 
information on  the mass of the intermediate W-boson.
Without proper guidance on              %the form factors,
the W-mass, these effects were expected to be visible already in the few-GeV range, and 
actually this was one of the main motivations to build the first underground neutrino detectors
\cite{Zheleznykh-2008}. More generally, the availability
of neutrinos with energies beyond what could realistically be expected
from accelerator beams was recognized as a tempting method to search for
phenomena beyond the standard model; however not by everybody! F.\,Reines
notes in his summary of the Neutrino-81 conference  \cite{Reines-1981}:
\begin{quotation}
Estimates of the atmospheric flux suggest that interactions of this
source of $\ge$ 1 TeV neutrinos might be usefully observed, although
our accelerator-based colleagues are not keen on this as a source
of new information. 
\end{quotation}
Actually the attitude of the broader
particle physics community with respect to the physics potential of atmospheric neutrinos changed
only together with the detection of neutrino oscillations in the nineties.

Before studying atmospheric neutrinos they had to be detected. 
This was achieved in 1965, almost simultaneously, by two groups. 
One was led by Frederick Reines 
(Case-Witwaters\-rand group, later Case-Witwatersrand-Irvine, CWI).
Three years earlier, Reines had asked Markov in Geneva, whether
 mines, tunnels or caverns existed in the Soviet Union which were suitable for
underground experiments \cite{Markov-1993}. Markov then charged
his colleague A.\,Pomanski to investigate this case, with the result, that the were 
no such locations found. 
In the seventies Russia solved the problem in a unique brute-force attempt, by digging a 
special tunnel in the Baksan valley in the Caucasus.

The Reines group operated two walls of segmented liquid scintillator 
in the East Rand gold mine in South Africa, at a depth of 8800\,m water equivalent.
The configuration was chosen to identify horizontal muon tracks
\footnote{The best signature to detect muon neutrinos is 
identifying {\it upward} moving muons, since they must be due to a particle
being able to cross the Earth (see Fig.\ref{sources}). Since the
Reines detector could measure the direction but not the sense of the
direction, Reines had to focus to {\it nearly horizontal} tracks.
At 8800\,m depth water equivalent, the
slant depth then is so large that
it can be passed only by neutrinos, irrespective of the sense of the direction.}. 
Between February and July 1965, seven such tracks were recorded,
with a background of less than one event from muons not induced by neutrinos.
It is interesting to note that the first of these tracks was recorded at February 23, 1965,
exactly 22 years before
the neutrinos from supernova SN1987A reached the Earth (23/2/1987).
Some personal reminiscences of this experiment can be found in
\cite{Kropp-1991}.

The detector of the other group (a Bombay-Osaka-Durham collaboration)
was operated in the Indian Kolar Gold Field (KGF) mine,
at a depth of 7500\,m water equivalent. It consisted of  two walls of plastic
scintillators and flash tubes. The KGF group started 
data taking nearly six months after the CW group, saw the first of three
neutrino candidates two months later than Reines (20/4/1965), but published 
two weeks earlier than the CW group:
KGF at August 15, 1965 (submitted 12/7/1965 \cite{KGF-1965}), CW at
August 30, 1965 (submitted 26/7/1965 \cite{Reines-1965}).
So, indeed Reines recorded the first cosmic neutrino ever, but the  
formal priority is with the KGF group; a historic race
which had no losers but two winners.

\begin{figure}[h]
\sidecaption
\includegraphics[width=6cm]{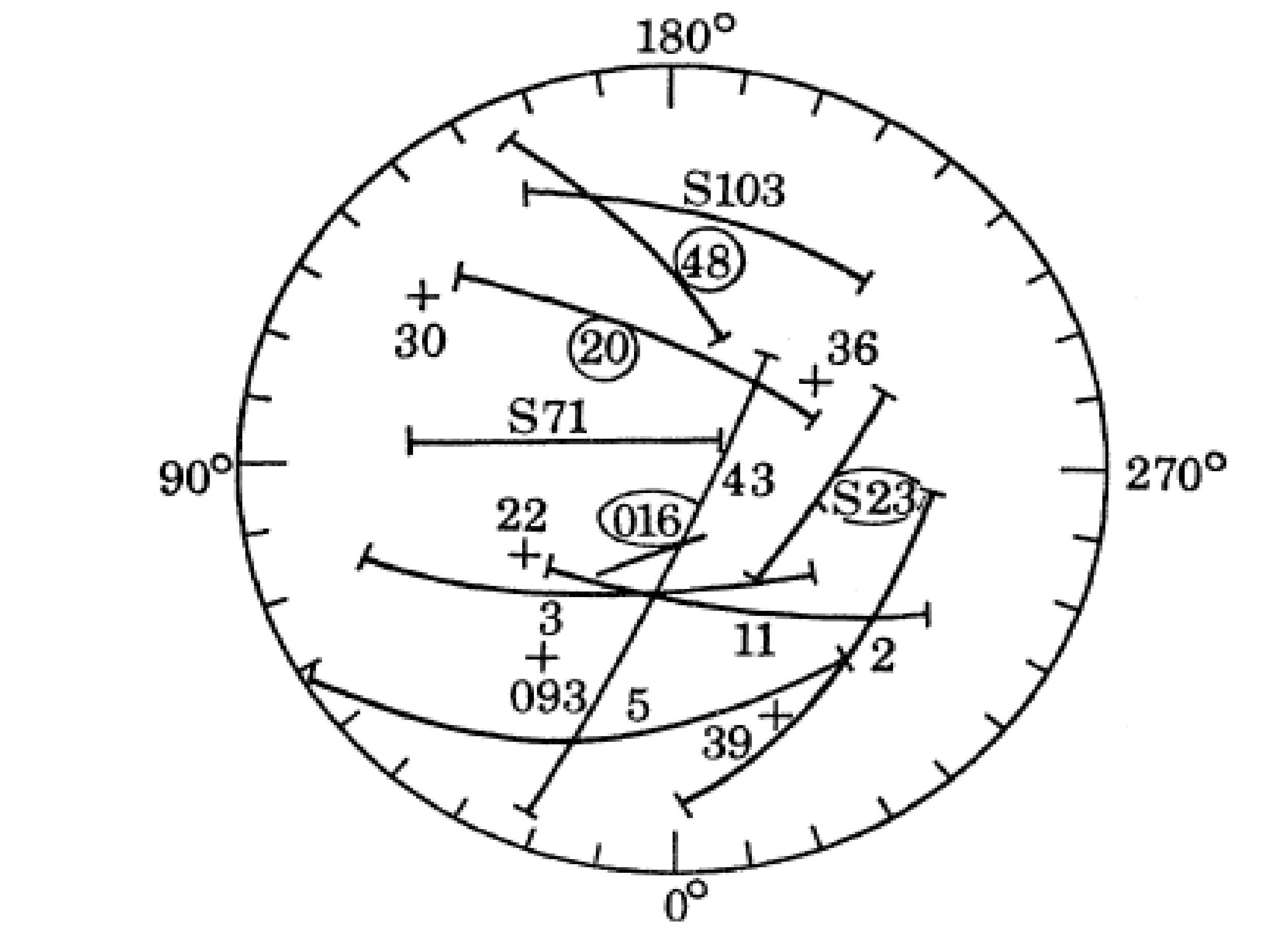}
\caption{
The first neutrino sky map with the celestial coordinates of 18 
KGF neutrino events \cite{Krishnaswamy-1971}.
Due to uncertainties in the azimuth, the coordinates for some events are
arcs rather than points. The labels reflect the numbers and registration mode
of the events (e.g. "S" for spectrograph). Only for the ringed events
the sense of the direction of the registered muon is known.
}
\label{KGF-skymap}
\end{figure}

With improved  detectors, the two groups continued measurements for 
many years \cite{Crouch-1978,Krishnaswamy-1971},  
collecting a a total sample of
nearly 150 neutrino events. The KGF group was the first to release a sky map
(see Fig.\,\ref{KGF-skymap}). A comprehensive review of particle physics
activities  in the Kolar Gold Field mine, from the sixties to the nineties, is given in
\cite{Narisimham-2004}.

A third underground detector was operated from 1967 on by a group in Utah
\cite{Bergeson-1967}. It consisted of spark chambers and
Cherenkov counters. The anomalies in muon depth-intensity
which they reported in 1967-71 provided one of the initial motivations
to build underwater detectors. In 1973, the Utah group reported also first results on
neutrinos \cite{Bergeson-1973}. Combining their energy spectra with
those of the KGF group they concluded that the spectrum is consistent
with a linear growth of the neutrino cross-section up to high energies and
that the W-boson mass has to be large. 
  
In 1978, the Baksan Neutrino Telescope (BNT) in the Caucasus started (partial) operation. 
It was followed by a phalanx of new detectors
in the eighties, which mostly were motivated by the search for proton decay.
The largest of them was the IMB detector which produced the first neutrino
sky map of reasonable quality, with 187 events from 396 live-days
\cite{IMB-1987}. 

The study of atmospheric neutrinos and of  MeV-neutrinos from a galactic
Supernova seemed to be feasible with detectors of a few hundred or thousand tons, 
with the main unknown being the rate at which those Supernovae occur.  Predictions
ranged from a several per millennium up to a few per century. Therefore a Supernova
neutrino detector had better to be a multipurpose device with
alternative goals, for instance atmospheric neutrinos or cosmic ray studies as reliable
aims, combined with the high-risk aim to search for proton decay. Actually, 
the two water Cherenkov detectors which detected neutrinos from the
supernova 1987A, IMB (USA) and Kamiokande (Japan), had both been funded
for different primary purposes, most notably proton decay.

In contrast to investigating atmospheric and supernova neutrinos,
the study of high-energy extraterrestrial neutrinos had the inherent risk that
no reliable predictions for the expected fluxes could be made. Under these circumstances
it appeared logical to tackle this problem with the largest devices conceivable,
with underwater detectors of the kind which M.\,Markov had proposed in 1960.

Actually, a few proposals with elements of Markov's idea were made
in the sixties.  One of them in 1965 suggested to detect upward moving muons from neutrino
interactions by observing a 10\,m thick water target ({\it "possibly in ocean or a lake"}) 
with downward looking photomultipliers \cite{Uberall-1965}.
Another one \cite{Bogatyrev-1971} ambitiously proposed three detectors, each 
with $10^7$ tons  of distilled  water a several km depth. The detectors
should be widely spaced in order to allow for Supernova triangulation, 
i.e. determining the arrival direction
of the neutrino swarm by measuring differences in arrival times at very
distant locations \footnote{The principle of Supernova triangulation
has been first proposed in a paper of Grigorij Domogatsky and
Georgij Zatsepin  \cite{Domogatsky-1965}.}. 
Instead of housing photomultipliers in pressure resistant
glass spheres of cm-thickness (the accepted method for all later
deep underwater experiments), the author proposed to construct large
pressure resistant phototubes and, moreover, to use wavelength shifters. With these
two 1971 ideas he anticipated options which later were considered,
although eventually abandoned, for the Baikal experiment, for DUMAND and for IceCube.  

The first step from conceptual ideas to large-scale experimental
efforts was done by the DUMAND project which will be described
in the next section.

\section{DUMAND:  from the first workshop to the DUMAND-II proposal}
\label{sec-dumand}

\subsection{The early years}

The history of underwater neutrino telescopes starts with a project which
eventually was cut off but left an incredibly rich legacy of ideas and technical
principles: The DUMAND project. DUMAND stands for Deep Underwater Muon and
Neutrino Detector. Its early history is excellently covered in a {\it Personal
history of the DUMAND project} by Arthur Roberts \cite{DUMAND-Roberts}. 

At the 1973 International Cosmic Ray Conference (ICRC), a small group of physicists
including F.\,Reines, J.\,Learned, H.\,Davis, P.\,Kotzer, M.\,Shapiro (all USA),
G.\,Zatsepin (USSR) and S.\,Miyake (Japan)
discussed a deep-water detector to clarify puzzles in muon depth-intensity
curves. The anomalies reported by the group of
W. Keuffel in Utah ("Keuffel effect")
\cite{Bergeson-1967} faded away, but it was obvious that such a
detector could also work for neutrinos. An informal group of people
to study such a detector was assembled, led by Reines, Roberts, 
Miyake and Learned. 

The year 1975 saw the first of a -- meanwhile legendary -- series of DUMAND 
Workshops, this one at Washington State University \cite{DUMAND-1975}. A survey of possible
sites converged on the Pacific Ocean close to Hawaii, since it offered
deep locations close to shore. A year later, a two-week workshop took
place in Honolulu \cite{DUMAND-1976}.
At that time, three options for a deep sea array were discussed:

\begin{itemize}
\item UNDINE (for "UNderwater Detection of Interstellar Neutrino Emission") was 
intended to detect neutrinos from supernova collapses from far beyond our
own Galaxy (leaving the Galactic Supernovae to underground detectors)
\footnote{Recently this idea was revived (without
knowledge of the early DUMAND concept)  
with the proposal to drastically increase the photomultiplier density of a part of IceCube  
in order to monitor the full Virgo cluster and record several SN collapses per year.}.
Based on overoptimistic assumptions of the neutrino energy spectrum,
it was soon discarded.
\item ATHENE (for "ATmospheric High-Energy Neutrino Experiment") was tailored
to high-energy particle physics with atmospheric neutrinos. 
\item UNICORN (for "UNderwater Interstellar COsmic-Ray Neutrinos") had the
primary goal to search for high-energy extraterrestrial neutrinos.
\end{itemize}

At the 1976 workshop and, finally, at the 1978 DUMAND workshop at the Scripps Institute
in La Jolla \cite{DUMAND-1978} the issue was settled in favor of an array
which combined the last two options, ATHENE and UNICORN.  It is noteworthy that
concepts of UNDINE were revived in 1978 for IMB and later for Kamiokande,
both in mines (see section \ref{subsec-Underground}).

\begin{figure}[ht]
%\sidecaption
\hspace{0.5cm}
\includegraphics[width=4.5cm]{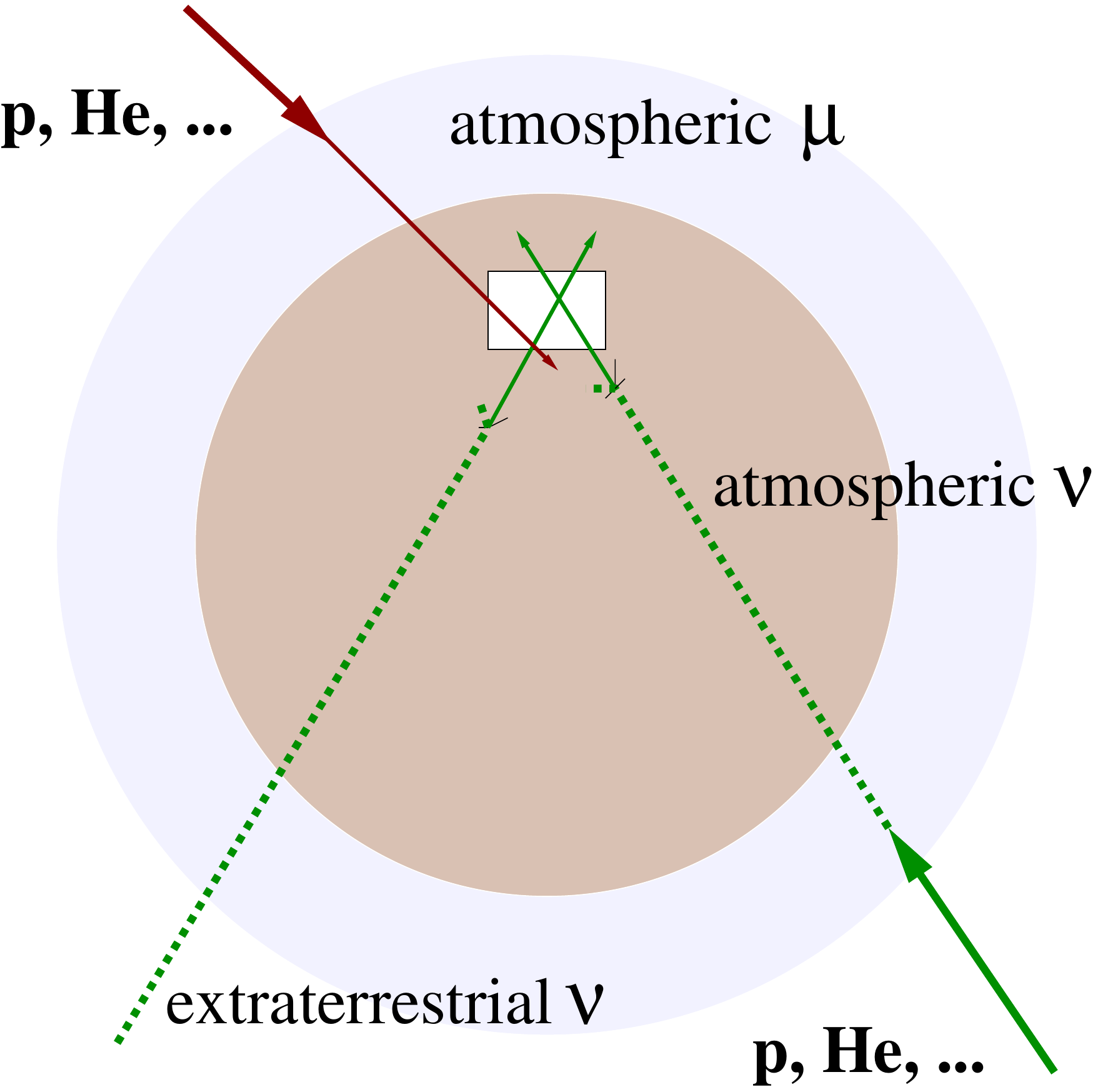}
\hspace{1cm}
\includegraphics[width=5.5cm]{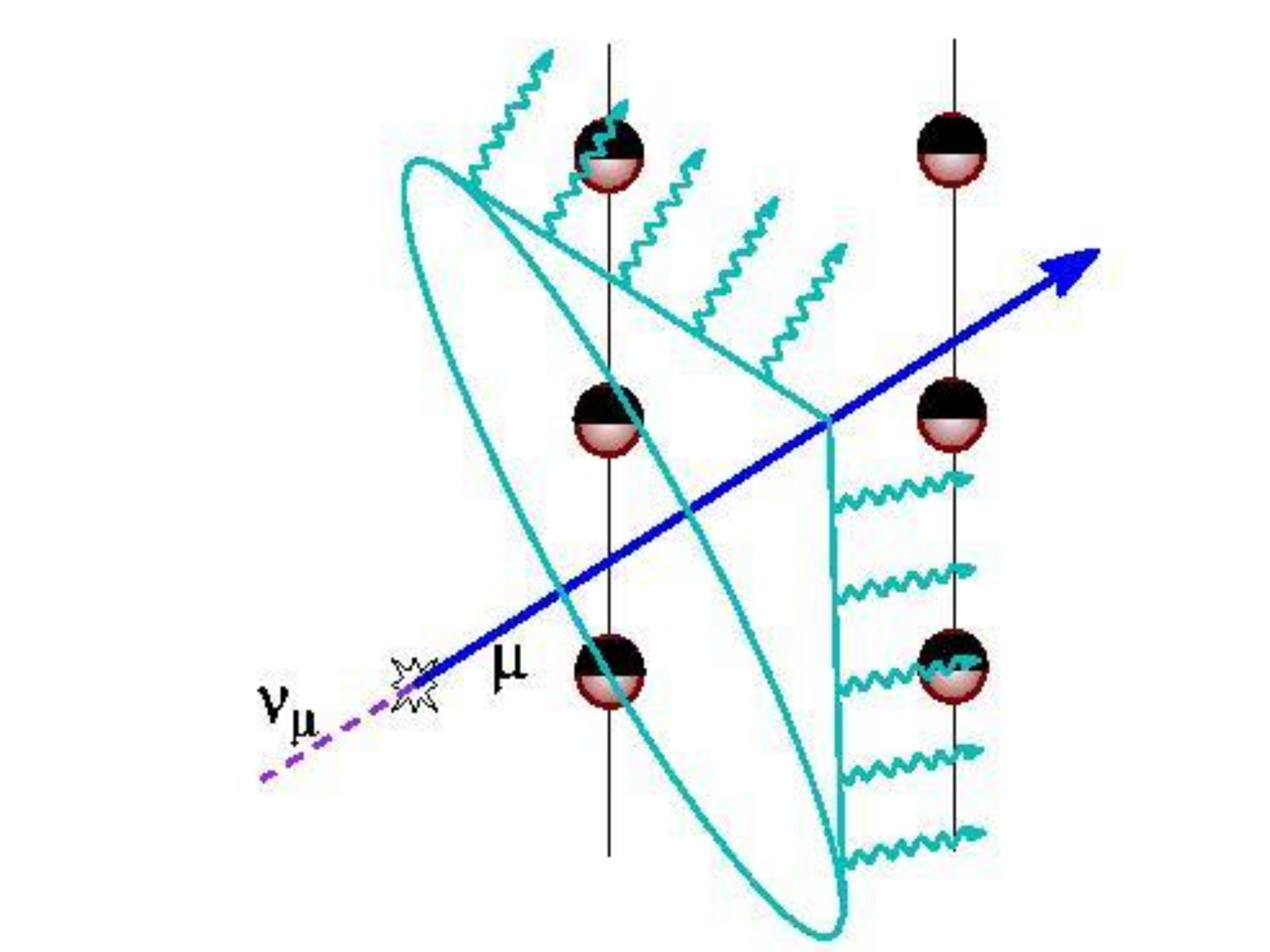}
\caption{   %\texttt{
{\it Left:} Sources of muons in deep underwater/ice detectors. Cosmic nuclei -- protons (p),
$\alpha$ particles (He), etc.\ -- interact in the Earth atmosphere
(light-colored). Sufficiently energetic muons produced in these interactions
("atmospheric muons'') can reach the detector (white box) from above. 
Upward-going muons must have been produced in neutrino interactions.
{\it Right:} Detection principle for muon tracks.  %}
\label{sources}
}
\end{figure}

The principle of the detector was to record upward-traveling muons generated in 
charged current muon neutrino interactions.
Neutral current interactions which produce no muons had been only discovered in 1973. 
They result in  final-state charged particles of rather low energy and did not play a role
for the design studies 
\footnote{Muon-less events from charged current interactions
of electron and tau-neutrinos were assumed to contribute only 1/3 of
the flux (oscillations leading to an 1:1:1 ratio have not yet been established at that time!)
and much less with respect to the event rate. Moreover they have a less characteristic signature
and were also not prioritized.}. 
The upward signature guarantees the neutrino origin of
the muon since no other particle can traverse the Earth. Since the sixties,
a large depth was recognized as necessary in order to suppress
downward-moving muons 
\footnote{
The ratio of downward to upward muons is
of the order of $10^6$ at a depth of 1 km.w.e. (km water equivalent)
but only of the order $10^4$ at 4 km.w.e.}
which may be mis-reconstructed as upward-moving ones
(Fig.\,\ref{sources}, left).
 Apart from these, only one irreducible background to
extra-terrestrial neutrinos remains: neutrinos generated by cosmic ray
interactions in the Earth's atmosphere (``atmospheric neutrinos''). This
background cannot be reduced by going deeper. On the other hand, it provides a
standard calibration source and a reliable proof of principle.

\begin{figure}[ht]
%\sidecaption
\begin{center}
\includegraphics[width=13cm] {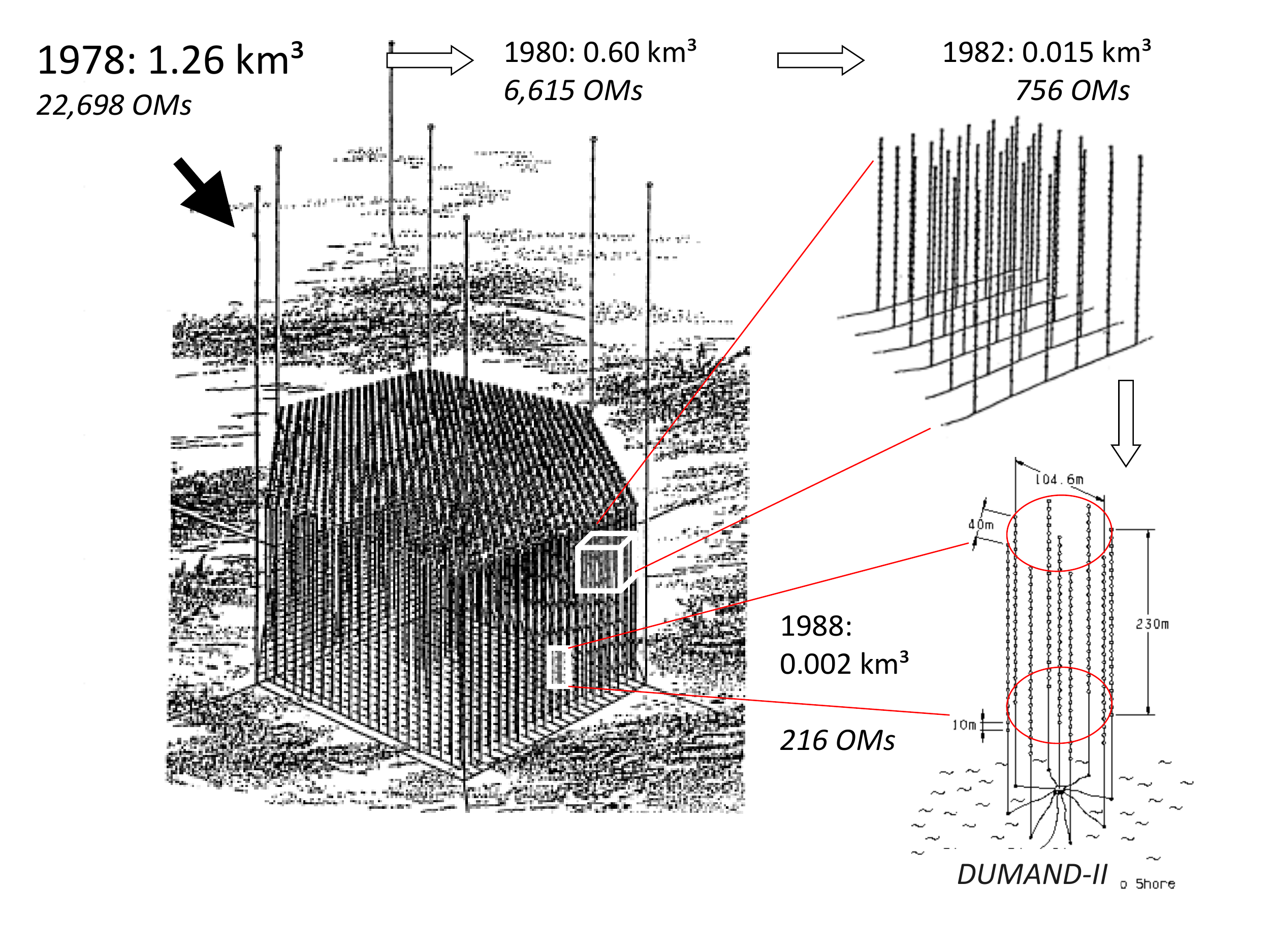}
\end{center}
\caption{   %\texttt{
The originally conceived DUMAND cubic kilometer detector and the
phased downgrading to the 1988 plan for a first-generation underwater neutrino telescope
DUMAND-II. See text for explanations.}   %}
\label{DUMAND}
\end{figure}

The DUMAND design envisaged an array of photomultiplier tubes
(PMTs) housed in transparent pressure spheres spread over 
a cubic kilometer (see Fig.\ref{DUMAND}, left). The PMTs would record arrival time and
amplitude of Cherenkov light emitted by muons or particle cascades. 
The spheres were to be attached to strings moored at the ground and held
vertically by buoys. 
From the arrival times, the direction of the muon track can be reconstructed,
and it turned out that a directional accuracy of 1\,degree is achievable. This
is of a similar size as the kinematic smearing between neutrino and muon direction 
and allows for {\it neutrino tracing}, i.e. for neutrino astronomy (see Fig.\,\ref{sources}, right). 
I refer to \cite{Spiering-2011} for a detailed description of the functional
principle, the technical realizations and the various event signatures in underwater/ice
neutrino telescopes.

%\begin{figure}[ht]
%\sidecaption
%%\center{
%\epsfig{file=figs/Ch-Cone.pdf,width=6cm}
%\caption{Detection principle for muon tracks}
%\label{principles}
%%}
%\end{figure}

Naturally, the idea to construct a cubic-kilometer detector with more than
20\,000 large-size photomultipliers (see Fig.~\ref{DUMAND}) challenged
technical and financial possibilities. A.\,Roberts remembers \cite{DUMAND-Roberts}: 
\begin{quotation}
The 1978 DUMAND
Standard Array, on closer examination, assumed more and more awesome proportions.
... 1261 sensor strings, each with 18 complex sensor modules ...
to be deployed on the ocean bottom at a depth of 5 km! The oceanographers were
amazed -- this project was larger than any other peacetime ocean project by a factor
of the order of 100. The size of the array was based on relatively scant information
on the expected neutrino intensities and it was difficult to justify in detail; the general
idea was that neutrino cross section are small and high-energy neutrinos are
scarce, so the detector had better be large.
\end{quotation}
Confronted with the
oceanographic and financial reality, the 1.26 km$^3$ array was abandoned. 
A half-sized configuration (1980) met the same fate, as did a much smaller array with 756 phototubes
(1982). The latter design was comparable in size to the %meanwhile decommissioned
AMANDA detector at the South Pole (see Sect.~\ref{sec-amanda}) and the
ANTARES telescope in the Mediterranean Sea, close to Toulon (see
Sect.~\ref{sec-medi}). What finally emerged as a technical project was a
216-phototube version, dubbed DUMAND-II or "The Octagon'" (eight strings at the
corners of an octagon and one in the center), 100\,m in diameter and 230\,m in
height \cite{DUMAND-Project} (see Fig.\,\ref{DUMAND}).
The plan was to deploy the detector 30\,km off the coast of Big Island, Hawaii, 
at a depth of 4.8\,km.

\subsection{Flux predictions}

The evolution of the detector design, largely following
financial and technological boundary conditions, was the one side
of the story. What about the flux predictions?

At the 1978 workshop first 
investigations on neutron star binary systems as point
sources of high energy neutrinos were presented, specifically
Cygnus X-3 (D.\,Eichler/ D.\,Schramm  and D.\,Helfand in 
\cite{DUMAND-1978}). The connection to the indications
for sources of TeV-$\gamma$-ray (none of them significant at that time!)
was discussed by T.\,Weekes. At the same time, the possibilities for
diffuse source detection were disfavored (R.\,Silberberg, M.\,Shapiro,
F.\,Stecker). 

The gamma-neutrino connection was discussed further by
Venjamin Berezinsky at the 1979 DUMAND Workshop in Khabarovsk and Lake
Baikal (see \cite{DUMAND-1979}). He emphasized the concept of "hidden"
sources which are more effectively (or {\it only}) detectable by neutrinos
rather than by $\gamma$ rays. Among other mechanisms, Berezinsky also investigated the
production of neutrinos in the young, expanding shell of a supernova
which is bombarded by protons accelerated inside the shell  
("inner neutrino radiation from a SN envelope") \cite{Berezinsky-1977}. 
He concluded that
a 1000\,m$^2$ detector should be sufficient to detect
high-energy neutrinos from a galactic supernova over several weeks or
months after the collapse. Naturally, ten years later, in 1987, 
much attention was given to this model in the
context of SN1987. But alas! -- this supernova was at about 50 kpc distance,
more than five times further than the Galactic center. Moreover 
all underground detectors existing in 1987 had areas much smaller
than 1000\,m$^2$. Therefore the chances to see 
inner neutrino radiation from the envelope were rather small, and actually 
"only" the MeV burst neutrinos and no high energy neutrinos have been recorded.

A large number of papers on expected neutrino
fluxes was published during the eighties. The   
fluxes were found to depend strongly {\it a)} on the energy spectrum of the $\gamma$-ray
sources which could only be guessed since the first uncontroversial TeV-$\gamma$ observation
was the Crab nebula in 1989 \cite{Whipple-1989}, and {\it b)} on the
supposed $\nu/\gamma$ ratio which depends on the unknown thickness of matter
surrounding the source.
 
The uncertainty of expectations is reflected in Table \ref{table-event-numbers} 
which is taken from the DUMAND-II proposal \cite{DUMAND-Project}.

\begin{table}[ht]
{\tiny
\begin{center}
\begin{tabular}{|l|r|r|r|r|r|r|r|r|r|}
\hline
Source & RA & Dec & Dist & $\gamma$ En. & $\gamma$ at Earth & Luminos. & Assumed & 
\multicolumn{2}{c|}{$\mu$/yr in DUM-II}  \\\cline{9-10}
Name & (hh: & (deg) & (kpc) & (TeV)  & (cm$^{-2}$ s$^{-1}$)  & (erg s$^{-1}$) & Spectr. & $\varepsilon_{\nu/\gamma}$=1  
& $\varepsilon_{\nu/\gamma}$=30  \\
& mm) &  & &  &  &   & Index $\gamma$ & Min.$\gamma$  & Max.$\gamma$ \\
\hline
\hline
Vela PSR & 08:33 & -45 & .5 & 5 & $1.8\times10^{-12}$ &
$3\times10^{32}$ & 2.0-3.5 & .1  & 1506 \\
\hline
Vela X-1 & 09:00 & -40 & 1.4 & 1 & $2\times10^{-11}$  & $2\times10^{34}$ &
2.0-4.0 & .2 & 126 \\
\hline
Crab SNR & 05:33 & +22 &  2 & 2 &$1.1\times10^{-11}$  & $2\times10^{34}$  & 2.0-4.0 &
.2  & 438 \\
\hline
Crab PSR &  05:31 &  +21 &  2 &  1 &$7.9\times10^{-12}$  & $6\times10^{33}$  &
2.0-4.0  &  .06 & 38 \\
\hline
Geminga & 06:49 & +18 &  .5-2.1 &  6 & $9.5\times10^{-12}$  & $3\times10^{33}$  &
2.0-3.2 & .49  & 1506 \\
\hline
4U 0115 & 01:15 & +63    & 5  & 1 & $7.0\times10^{-11}$  & $6\times10^{35}$  &
2.0-4.0 & .47  & 273 \\
\hline
Her X-1 & 16:57  & +35  & 5 & 1  & $3\times10^{-11}$  & $3\times10^{35}$  &
2.0-4.0 & .24  & 141 \\
\hline
SS433 & 19:09 & +05 & 5 & 1 & $  <10^{-10}$  & $ <4\times$10$^{35}$  &
2.0-4.0 & $ < .88$ &  $ < 510$ \\
\hline
Cen X-3 & 11:19 & -60  & 5-10 & 1 & $<5.2\times$10$^{-12}$  & $ <2\times$10$^{34}$  &
2.0-4.0 & $ <.08$  & $ <48$ \\
\hline
Cyg X-3 &  20:32 & +41 & $\ge 11$ & 1 &$ 5.0\times10^{-11}$  & $3\times10^{36}$  &
2.1-4.0 & .4  & 234 \\
\hline
LMC X-4 & 05:32  &  -66 &  55 &  $10^4$ & $5\times10^{-15}$  & $1\times10^{38}$  &
2.0-4.0 &  0.0001 &   0.048   \\
\hline
M 31 & 00:41 & +41 & 670 &  1 &$2.2\times10^{-10}$  & $2\times10^{40}$  &
2.0-4.0 & 1.8 & 1050 \\
\hline
Cen A &  13:24 &  -43 &  4400 &  .3 & $4.4\times10^{-11}$  & $3\times10^{40}$  &
2.0-4.0 & .14 & 6 \\
\hline
3C 273 &  00:12 &  +02 &  6$\times$10$^5$ & 5 &$ <9\times10^{-12}$  & $<3\times$10$^{45}$  &
2.0-3.3 & $ <.4$ & $ <1506$ \\
\hline
\end{tabular}
\caption{
Tabulation of various $\gamma$-ray sources used for neutrino estimates
(taken from \cite{DUMAND-Project}).The last two columns give the
number of upward muon neutrino events per year in DUMAND-II.
The minimum numbers are obtained for a   $\nu/\gamma$ ratio of 1
and flat gamma source spectra (Min\,$\gamma$), the maximum  numbers for
$\nu/\gamma = 30$ and steep gamma spectra (Max\,$\gamma$). %(typically $E_{\nu}^{-2}$)
The spectral index $\gamma$ is the differential one.
}
\label{table-event-numbers}
\end{center}
}
\end{table}

%\begin{figure}[ht]
%\sidecaption
%\includegraphics[width=13cm]{figs/Expectations-1988.pdf}
%\caption{
%\label{Expectations-Table}
%\end{figure}

One sees that pessimistic and optimistic numbers differed by 2-3 orders of
magnitude and left it open whether DUMAND-II would be able to detect
neutrino sources or whether this would remain the realm of a future cubic kilometer
array.  Two years later, V.\,Berezinsky reiterated his earlier estimates that
for neutrinos from a fresh neutron star a detector with an effective
area of 1000\,m$^2$ (i.e. a large underground detector) would be sufficient,
but that the detection of extragalactic sources would require detectors
of 0.1-1.0 km$^2$ size \cite{Berezinsky-1990}. DUMAND-II, with 25,000 m$^2$ area, fell just
below these values. Again citing A.\,Roberts \cite{DUMAND-Roberts}:
\begin{quotation}
These calculations 
serve to substantiate our own gut feelings. I have myself watched the progression
of steadily decreasing size ... at first with pleasure (to see it become more practical),
but later with increasing pain. ... The danger is, that if DUMAND II sees no
neutrino sources, the funding agencies will decide it has failed and, instead of
expanding it, will kill it.
\end{quotation}

In 1980, a feasibility study was started and supported by the University Hawaii.
As a part of the study, a series of deep-ocean experiments was performed
with the aim to test the detection concepts and measure the relevant 
water parameters. Actually, first site surveys had started already in 1978.
For the first time, the attenuation length of sea water was measured with
reasonable precision and the ambient light noise determined, including
the study of stimulated bio-luminescence. 
Tests and investigations made during many cruises pushed ahead of
oceanographic practice at that time. However, the ocean turned out to be a hostile
place: in 1982 a test string was lost as its support cable broke in heavy seas,
in 1984 another
string failed to be released from its anchor at the bottom of the sea (but was 
recovered 18 months later). 

After the initial series of site surveys and technical tests,  in 1983 the Department of Energy (DOE) 
approved the funding for DOE-supported U.S. groups to deploy the "Short Prototype
String" (SPS). With additional support from NSF, ICRR in Japan and the
University of Bern in Switzerland, the SPS was conceived to
develop and test the basic detector techniques, to further study the
environmental effects, to demonstrate that muons can be reconstructed 
and to measure the muon vs. depth dependence.
In 1987, this 7-phototube test string was successfully deployed for some hours from a 
a high-stability Navy vessel 
\cite{DUMAND-Babson}. It provided the measured muon intensity as a function of
depth. 
Ironically, this most sustaining physics result from DUMAND  returned, in a sense, 
to the initial idea of the 1973 ICRC.

After the successful SPS test, in 1988 the DUMAND-II proposal was submitted to DOE and NSF.
The collaboration groups of this proposal were: UC Irvine, CalTech, U.\,Hawaii, Scripps
Inst.\,Oceanology, U.\,Vanderbilt, U.\,Wisconsin (USA), 
U.\,Kinki, ICRR Tokyo (Japan), TH Aachen, U.\,Kiel (Germany)
and U.\,Bern (Switzerland). 
DUMAND-II with its 100\,m diameter and 230\,m height, would have detected
three down going muons per minute and about 3500 atmospheric neutrinos
per year.

\subsection{Technological solutions}

A wealth of technological solutions was found within the design study for the SPS,
and many  remained as a legacy for other neutrino telescopes. 
Some of the innovative  
solutions were only possible since basic  technologies had only 
recently appeared on the market. Here I highlight two of them: {\it a)}
the development of a highly sensitive photo-sensor and {\it b)} the development
of fast data transmission using optical fibers.

The photomultiplier tube (PMT)
had to be large (to collect much light), fast (to allow for
fast timing and good muon angular resolution) 
and to have a good amplitude resolution (allowing identification
of the 1-photoelectron (PE) peak and separation from noise and 
possibly from the 2-PE signals). 1-to-2 PE separation had the potential
to separate hits due to light from $^{40}$K decays (nearly entirely 1-PE)
from those from charged muons and is accomplished
by PMTs with a high-gain first dynode. 

\begin{figure}[ht]
\sidecaption
%\begin{center}
\includegraphics[width=6cm] {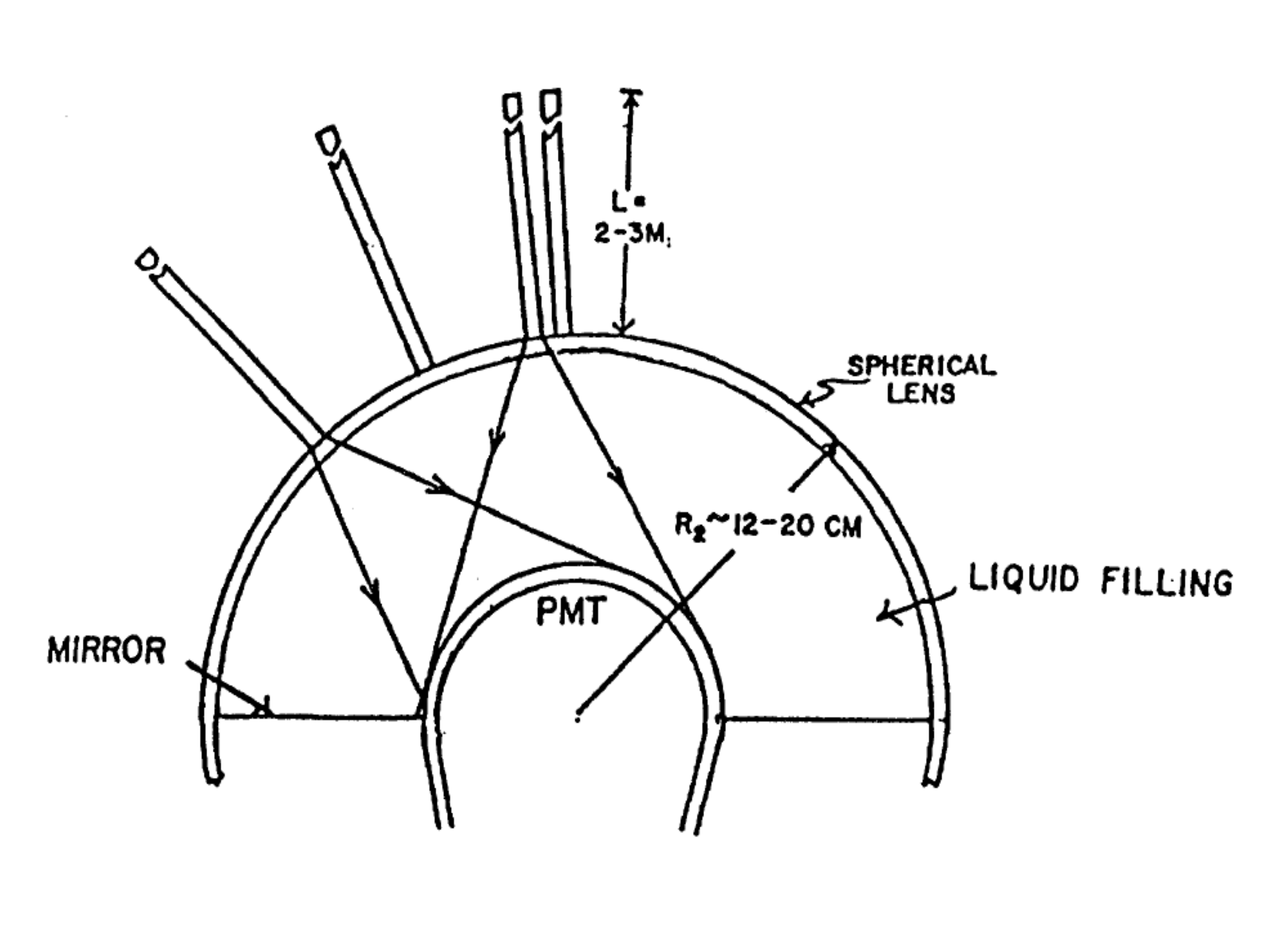}
%\caption{ 
\caption{The Sea-Urchin (Figure taken from \cite{DUMAND-Roberts}).
See text for explanation.
}  
\label{Urchin}
\end{figure}

A first approach to collect a maximum amount of light -- long before
the SPS phase -- 
was the "Sea Urchin": a medium-sized PMT was foreseen to be centered
in a commercially available 17-inch glass pressure sphere filled with liquid. 
Spines made from radial
glass tubes about 2\,cm diameter and 3\,m length and doped with wave length
shifter would guide the light to the sphere -- see Fig.\,\ref{Urchin}. 
However, such a design is
highly fragile and turned out not to be feasible for underwater operation. 

Various innovative attempts were made but eventually discarded
after Hamamatsu Comp. (Japan) committed to develop
a spherical 15-inch PMT R2018, following a computer-simulated design
developed by J.\,Learned. This PMT fitted
into a  17-inch commercial pressure sphere which was
tested down to 6700\,m depth (Benthos company, USA). 
The PMT was embedded in silicone gel and shielded by a 
mu-metal grid against the Earth magnetic field.
The only other practicable design for a light sensor was the
PHILIPS "smart" photomultipler XP2600  \cite{Philips-1983}.
This PMT had a better amplitude resolution and was rather 
insensitive against the Earth magnetic field. 
I will sketch its operation principle  
in the context of the Baikal neutrino telescope, where a similar tube 
was developed.

Another example for applying brand-new techniques is the use of optical fibers
for data transmission.
The processed signal is sent as optical pulse through a multi-mode fiber cable.
Fibers for use in undersea cables had been become available just in the late
seventies. This was  the second fortunate event with remarkable consequences 
since it removed the
low-data-rate barrier imposed by shore cables with copper lines of 40\,km length.

To feed the optical signal through the glass sphere, a special
penetrator was designed by the collaboration since no connector matching
the requirements was on the market
(production by Diamond company, Switzerland).  Remarkably, a decade later
this connector had a revival, when the AMANDA experiment decided to replace 
the analog data
transmission via electrical cables by optical fiber transmission (see below).

\subsection{Acoustic detection of neutrinos}
\label{subsec-acoustic-DUMAND}

\begin{figure}
\sidecaption
\includegraphics[width=4.5cm]{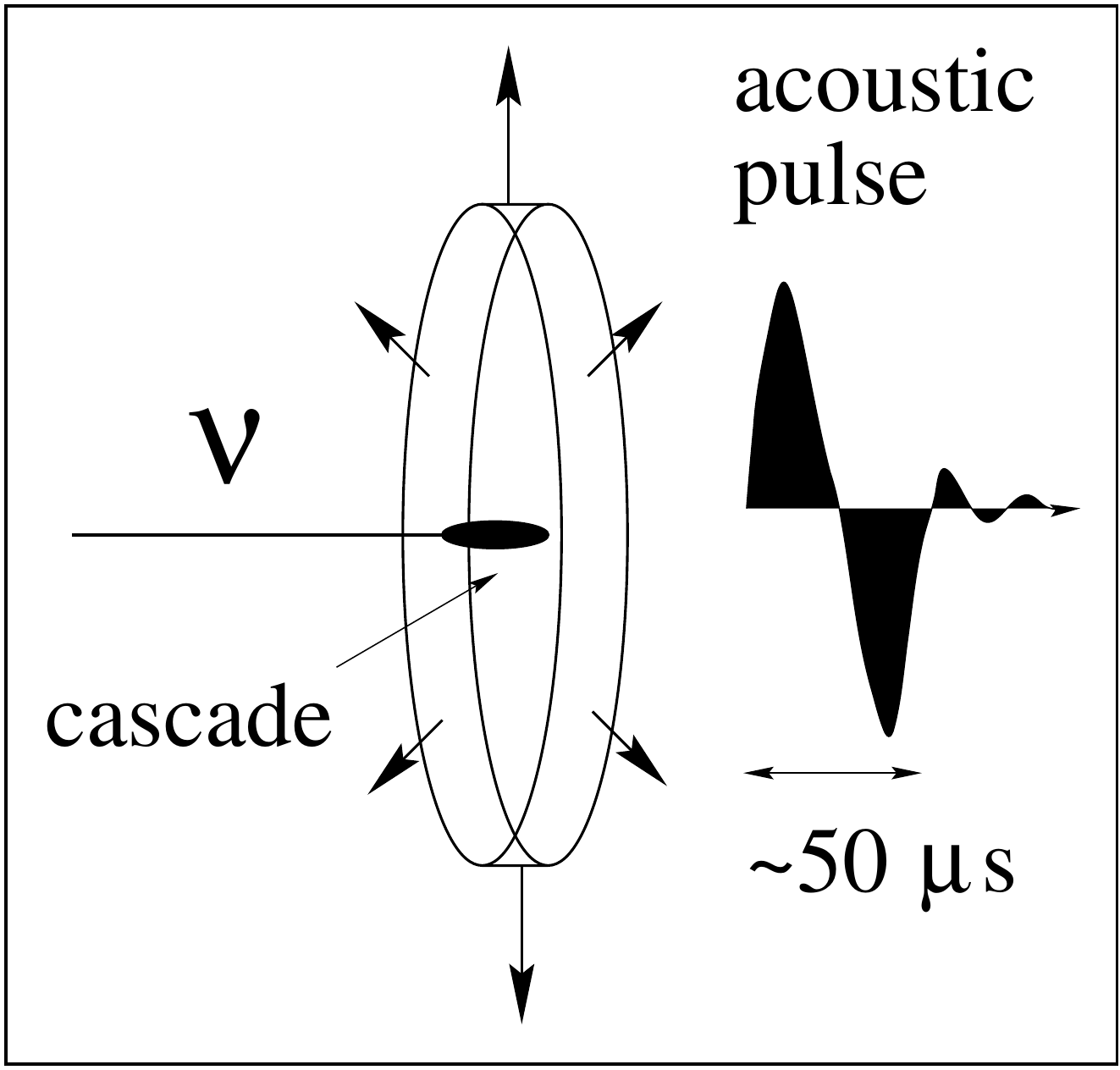}
\caption{
Acoustic emission by a particle cascade.}
\label{acoustic}
\end{figure}

In 1957, the Soviet physicist Gurgen Askaryan had predicted the production 
of pressure waves by charged particles depositing energy in
liquids or in solid media \cite{acoustic-Askaryan-1957}. 
Already at the
second DUMAND Workshop {DUMAND-1976} it was discussed whether
this method could complement the detection of charged particles 
via their Cherenkov light.
In the case of a particle cascade, its entire
energy is deposited into the medium, mostly through ionization, and converted to
heat on a time scale that is very short compared to the typical time scales
relevant for generation and propagation of acoustic pulses. The effect is a fast
expansion, generating a bipolar acoustic pulse with a width of a few ten
microseconds in water or ice (see Fig.~\ref{acoustic}), corresponding to a peak
signal power at 20\,kHz. Transversely to the pencil-like cascade, the acoustic
pulse propagates into the medium within a disk-shaped volume with a thickness
corresponding to the cascade length of about 10\,m.  

Since the attenuation
length of sound in water is larger than that of light, the hope was that with
comparably cheap hydrophones much larger volumes could be instrumented  than with
expensive photomultipliers. Therefore already in 1977, a special "Workshop on Acoustic 
Detection of Neutrinos" was organized in La Jolla \cite{DUMAND-1977}.
Two years later, in 1979, the principle was experimentally proven with high-intensity proton 
beams at accelerators mimicking particle cascades \cite{acoustic-Learned-1979}. 
However, it was rather quickly concluded that the 
acoustic method while possibly offering great volumes, had thresholds
in the range of 100 PeV, higher 
than the only guaranteed source, namely atmospheric neutrinos.
So no signal at all would be likely.  

Since the late seventies,
signal predictions have been considerably scaled down. Adding to this the
meanwhile better understood backgrounds of the ambient and
intrinsic noise, today the threshold for a reasonable spacing 
is estimated in the range of several EeV ($10^9$ GeV), see section \ref{new-methods}.

\section{The evolution of the Baikal project}
\label{sec-baikal}

\subsection{The first years}
\label{sec-first-years}

Russian participation in the DUMAND project was strong from the beginning and
represented by names like A.\,Chudakov, V.\,Berezinsky, L.\,Bezrukov, B.\,Dolgoshein,
A.\,Petrukhin and I.\,Zheleznykh.  In 1979, the DUMAND Workshop
was held in Kha\-ba\-rovsk and at Lake Baikal \cite{DUMAND-1979}.
However, in the context of the Soviet
invasion in Afghanistan, in 1980 the Reagan administration terminated the cooperation.
As A.\,Roberts remembers \cite{DUMAND-Roberts}:
\begin{quotation}
The severing of the Russian link was done with
elegance and taste. We were told, confidentially, that while we were
perfectly free to choose our collaborators as we liked, if perchance
they  included Russians it would be found that no funding was
available.
\end{quotation} 

Although A.\,Chudakov 
(together with F.\,Reines, O.\,Allkover, S.\,Myaki, J.\,Learned and M.\,Shapiro) 
still signed an open letter from July 26, 1980 which declares the formation of
a DUMAND collaboration board \cite{Markov-1993}, it was clear that Russia had to find its own way.

Also in 1980, Chudakov proposed to use the deep water of Lake Baikal in Siberia
as the site for a "Russian DUMAND".  The advantages of Lake Baikal seemed
obvious: it is the deepest freshwater lake on Earth, with its largest depth
at nearly 1700 meter, it is famous for its clean and transparent water,
and in late Winter it is covered by a thick ice layer which allows 
installing winches and other heavy technique and deploying  
underwater equipment without any use of ships. 

In 1981, first shallow-site experiment with small PMTs started.  
The chair of a dedicated laboratory at the Moscow Institute of Nuclear Research,
Academy of Science of USSR (INR) became Grigorij Domogatsky, a theoretician,
flanked by Leonid Bezrukov as leading experimentalist.

Soon a site in the Southern part of Lake Baikal was identified as suitable.
It was about 30 km South-West from the outflow of Lake Baikal into
the Angara river and approximately 60 km from the large city Irkutsk.  A site at a distance of
3.6\,km to shore and at a depth of  about 1370\,m was identified as the
optimal location for a detector which would be
installed at a depth of about 1.0-1.1 km.  Detectors could be installed
in a period between late February and early April from the ice cover,  and operated
over the full year via a cable to shore.

In the US, these efforts were noticed but obviously not understood as
a competition to DUMAND. V.\,Stenger, who was the leading Monte-Carlo expert
of the DUMAND project, repeatedly expressed his doubts that one could
separate neutrinos from background in Lake Baikal. He argued that the lake was too shallow
and the background of downward-going muons much too high. Therefore,
the necessary cuts to reject the background would inevitably also 
strongly diminish the signal of upward going muons from neutrino interactions,
with the exception of rather small and dense arrays. 

\begin{figure}[ht]
%\center{
\sidecaption
\includegraphics[width=7cm]{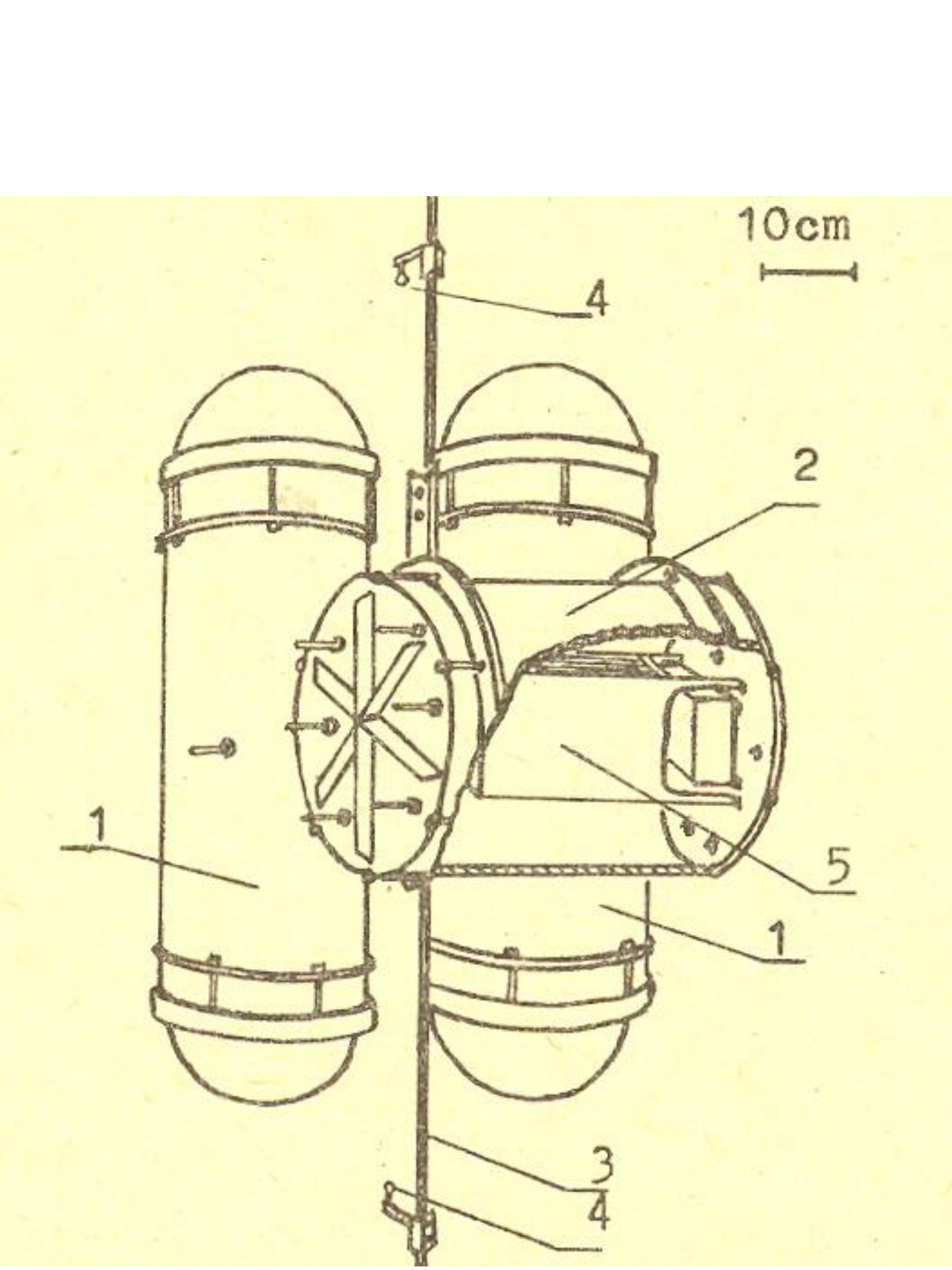}
%\caption
%\caption{ A 
\caption{A {\it "svjaska"} (Russian for "bundle") as used in GIRLANDA-86 for the monopole search,
with 4 PMTs in 2 cylinders (1) covered by hemispheric plexiglass caps.
The "backpack" module (2) contains electronics (5). Calibration LEDs (4) are
fixed at the geophysical cable (3). 
}
\label{svjaska}
\end{figure}

After operation of first underwater modules with a 15-cm PMT 
in 1982, in the following year a small {\it string} was operated for several days. 
In 1984, a first {\it stationary} string was deployed (Girlanda-84) and
recorded downward moving muons \cite{Baikal-1984}. It consisted of
three floors each with four PMTs in two pressure-tolerant cylinders
of glass-fiber enforced epoxy. At that time, no pressure tight
glass spheres where available in the USSR. The end of the cylinders
were closed by caps of plexiglass (see Fig.\,\ref{svjaska}).
The PMT was a Russian tube (type FEU-49)
with a 15\,cm flat photocathode and modest
amplitude and time resolution. An electrical cable connected the
string to the shore.

%\vspace{-1cm}

The 1984 string was followed by another stationary string in 1986 
(Girlanda-86). Data from this string were used to
search for slowly moving bright particles like magnetic monopoles 
\cite{Baikal-1986}.
These particles had been introduced by P.\,Dirac
in 1931 \cite{Dirac-1931}. In the seventies they
%After decades of unsuccessful monopole searches at
%accelerators and in cosmic rays, the efforts gained new momentum when monopoles
turned out to be a consequence of most variants of Grand Unified Theories (GUTs).
A phase transition in the early Universe might have filled the Universe with a 
significant amount of monopoles.
Typical GUT versions predict monopoles with masses $10^{16}$ GeV and more. 
These monopoles would have typical velocities of
$v/c = 10^{-4} - 10^{-3}$. They might catalyze baryon decays along their path
\cite{Rubakov-1981} which could be detected via the Cherenkov light from the decay particles. 
%Girlanda-86
%consisted of six floors at the rather large vertical distance of 50\,m. It was optimized to
%detect tracks of non-relativistic particles much brighter than muon tracks.
Other slowly moving exotic particles would be {\it nuclearites} (or "strangelets'') ,
hypothetical aggregates of $u$, $d$ and $s$ quarks combined with electrons
%to
%adjust electric neutrality. They might be stable for baryon numbers ranging from
%those of usual nuclei up to those of neutron stars ($A\approx10^{57}$)
\cite{Rujula-1984}. Nuclearites could have been produced in the
primordial Universe or in certain astrophysical processes like the collision of
neutron stars. They would induce a thermal shock wave along their path,
the heated medium would emit Planck radiation and the effect would be a similar 
light pattern as for GUT monopoles. Data from Girlanda-86
set stringent limits on the flux of magnetic monopoles with high catalysis 
cross section \cite{Baikal-1986} and  on the flux of nuclearites \cite{Bainuc}.

\subsection{Towards NT200}

Girlanda-84 took data for a total of 50 days and then sank down due to leaking buoys
which held the string in vertical position. But also the cable penetrators through
the epoxy cylinders as well as the cap-to-cylinder hermetic connection
tended to leak and were a notorious source of headaches. Moreover, it was clear that 
the PMT used was much too small and too slow for a neutrino telescope. Therefore 
a technology using glass spheres and a new type of photo-sensor were developed. 

\begin{figure}[ht]
\sidecaption
\includegraphics[width=6cm]{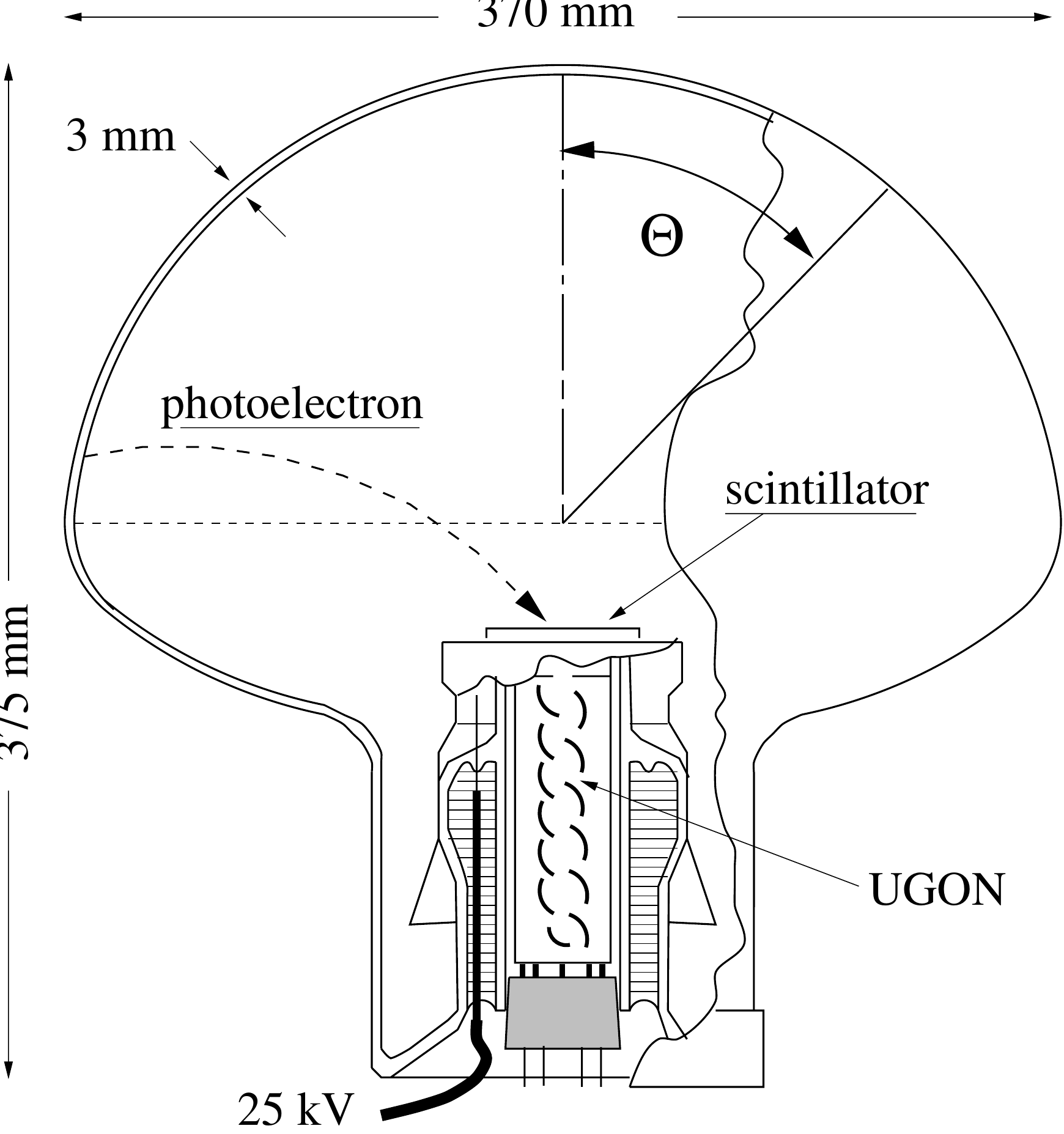}
%\hfill
%\epsfig{file=figs/OM-Baikal.pdf,width=9cm}
\caption{   %\texttt{
The QUASAR-370 phototube}   %}
\label{Quasar}
\end{figure}

The Russians started with testing the "smart" XP2600 from PHILIPS (see section \ref{sec-dumand}) 
in Lake Baikal \cite{Bezrukov-1988a}.  In parallel, the development of
an equivalent Russian device, the QUASAR, was tackled, in cooperation with 
the EKRAN company in Novosibirsk. 
The QUASAR (Fig.\ref{Quasar}) is a hybrid device similar to the PHILIPS 2600
developed for the DUMAND project.
Photoelectrons from a 370\,mm diameter cathode (K$_2$CsSb)
are accelerated by 25\,kV to a fast, high-gain scintillator 
placed near the center of the glass bulb. The light from the scintillator is
read out by a small conventional photomultiplier (type UGON). One photo-electron
from the hemispherical photocathode yields typically 20 photoelectrons in the
small photomultiplier. This high multiplication factor results in an
excellent 1-PE resolution, clear distinction between 1-PE and 2-PE pulses,
a time jitter as small as 2\,ns and negligible sensitivity to the
Earth's magnetic field \cite{Baikal-OM-Bagduev-1999}.

In 1988, the Baikal experiment was approved as a long-term direction of research
by the Soviet Academy of Sciences and the USSR government which included
considerable funding. A full-scale detector (yet without clear definition of its size)
was planned to be built in steps of intermediate detectors of growing size.
In the same year 1988, the author's group from the East German Institute of High Energy Physics
(IfH) in Zeuthen joined the Baikal experiment
 \footnote{As a result of the German unification, in
1992 IfH became part of
the Deutsches Elektronensynchrotron, DESY.}.

After German unification in 1990,
the Zeuthen group had access to the Western market and contributed with Jena glass
spheres and some underwater connectors to the strings which were deployed
in 1991 to 1993 (see below).
In parallel, Russian spheres were developed in collaboration with industry,
as well as penetrators and connectors which tolerated water depths down to
2\,km -- not suitable for large-depth Ocean experiments but 
sufficient for Lake Baikal.

In 1989, a preliminary version of what later was called the NT200 project 
\cite{baikal-web} was
developed, an array comprising approximately 200 optical modules.  The
final version of the project description was finished in 1992 \cite{Baikal-Project}.
At this time, the participating groups came from INR Moscow, Univ.\,Irkutsk,
Moscow State Univ., Marine Techn.\,Univ.\, St.\,Petersburg, Polytechnical Institutes in
Niszhni Novgorod and Tomsk, JINR Dubna, Kurchatov Inst.\,(Moscow),  Limnological Inst.\,
Irkutsk (all Russia), DESY-Zeuthen (Germany) and KFKI Budapest (Hungary).

\begin{figure}[ht]
%\sidecaption
\hspace{1cm}
\includegraphics[width=13cm]{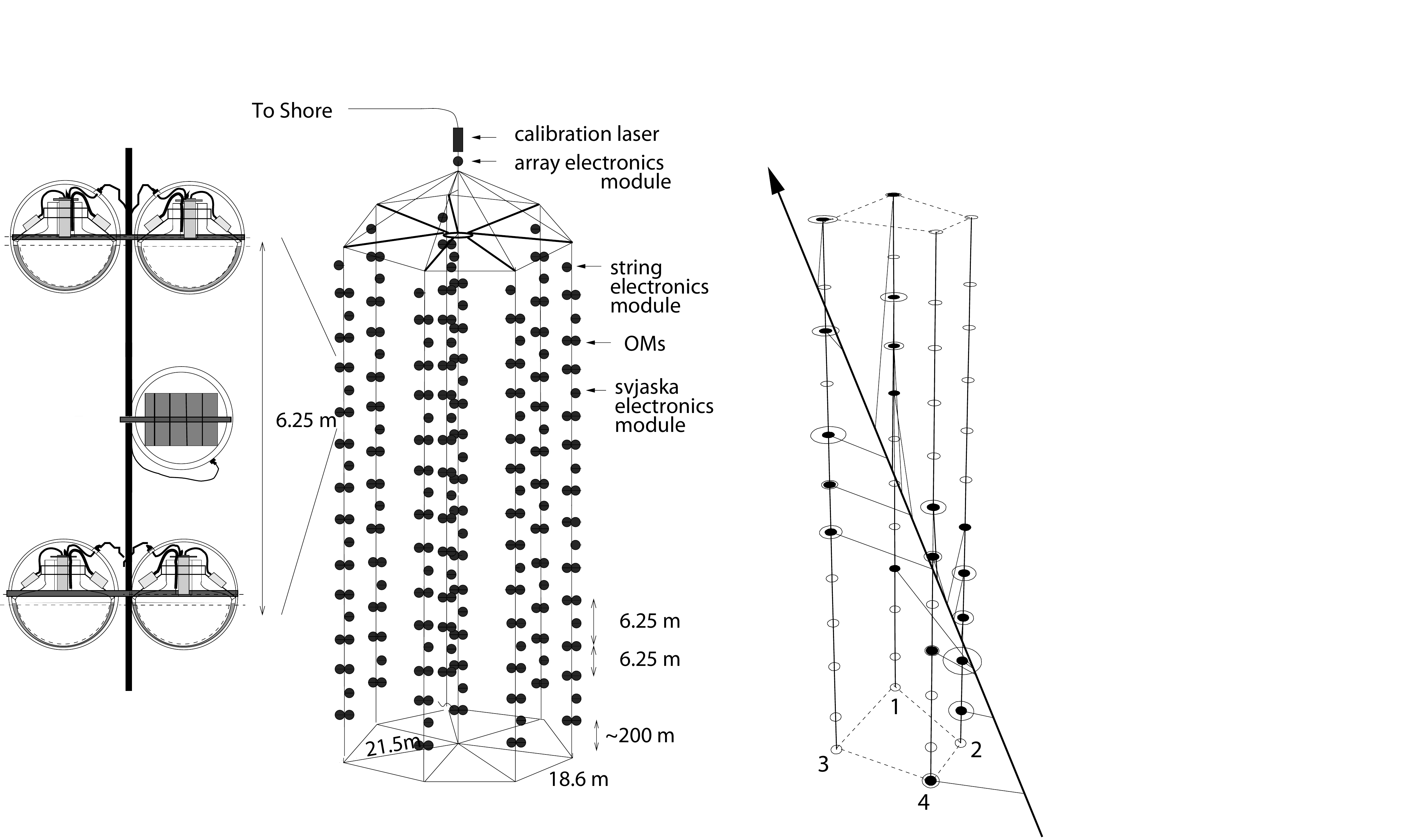}
\caption{    %\texttt{
Left: The Baikal Neutrino Telescope NT200. Right: One of the first upward moving
muons from a neutrino interaction recorded with the 4-string stage of the
detector in 1996 \cite{Baikal-atm-Balkanov-1999}. The Cherenkov light from the
muon is recorded by 19 channels.}  %}
\label{Baikal}
\end{figure}

NT200 (Fig.\,\ref{Baikal},\,left) is an array of 192 optical
modules carried by eight strings which are attached to an umbrella-like
frame consisting of 7 arms, each 21.5\,m in length. The strings are anchored by
weights at the lake floor and held in a vertical position by buoys at various
depths. The configuration spans 72\,m in height and 43\,m in diameter.
The finely balanced mechanics of this frame, with all its buoys, anchor weights and 
pivoted arms is another stunning feature of the Baikal experiment.   
The detector is deployed (or hauled up for maintenance) within a period of about
6~weeks in February to April, when the lake is covered with a thick ice layer
providing a stable working platform. It is connected to shore by several copper
cables on the lake floor which allow for operation over the full year.
 
The optical modules with the  QUASAR-370 phototubes 
are grouped pair-wise along a string. In order
to suppress accidental hits from dark noise (about 30\,kHz) and bio-luminescence
(typically 50\,kHz but seasonally raising up to hundreds of kHz), the two
photo-multipliers of each pair are switched in coincidence.
%, defining a {\it channel}, with typically only $100$\,Hz noise rate.  
The time calibration 
is done using several nitrogen lasers in pressure-tight glass cylinders. 

The construction of NT200 coincided with the decay of the USSR and an
economically desperate period. Members of the collaboration and even some 
industrial suppliers had to be supported by grants from Germany; nevertheless
many highly qualified experimentalists left the collaboration and
tried to survive in the private sector. Over a period of three years, a large part of
the food for the winter campaigns at Lake Baikal had to be bought in Germany and
transported to Siberia.
%(another story of bad surprises and adventures). 
Still, a nucleus of dedicated Russian physicists heroically continued to work
for the project. Under these circumstances, 
the construction of NT200 extended over more than five years.
It started with the deployment of a 3-string array \cite{NT-36} with 36 optical modules
in March/April 1993.
% (three strings are the minimum for full spatial reconstruction).
%At this time, the DUMAND collaboration was working towards installation
%of their first three strings, whereas the AMANDA collaboration was planning
%to deploy four strings at the South Pole in late 1993. 
The first two upward moving muons, i.e. neutrino candidates, were 
separated from the 1994 data.
%\cite{Domogatsky-1996}.
In 1996, a 96-OM array with four NT200 strings was 
operated \cite{Baikal-atm-Balkanov-1999} and provided the the first  textbook neutrinos 
like the one shown in  Fig.\ref{Baikal}\,(right).

%\vspace{1cm}

NT200 was completed in April 1998 and has taken data since then. 
The basic components have been designed and built in Russia, notably the
mechanics of the detector, the optical module, the underwater
electronics and the cabling. The non-Russian
contributions all came from DESY: the laser time calibration system, a transputer
farm in the shore station for fast data processing, an online monitoring system
%\cite{thomas-thesis,shore} 
and a special underwater trigger tailored
to register slowly moving very bright particles (as GUT monopoles), 
not to mention the supply of Western electronics and glass spheres.

The small spacing of modules in NT200
leads to a comparably low energy threshold of about 15\,GeV for muon detection. 
About 400 upward muon events were collected over 5~years. This comparatively
low number reflects the notoriously large number of
failures of individual channels during a year rather than
what would correspond to the effective area. Still, NT200 could
compete with the much larger AMANDA for a while by searching for high energy
cascades {\it below} NT200, surveying a volume about ten times as large as NT200
itself \cite{Baikal-diff-Aynutdinov-2006,Baikal-diff-Avronin-2009}. In order to improve pattern
recognition for these studies, NT200 was fenced in 2005--2006 by three
sparsely instrumented outer strings (6 optical module pairs per string). This
configuration is named NT200+ \cite{Baikal-NT200+}, but suffered from several
problems (from the new strings as well as from the meanwhile
antiquated NT200 itself), so that no satisfying statistics and no convincing results 
have yet been obtained.

\section{The Venice Workshops and the 1992 Hawaii Workshop}
\label{sec-Venice}

In 1988, Milla Baldo-Ceolin from University Padova
organized the first "International Venice Workshop on 
Neutrino Telescopes", a conference series which developed to a discussion forum
for new ideas and results in the field and a valuable archive of the stormy
development of this branch of physics.  

%This first workshop was still under the immediate impression of 
%the detection of neutrinos from supernova SN1987A which was
%discussed by A.\,Mann from Pennsylvania University, a member
%of the Kamikokande collaboration. 
At the first meeting, R.\, March from Wisconsin explained a brand-new idea
of F.\,Halzen (Wisconsin) and J.\,Learned (Hawaii) to detect neutrinos
in ice (see section \ref{subsec-ice}).
Several talks presented new ideas on neutrino detectors at the surface 
which at that time seemed a promising possibility to avoid the difficulties of
deep water detectors (see section \ref{subsec-surface}).
M.\,Koshiba (Tokyo University) made a strong case for imaging water
detectors like Kamiokande and IMB, to which he attributed a firmer
perspective than to the deep water projects like Baikal and DUMAND
which used no imaging but timing technique.
D.\,Samm from Aachen presented test results from the hybrid
PHILIPS photomultiplier which at the same time already was being improved by
Russian physicists (the QUASAR tube) and then constituted the basic element 
of the Baikal experiment (see sections \ref{sec-dumand} and \ref{sec-baikal}).

Two years later, at the second Venice Workshop, results from 
data taken with the first of six modules the MACRO 
underground experiment in the Italian Gran Sasso Laboratory
were reported. This experiment was going to be, together with
Super-Kamiokande in Japan, the largest underground experiment
(see section \ref{subsec-Underground}).
J.\,Learned (Hawaii) 
and T.\,Gaisser (Delaware) discussed {\it The Future of High Energy
Neutrino Astronomy} \cite{Learned-1990} and {\it Prospects for
High Energy Neutrino Astronomy} \cite{Gaisser-1990}.
While Moisei Markov and
Frederick Reines, the visionary fathers of the field, had retired
from active physics, 
Venyamin Berezinsky, Francis Halzen and  John Learned
became the driving and propagandistic motors of high-energy
neutrino detection. Emerging from a round table discussion
at the 1993 Venice Workshop, Halzen and Learned prepared a paper
{\it High Energy Neutrino Astronomy: Towards a 1 km$3$ Detector}
\cite{Halzen-1993}.
In the summary they write:

\begin{quotation}
Yet, we must acknowledge that those instruments now building for
operation by 1995 \footnote{i.e. Baikal NT200, DUMAND-II and in
addition (see below)
AMANDA at the South Pole and NESTOR in the Mediterranean Sea}
will probably not be large enough to really undertake neutrino astronomy.
It will require another step of about two orders of magnitude to be
well into business, and for that it is not too early for dreams,
plans, and studies.
\end{quotation}

Learned already in his 1990 talk in Venice \cite{Learned-1990}
had pointed to the need
for a cubic kilometer detector to do real astronomy but 
somewhat less reluctant supposed that

\begin{quotation}
... in order to proceed with such grand visions we must have success
in detecting the first astrophysical point sources of high energy neutrinos 
in the shorter term.
\end{quotation}

Over more than two decades, the Venice Workshops gathered hundreds
of physicists to discuss neutrino physics, astrophysics and cosmology.
The strong role of the workshop for the development of
neutrino astronomy cannot be underestimated. 

Another key meeting of the nineties was the Workshop on High-Energy
Neutrino Astrophysics in Honolulu, Hawaii
\cite{Hawaii-1992}. It was attended by more than
sixty experts in the field and was particularly important since 
it included a large number of astrophysicists  which
helped to get a better picture about the expected 
neutrino fluxes from astrophysical sources.

The workshop was focused to extragalactic rather than galactic
sources. J.\,Learned and V.\,Stecker presented the
number of muon and cascade events per year to be expected in
DUMAND-II from unresolved and resolved Active Galactic Nuclei,
see Table \ref{AGN-predictions}. 
The estimates from different models on unresolved AGN 
seem to differ by less than an order of magnitude 
and suggest a certain reliability of the predictions. 
This appears misleading since some models consider
neutrino production in the core of AGN, some in the
jets. Some models assume pion production
in proton-proton collisions, others in proton collisions
with ambient photon fields (p$\gamma$ collisions),
some include both collision types. 
By today, all of these models have been excluded. 

\begin{table}[ht]
\begin{center}
\begin{tabular}{|l|l|l|l|l|l|}
\hline
Source & $E_{\mu} > $ & $E_{\mu} > $   & 
$E_{\mbox{cas}} > $ & $E_{\mbox{cas}} > $ & Model  \\
 & 100 GeV & 10 TeV & 1 TeV & 100 TeV & \\
\hline
\hline
Unresolv. AGN & 154 & 66 & 276 & 264 & Stecker et al. \\
\hline
   & 109 & 23 & 113 & 52 & Protheroe \\
\hline
   &  366 & 75  & 379 &  172 &  Biermann \\
\hline
   &  897 &  148 & 680 & 125 & Sikora \& \\
   & & & & & Begelmann \\
\hline
NGC 4151 & 5.0 & 1.1 & - & - & Protheroe \\
\hline
3C279 & 20.3 & 4.2 & - & - &  $\nu = \gamma$ \\
  & & & & & (extrap. from GRO) \\
\hline
 & 0.054 & 0.013 & - & - & Protheroe \\
\hline
Mkr-421 & 3.2 & 0.67 & - & - & $\nu = \gamma$ \\
 & & & & & (obs. at 1 TeV) \\
\hline
 & 0.8 & 0.19 & - & - &  Protheroe \\
\hline
3C273 & 0.8 & 0.19 & - & - & Protheroe \\
\hline
atm. $\nu$ & 2950 & 22.8 & 3435 & 5 & Volkova \\
\hline
\hline
\end{tabular}
\caption{
The number of muon and cascade events per year expected in
DUMAND-II from unresolved and resolved AGNs, according to the
various independent model calculations reported in  
proceedings of the 1992 Hawaii Workshop.
Also shown are the rates expected from two sources observed
by GRO, on the assumption that the neutrino flux equals the measured
$\gamma$-ray flux (table taken from \cite{Stenger-1992a}, see
there also for references).
\label{AGN-predictions}
}
\end{center}
\end{table}

The table also demonstrates the fact that the ratio of
AGN neutrinos to atmospheric neutrinos improves
with energy since the first have a harder spectrum than 
the latter. According to the Stecker model,
an energy cut of 10 TeV in the diffuse spectrum
would have yield 66 
AGN neutrinos on a background of 23 atmospheric neutrinos.
With the much lower exclusion limits for diffuse fluxes
of the year 2011, this
cut value has moved to the region of some hundred TeV.

The expected fluxes for individual AGN (with the exception
of the GRO extrapolation for 3 C279) are close or below the
atmospheric background in a 2 degree search bin 
(defined by the angular resolution of about 1 degree at 1 TeV).
This fact is reflected
in the 1993 comparatively reluctant formulation of Halzen and Learned 
with respect to the 
capability of the telescopes to come in the mid-nineties.

\section{The community broadens}
\label{sec-broad}

\subsection{Surface detectors}
\label{subsec-surface}

The high threshold required to get a detector working in a hostile environment
such as the deep Pacific or the harsh conditions on the frozen Lake Baikal, resulted
in apparently long preparatory periods of both DUMAND and Baikal.
This led others to think about detectors near surface (for a review see \cite{shallow}). 
The advantages seemed tempting:
much easier access and a less challenging environment. Moreover,
proven techniques like tracking chambers or %imaging 
Cherenkov techniques $\grave{a}$ la Kamiokande could be used.  
The author still remembers the insistence with which Masatoshi Koshiba, the father
of Kamiokade, asked him  
in the mid nineties, at a meeting at the Jet Propulsion Laboratory /Pasadena: 
"Why don't you use the proven
technique to see Cherenkov rings from tracks with the help of  areas of phototubes like
we do?" But the disadvantages
were equally striking: the huge background from muons generated in
cosmic rays showers, and the difficulty to expand these arrays much beyond
a few $10^4$\,m$^2$. Actually, my answer to Koshiba focused to the latter fact.

\noindent
Several instruments were discussed in the late eighties and early nineties, most notably:

\begin{itemize}
\item 
GRANDE (Gamma-Ray and Neutrino DEtector), a dense phototube 
water Che\-ren\-kov array 
of about 30,000 m$^2$ area at 0-50\,m depth, to be installed in Arkansas
\cite{GRANDE}.
GRANDE was proposed by H.\,Sobel, a leading scientist of the IMB 
proton decay detector (see section \ref{subsec-Underground}). It included 
twelve US and one Polish institution.
\item
LENA (Lake Experiment on Neutrino Activities), a large area water Cherenkov
detector, also with a high coverage of the walls with PMTs and the aim
to study neutrino oscillations with atmospheric and accelerator neutrinos
and to  do gamma and neutrino astronomy \cite{LENA}.
\item
NET, a three layer water Cherenkov detector on top of Gran Sasso,
essentially an Italian-French collaboration \cite{NET}. 
\item
PAN (Particle Astrophysics in Norrland), a Swedish initiative to deploy 
a shallow detector in one of the clear
North-Swedish lakes, discarded later in favor of Swedish participation in AMANDA
\cite{PAN}.
\item
SINGAO (Southern Italy Neutrino and Gamma Astronomy Observatory),
a sampling array for air showers.  10,000 m$^2$ of resistive plate
chambers in the center were expected to be able separating
upward going neutrino-induced muons from the huge background of downward muons\,
\cite{SINGAO}. SINGAO was an Italy/UK collaboration. 
\end{itemize}

None of these projects were realized, be it by financial reasons, by 
the failure to convincingly demonstrate the background rejection
capabilities, or since shallow lake water parameters turned out to
be worse than expected. 
However, the legacy of these projects can be found in two successful
gamma detection arrays. That of the  first four projects, all
using Cherenkov detection, in the water pool detector MILAGRO in the US
(now terminated and being replaced by the much bigger HAWC detector
in Mexico); that of SINGAO in the resistive plate chamber detector ARGO/YBJ in Tibet.
See for these detectors for charged cosmic rays and gamma rays the
review articles of K.-H.\,Kampert and A.\,Watson \cite{Kampert-2012}
and E.\,Lorenz and R.\,Wagner \cite{Lorenz-2012}.

\subsection{Underground detectors}
\label{subsec-Underground}

Whereas deep underwater and surface neutrino detectors were plagued with the
problems described in the previous sections, some of the smaller underground detectors
moved from success to success. Remarkably, two of these successes had not been
on the top priority list of the experiments: neutrino oscillations (since the
trust in their existence was low in the eighties) and neutrinos from supernova
SN1987A (since Galactic or near-Galactic supernovae are rare).
Table \ref{table-underground} lists the large underground experiments 
operating between 1978 and 1998
with high energy neutrino detection capability. 

\begin{table}[ht]
\begin{center}
%sidecaption
\begin{tabular}{|l|l|l|l|l|}
\hline
Detector, Location & $\mu$ area & Technique & Primary purposes & Operation  \\
 & (m$^2$) & & & \\
\hline
\hline
KGF, South India & 20  &  ST &  p-decay, atm. $\nu$  & 1980  - 1992 \\
\hline
Baksan, Caucasus &  250  &  LS tanks  &  atm. $\nu$, SN & 1978 -  now \\ 
\hline 
%Artemovsk, Ukraina &  ??  &  LS   &  , SN  & ???  \\ 
%\hline 
IMB, Ohio & 400  &  WC  &  p-decay & 1982 - 1991 \\
\hline 
%HPW, Utah &  100 &  WC  &  p-decay  & 198x - 198x\\
%\hline 
Kamiokande, Japan & 120   & WC  &  p-decay & 1983 - 1996\\
\hline 
NUSEX, French Alps & 10   & ST + Fe   & p-decay  & 1982 - 1998\\
\hline 
Frejus, Alps &  90  &  GT/ST+Fe  &  p-decay & 1984 - 1988\\
\hline 
Soudan I, USA  &  10  & ST+concrete  & p-decay  & 1981 - 1990\\
\hline 
Soudan II, USA  & 100  & DT+concrete  &  p-decay & 1988 - 2001\\
\hline
MACRO, Gran Sasso & 1100   &  LS + ST &  monopoles, CRs & 1988 - 2001\\
\hline
LVD, Gran Sasso &  800  &  LS + ST  & SN $\nu$  & 1992 now\\
\hline
Super-K, Japan &  1000 &  WC  &  p-decay,  solar+atm. $\nu$   &  1996 - now\\
\hline
\hline
\end{tabular}
\caption{Underground experiments with high energy capabilities, 1978-1998 
(modified and updated from \cite{Learned-1990}).  
WC = water Cherenkov, GT = Geiger Tubes, ST = streamer tubes, DT = drift tubes, LS = liquid scintillator}
\label{table-underground}
\end{center}
\end{table}

The first massive detectors (1\,kton or more) of the eighties were primarily 
constructed to detect proton decay, with a lifetime less than $10^{32}$ years,
as predicted by early Grand Unified Theories. Proton decay was not found,
instead  three of the detectors recorded MeV-neutrinos from the supernova SN-1987A
(Kamiokande 11 events, IMB 8 events and Baksan 5 events). 
Moreover, Kamiokande,
together with the radio-chemical experiments ClAr (Homestake mine, USA),
SAGE (Baksan laboratory) and GALLEX (Gran Sasso Laboratory) measured
the solar neutrino flux.
I refer to \cite{McDonald-2004} for an overview on solar and supernova neutrino 
experiments and to \cite{Bahcall-1994} for a compilation of 
historically relevant articles on solar neutrinos.

The solar neutrino data strongly supported the oscillation hypothesis, with the
final confirmation by the SNO experiment in 2001 \cite{SNO-2001}. 
Neutrino oscillations denote the transformations between the three types ("flavors") of neutrinos: 
electron neutrinos ($\nu_e$) muon neutrinos ($\nu_\mu$) and tau-neutrinos ($\nu_\tau$). 
They are only possible for massive neutrinos.
According to this concept, the weak flavor eigenstates
$\nu_e, \nu_\mu, \nu_\tau$ are linear combinations of mass
eigenstates $\nu_1, \nu_2, \nu_3$. 
For the simplified case of
two flavors $\nu_\mu, \nu_\tau$ and 
two mass eigenstates $\nu_i, \nu_j$ with different masses
%\begin{eqnarray}
% \nu_\mu   =   \nu_2 \cos \theta_{23} + \nu_3 \sin \theta_{23} \\
%   \nonumber 
%  \nu_\tau  = - \nu_2 \sin \theta_{23} + \nu_3 \cos \theta_{23}. 
%   \label{numunutau} 
%\end{eqnarray}
%If the masses $m_2$ and $m_3$ are different, 
quantum mechanical time evolution
of an initial $\nu_\mu$ state induces a non-zero transition probability to
$\nu_\tau$. The survival probability for the muon neutrino is

\begin{equation}
 P(\nu_\mu\to\nu_\mu) = 
 1-\sin^2(2\theta_{ij})\cdot\sin^2\left(\frac{1.27 \Delta m^2_{ij} \cdot L }{E_\nu}\right)\;,
 \label{survival}
\end{equation}
where $L$ (in km) is the distance traveled by the neutrino, $E_{\nu}$ (in
GeV) its energy and $\Delta m^2_{ij} = m_i^2 - m_j^2$ (in eV$^2$). The
three-flavour case is governed by two independent differences of mass squares
and three mixing angles. The best-fit oscillation parameters derived from
present data are:

%\begin{eqnarray}
%\nonumber
%\mid\Delta m_{31}^2 \mid & = 2.4  \times10^{-3}\mbox{eV}^2  \approx|\Delta m_{23}^2|
%\qquad&
%\sin^2 2\theta_{23} & \simeq 1 \\
%\nonumber
%\Delta m_{21}^2   & = 7.65 \times10^{-5}\mbox{eV}^2 
%\qquad&
%\tan^2 \theta_{12}  & = 0.304 \\
%&&
%\sin^2 2\theta_{13} & \le 0.056\;. 
%\label{bestvalues}
%\end{eqnarray}

\begin{eqnarray}
\nonumber
\mid\Delta m_{31}^2 \mid  = 2.4  \times10^{-3}\mbox{eV}^2  \approx|\Delta m_{23}^2| 
%\nonumber
\hspace{1cm}
\Delta m_{21}^2   = 7.65 \times10^{-5}\mbox{eV}^2  \hspace{0cm} \hfill \\
\hspace{-1cm}
\sin^2 2\theta_{23}  \simeq 1 \hspace{1.55cm} \tan^2 \theta_{12}  = 0.304 
\hspace{1.55cm} \sin^2 2\theta_{13} \sim 0.09 \hfill
\label{bestvalues}
\end{eqnarray}

The  mass difference $\Delta m_{21}^2   = 7.65 \times10^{-5}\mbox{eV}^2$
has been derived from solar neutrino experiments. That the difference of the
squared masses itself is known, rather than
only its absolute value, is due to the fact that the neutrinos undergo
not only the vacuum oscillations described by the formula above,
but also matter induced, resonant oscillations in the Sun. 
These oscillations are sensitive to the mass difference and not only to its squared value. 

Concerning the physics with neutrinos above a 100 MeV (i.e. clearly
above the energy of
supernova burst neutrinos), the spectacular observation was the deficit
of atmospheric neutrinos with increasing zenith angles. It pointed to the the second
mass difference $\mid\Delta m_{31}^2 \mid = 2.4  \times10^{-3}\mbox{eV}^2$
which is due to muon neutrinos oscillating to tau neutrinos on their way through
Earth. Interestingly this effect was controversial at the beginning. 

The first detector
which announced hints of oscillations of atmospheric neutrinos, i.e. a apparent deficit of atmospheric
neutrinos, was the Baksan Neutrino Telescope BNT \cite{Baksan-1981}. 
Given the uncertainties on the atmospheric
neutrino flux and the methodical uncertainties of measurement, the
deficit was not taken too seriously at that time. Actually, at the end of the
eighties, BNT was among those underground experiments which declared
{\it no} discrepancy between data and predictions.
Later in the eighties, a deficit was
reported by the water Cherenkov detectors IMB and Kamiokande but not by
Frejus and NUSEX with their sandwiches of iron and tracking detectors. 
In the nineties, however, also tracking devices (Soudan-2 and MACRO) confirmed
the effect. After a thorough study with Super-Kamiokande (with
the first clear data showing muon neutrino rates as a function
of zenith angle and energy, i.e. of $L/E$, see formula \ref{survival}),
 in 1998 it was eventually announced as discovery 
\cite{SuperK-1998}. The values of mass differences and mixing angles
derived from underground experiments were subsequently confirmed and further
iterated by experiments at nuclear reactors and accelerators and resulted in the values
shown in (\ref{bestvalues}) above. 

Massive neutrinos provide the first hint for physics beyond the
original standard model of particle physics. Moreover this phenomenon
 affects the predictions for
neutrinos fluxes from astrophysical objects, starting with the Sun and ending
with high-energy sources like Active Galactic Nuclei.

At  the same time, neutrinos from high-energy astrophysics sources were not
detected. Even the sky-maps from largest detectors with about 1000\,m$^2$ area (MACRO
and Super-Kamio\-kande) did not show any significant excess over atmospheric
neutrinos. 
%Figure \ref{underground-skymaps} reflects the progress over
%thirty years. It shows on the left side the 
%sky map from the 18 KGF neutrino candidates (published 1971) and on the right
%side the sky map from the final sample of 1365 neutrino-induced
%through- going neutrinos in MACRO (2001). The MACRO plot
%also shows the direction of some prominent source candidates.
In particular, none of the positions of notorious candidates such as Hercules-1, 
Geminga, the Crab Nebula, the binary systems SS-433 and Vela.
Also the supernova SN-1987A did not show any indication of a 
high energy signal. 
It had been included according to the prediction
of Berezinsky \cite{Berezinsky-1977} that the ejecta of Supernovae 
(such as that of 1987) can be a strong
source of neutrinos for months up to a few years. 

Seen from today, the search for
sources of high-energy neutrinos with detectors of 1000\,m$^2$ or less appears
to have been hopeless from the beginning, with the possible exception of certain transient
Galactic sources. But when these detectors had been constructed, this knowledge was not
common and the search for point sources appeared as a legitimate (although not
priority) goal.

\subsection{Neutrinos in ice?}
\label{subsec-ice}

In this situation, a new, spectacular idea appeared on stage. 
In 1988,  Francis Halzen from the University of Wisconsin gave a talk at the 
University of Kansas. At this occasion
he was contacted by Ed Zeller, a Kansas glaciologist. Zeller told him about 
a small test array of radio antennas at the Soviet Vostok station, close to 
the geomagnetic South Pole. The Russians were going to test whether secondary
particles generated in neutrino interactions could
be detected via their radio emission. The idea that showers
of charged particles would emit radio signals had been published back
in 1962 by the Soviet physicist  Gurgen Askaryan. 
Together with his colleagues Enrique Zas and Todor Stanev,
% and {\it using computer power not available to our Russian colleagues} 
Halzen  realized that the threshold for this method was
discouragingly high \cite{Halzen-icefishing}. Instead he asked himself whether the optical
detection via Cherenkov light, i.e. the DUMAND principle,
would be also feasible for ice. In \cite{Halzen-dreams} he
remembers: 

\begin{quotation}
I suspect that others must have contemplated the same
idea and given up on it. Had I not been completely ignorant about what was
then known about the optical properties of ice I would probably
have done the same. Instead, I sent off a flurry of E-mail messages to 
my friend John G.\,Learned,
then the spokesman of DUMAND. ... Learned immediately appreciated the 
advantages of an Antarctic neutrino detector.
\end{quotation} 

A few months later,
Halzen and Learned released a paper "High energy neutrino detection in deep
Polar ice" \cite{Halzen-1988}. With respect to the light attenuation length 
they proceeded 
\begin{quotation}
... on the hope that a simple test will confirm the belief
that it is similar to the the observed 25\,m attenuation length for blue to mid UV
light in clear water in ocean basins.
\end{quotation}

Bubble-free ice was hoped to be found
at depths smaller than 1\,km. Holes drilled into the ice 
and filled with water were supposed to 
refreeze
or, alternatively, to be filled with a non-freezing liquid. 

%What astonishes from today's perspective is the 
%proposed configuration of the array.
%Instead of making maximal use of the expensively drilled holes and
%arrange optical detectors closely spaced along a string, they envisaged
%a detector with only three layers of detection, e.g. strings with
%photomultipliers only at 500\,m, 1000\,m and
%1500\,m -- obviously inspired by the few-layer technique discussed for
%shallow detector {\bf (ask them!}. 

Learned sent,
still in May 1988, a "notice of intent" to NSF, asking for support
of a corresponding exploration program, but he got a negative
response: it seemed obvious that he had his hands full with DUMAND.
Halzen is a theorist, and so
both did not proceed  to do an experiment. But the
idea made it to  Buford Price's group at University of California, Berkeley. 
In 1989, two young physicists of the Price group,
Doug Lowder and Andrew Westphal, joined a Caltech group
drilling holes in Antarctic ice and tried to measure the ice
transparency using existing boreholes. It would take, however,
another year until  the first successful transparency measurement of natural
ice was performed  -- this time in Greenland. Bob Morse from
the University of Wisconsin and Tim Miller (Berkeley) lowered photomultipliers
into a 217\,m hole drilled by glaciologists \cite{Greenland-1990}.

In parallel to these first experimental steps, Buford Price, Doug Lowder and Steve Barwick
(Berkeley), Bob Morse and Francis Halzen (Madison) and Alan Watson (Leeds)
met at the International Cosmic Ray Conference in Adelaide and
decided to propose the Antarctic Muon and Neutrino Detection Array, AMANDA.

In 1991 and 1992, the embryonic AMANDA collaboration deployed photomultipliers
at various depth of the ice at the South Pole. Holes were drilled using a hot water
drilling technique which had been developed by glaciologists. 
Judging the
count rate of coincidences between photomultipliers (which are due to
downgoing muons), the light absorption length of the ice was estimated at
about 20\,m and scattering effects were supposed to be negligible \cite{Morse-1993}.
%25\, , which is the attenuation length of crystalline ice grown in the lab. 
It should turn out later, that this was a fundamental misinterpretation
of the rates. But exactly this interpretation encouraged the AMANDA
physicists to go ahead with the project.

\subsection{Neutrinos in the Mediterranean Sea?}
\label{subsec-med}

With ongoing activities in Hawaii and at Lake Baikal and the first ideas on a telescope
in polar ice, the exploration of the Mediterranean Sea as a site for an underwater
neutrino telescope was natural. First site studies along a route
through the Mediterranean Sea were performed in 1989
by Russian physicists who also measured the muon counting
rate as a function of depth \cite{Deneyko-91}.
In July 1991, a Greek/Russian collaboration led by
Leonidas Resvanis from the University of Athens performed a cruise
and deployed a Russian built
hexagonal structure made of titanium and carrying 10 photomultipliers
down to a depth of 4100\,m. The site was close to
Pylos at the West coast of the Peloponnesus. They measured the vertical muon intensity
and the angular distribution of downgoing muons.
This was the start of the NESTOR project \cite{nestor-web}, which was named
after the mythic king of Pylos who counseled the Greeks during the Trojan war.

The advantages of the Pylos site were obvious: The depth can be chosen
down to 5200\,m, dependent on the acceptable distance to shore,
deeper than any other candidate site. This would reduce the
background of downward muons.
% which with a certain
%probability are mis-reconstructed as upward
%muons and therefore fake neutrino interactions. The depths
%even allows looking slightly above horizon since. 
The water quality was
excellent and the bio-luminescence seemed to be lower than
at other Mediterranean sites.

In 1992, the collaboration included the University of
Athens, the Scripps Institute for Oceanography in San Diego (USA), 
the Universities of Florence (Italy), Hawaii, Wisconsin (USA), Kiel (Germany)
and the Institute for Nuclear Research Moscow. 
%The 
%membership of the NESTOR collaboration had a considerable
%fluctuation over the years. 
Naturally, more Greek institutes joined.
Several French Universities joined in 1994 and left again in 1997 to pursue their own
project ANTARES. More Italian groups from Rome and
Florence joined in 1994, but in 1998 also 
decided to follow their own project, NEMO, close to Sicily.
Since Italians had designed and constructed large parts
of the off-shore electronics, this
left the collaboration in a difficult state. Later, the Lawrence
Berkeley National Laboratory, LBNL,
got involved in the design of the electronics,
profiting from parallel developments for IceCube.
They provided essential electronics for the first
stationary hexagonal floor which was deployed in 2004.

\begin{figure}[ht]
\sidecaption
\includegraphics[width=6cm]{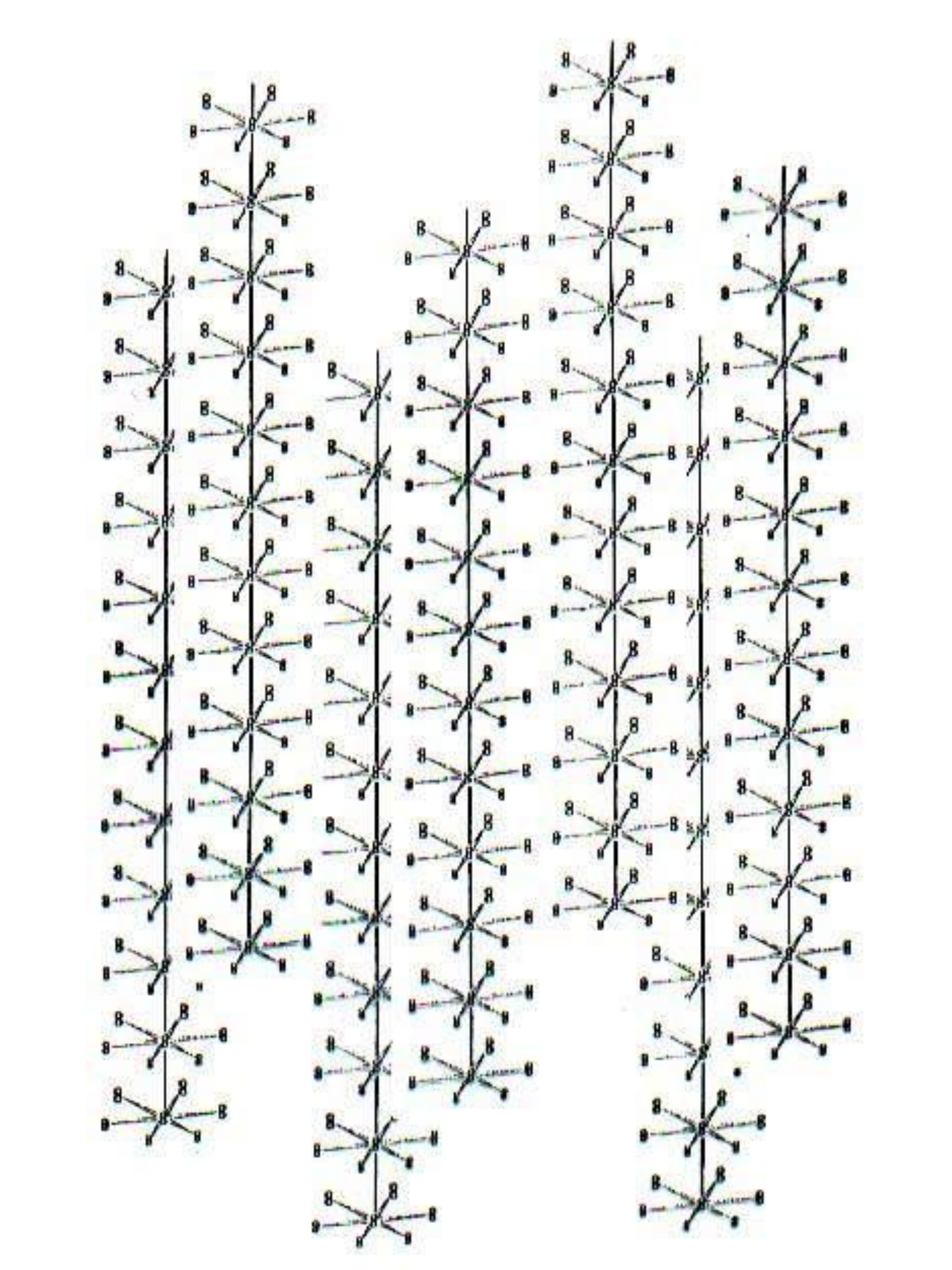}
\caption{     %\texttt{
The original NESTOR concept of a large array of towers, each
carrying 12 floors of six arms. The optical modules are fixed
at the end of the arms, plus one in the center.
   %}
\label{Nestor-towers}.
}
\end{figure}

The results of the early cruises and the concept for the
NESTOR detector were developed and presented during 
a series of Workshops in Pylos. 
NESTOR was conceived to consist of  seven "towers" 
(six on the edges of a hexagon and one in the center)
covering an area of about $10^5$\,m$^2$  as shown
in Fig.\ref{Nestor-towers}.

A single tower should carry 168 PMTs  on 12 hexagonal floors,
vertically spaced by 20-30\,m,
each with six omni-directional modules at the
end of 16\,m arms and one in the center \cite{nestor-1994}. The
"omni-directional module" contained two 15-inch PMTs, one upward,
one downward looking, in two 17-inch  
glass pressure spheres. 

The philosophy of NESTOR was to be not only sensitive to high energy
neutrinos (therefore the large area covered by seven towers) but also to study
atmospheric neutrino oscillations (hence the 5\,GeV
threshold inside the geometrical volume of a tower and
the omni-directionality of the modules).

It is noteworthy to recall that, similar to DUMAND, NESTOR had
a strong initial activity on acoustic neutrino detection which was
led by Igor Zheleznykh from INR Moscow.
The acoustical array was named SADCO
(Sea Acoustic Detection of Cosmic Objects), and was planned to consist
of several hundred autonomous hydrophones spread over
one cubic kilometer \cite{Sadco-Nestor}. However, given the enormous
challenges of NESTOR itself, the acoustic efforts
went dormant after a while.

%\subsection{JULIA}

Nearly forgotten is another effort in the Mediterranean Sea, 
pursued by Peter Bosetti and students
at Aachen \cite{Bosetti-1994}. Their plan was to develop JULIA, a nested
array, with an inner low-energy section (10 MeV threshold), moving to a layer
tuned for GeV energies, and an outer envelope for the TeV-range.
The preferred site was close to Rhodos, abouth 4\,km deep. 
However, the second Gulf war
prevented tests at this site. After a cruise to a site close to the Canaries
(with participation of two Zeuthen scientists) and deployment of
some equipment it became clear that manpower and 
funding were not sufficient to start an own project, and the project
was terminated. 

\section{The three-string race and the termination of DUMAND}
\label{subsec-3string}

In 1993 and 1994, three collaborations were going to deploy detectors with
three or more strings. Three strings are the minimum to achieve
full spatial reconstruction of events.  

The DUMAND collaboration was working
towards installation of the first three of the nine DUMAND-II strings.
Two of these strings were to be equipped with "Japanese Optical Modules"
(JOMs)  containing a 15-inch PMT "R2018" from Hamamatsu, and one 
string with "European Optical Modules" (EOMs)
containing the hybrid XP2600 from Philips
\cite{Wiebusch-1995}. This stage was christened
TRIAD. 

In 1992, the AMANDA collaboration was joined by the Swedish collaborators
from Stockholm and Uppsala. Steve Barwick, meanwhile
at University of California in Irvine, designed a four-string detector
with 80 PMTs which was going to be deployed between 800 and 1000\,m
depth. In Wisconsin, computer-controlled
hot water drills were developed, in close collaboration with the
Polar Ice Coring Office (PICO). 

However, the first three-string array (NT36) was deployed at Lake Baikal
in March/April 1993. 
It consisted of only 18 PMT-pairs at three strings of meager
40\,m length. But it served its purpose to demonstrate that 3-dimensional
reconstruction of muon tracks works as expected.
John Learned from DUMAND recognized the
importance of the Baikal achievement and sent an E-mail
to the author: 
{\it Congratulations for winning the 3-string race!}  Actually
the first two neutrino candidates were isolated from the data
taken with the same array in 1994.

%\subsection{The termination of DUMAND}

Meanwhile, in December 1993, a first string of the TRIAD, together with 
a string of environmental instruments was deployed and linked to shore via a
junction box placed on the ocean bottom and a shore cable which had
been laid a few months earlier. 
However, some pressure housings developed leaks. A short circuit in the junction
box (the central component for communication to shore)
did not clear due to a fuse failure, and soon the communication 
to shore failed \footnote{Deployment, string recovery and 
initial problem
analysis are summarized in \cite{Grieder-1994}}. 

The DUMAND progress had been slow, but had shown  
remarkable progress compared to ocean research at that time. 
This impressed oceanographers but not the
main funding organization, the Department of Energy (DOE), 
which was not used to a "try-and-try-again" mode of progress.  
Review committees without any ocean expertise judged
the project, following criteria typical for accelerator research.
Moreover, the termination of the Superconducting
Super Collider (SSC) by the US congress in 1993
created a strong risk aversion in DOE.
On the technical side, the reasons of the 1993 DUMAND 
failures had been identified and a redeployment was in preparation.
But in 1995, the mentioned circumstances regrettably
led to a termination of  the support for DUMAND. From now on
the goal to begin high energy neutrino astronomy was 
carried forward at the South Pole, in the Mediterranean Sea and 
in Lake Baikal.

\section{AMANDA}
\label{sec-amanda}

AMANDA is located several hundred meters from
the Amundsen-Scott station. Holes of 60\,cm diameter were drilled with
pressurized hot water; strings with optical modules were deployed in the 
water which subsequently refreezes. Installation operations at the South Pole
are performed in the Antarctic summer, November to February. For the rest of the
time, two operators (of a winter-over crew of 25--40 persons in total) maintained
the detector, connected to the outside world via satellite communication.

Already in 1990 in Greenland, the physicists were assisted by Bruce Koci 
who was working for the Polar Ice Core Office (PICO). He then became the
"drill master" of AMANDA. Without Koci, AMANDA
would certainly not have developed so fast and smooth as it did. 
{\it "The success of AMANDA 
rested on two pillars: the Amundsen Scott Station and Bruce Koci."} -- as
Francis Halzen formulated when Koci deceased in 2006.

The first AMANDA array with 80
optical modules on four strings was deployed in the austral summer 1993/94, at depths
between 800 and 1000\,m \cite{Amanda-shallow-1995}.  
Surprisingly, light pulses sent from one string to a neighbored string over 20\,m distance 
did not arrive after the expected 100\,ns, but were considerably delayed. 
The surprise was resolved at the 1994 Venice Workshop
on Neutrino Telescopes. Here, Grigorij Domogatsky informed Francis Halzen about 
results from an ice core extracted at the geomagnetic
South Pole where the Russian Vostok station is located. 
The data demonstrated, that air bubbles which remain from the original firn ice
at the surface did not yet disappear at 1\,km depth. The delay
was due to light scattering on the bubbles. Light would not travel straight
but via random walk and become nearly isotropic after a a few times the distance called
effective scattering length. The effective scattering
length was found to be between 40\,cm at 830\,m
depth and 80\,cm at 970\,m. The scattering by air bubbles trapped in the
ice made track reconstruction impossible. 

A great story might have been over before it really got started. AMANDA seemed to be 
"nothing other than a big calorimeter" -- as 
%one of our international competitors 
Leonidas Resvanis sarcastically commented -- without any real tracking capabilities. This could have been the point to give up the project. Nevertheless, the group from DESY joined. 
We were encouraged by a trend seen in the AMANDA data itself, as well as by ice core data taken at the Russian Vostok station: below 1300 meters bubbles should disappear. 

\begin{figure}[ht]
\sidecaption
\includegraphics[width=6.5cm]{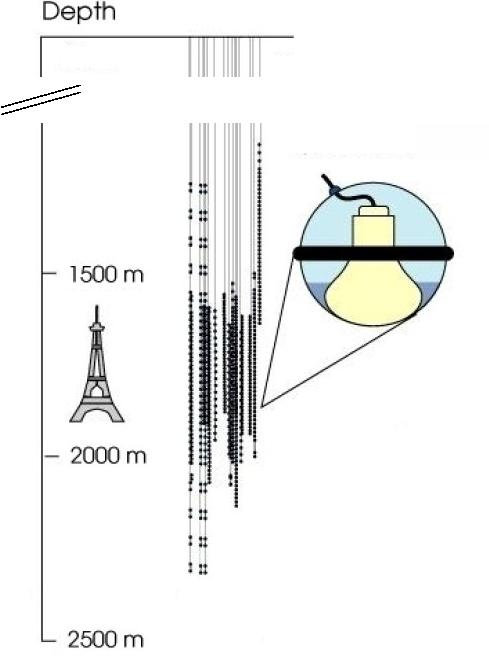}
\caption{    %\texttt{
The AMANDA configuration. The detector consisted of 677
optical modules at 19 strings. Three of the strings have been sparsely
equipped towards larger and smaller depth in order to explore ice properties,
one string got stuck during deployment at too shallow depth and was not used in
analyses. The Eiffel tower is shown to scale for size comparison.   %}
\label{AMANDA}
}
\end{figure}

This expectation was confirmed with a second 4-string array which was deployed in
1995/96. The remaining scattering, averaged over 1500--2000\,m depth,
corresponds to an effective scattering length of about 20\,m and is assumed to be due to dust. 
This is still considerably worse than for water but sufficient for track
reconstruction \cite{Amanda-ice-2006}. The proof that
it indeed {\it was} sufficient took some time, as well as the development
of the suitable reconstruction methods and selection criteria.
The identification of the first two neutrino events from the data
taken with the 4-string configuration(1996) is described in \cite{Andres-1999},
the sophisticated reconstruction and selection methods in \cite{amanda-2004b}.
The average absorption length
at 400\,nm is about 100\,m, much larger than
at water detector sites. The array was upgraded
stepwise until January 2000 and eventually comprised 19 strings with a total of
677 optical modules, most of them at depths between 1500 and 2000\,m.
Figure~\ref{AMANDA} shows the final configuration
of the AMANDA array.

\begin{figure}[ht]
\hspace{1cm}
\includegraphics[width=5cm]{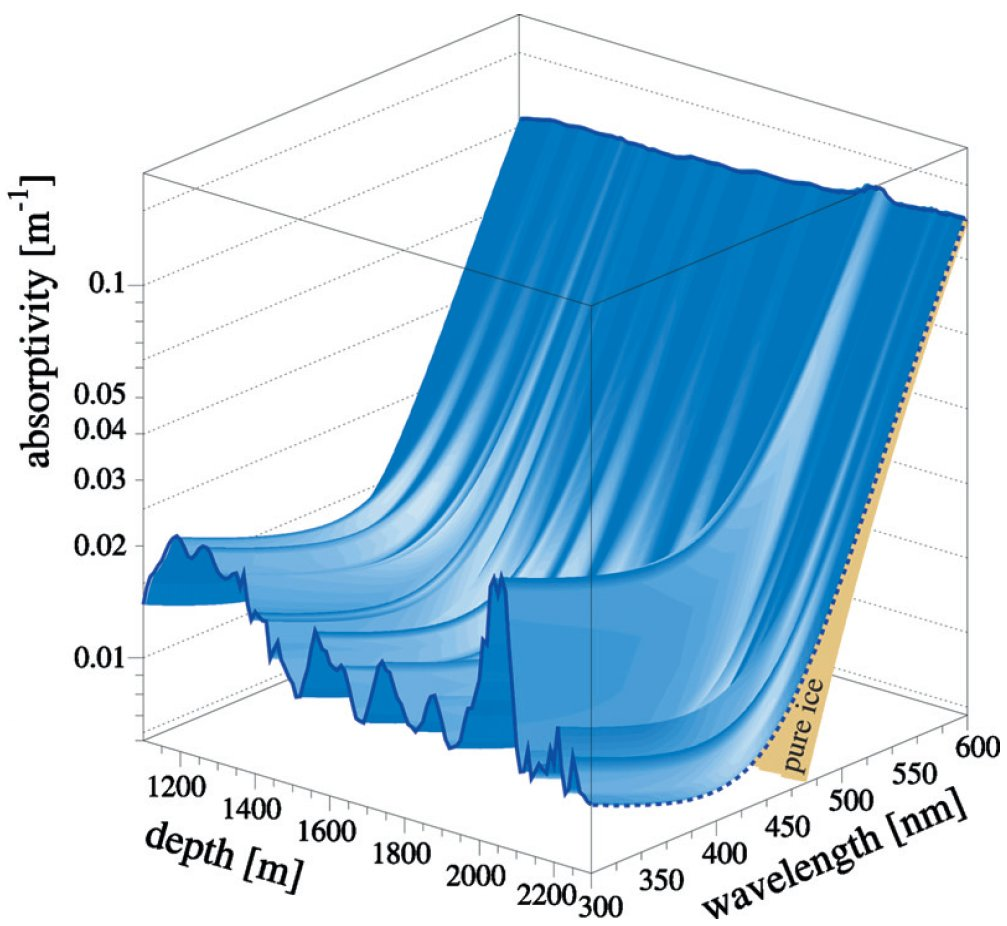}
\hfill
\includegraphics[width=5cm]{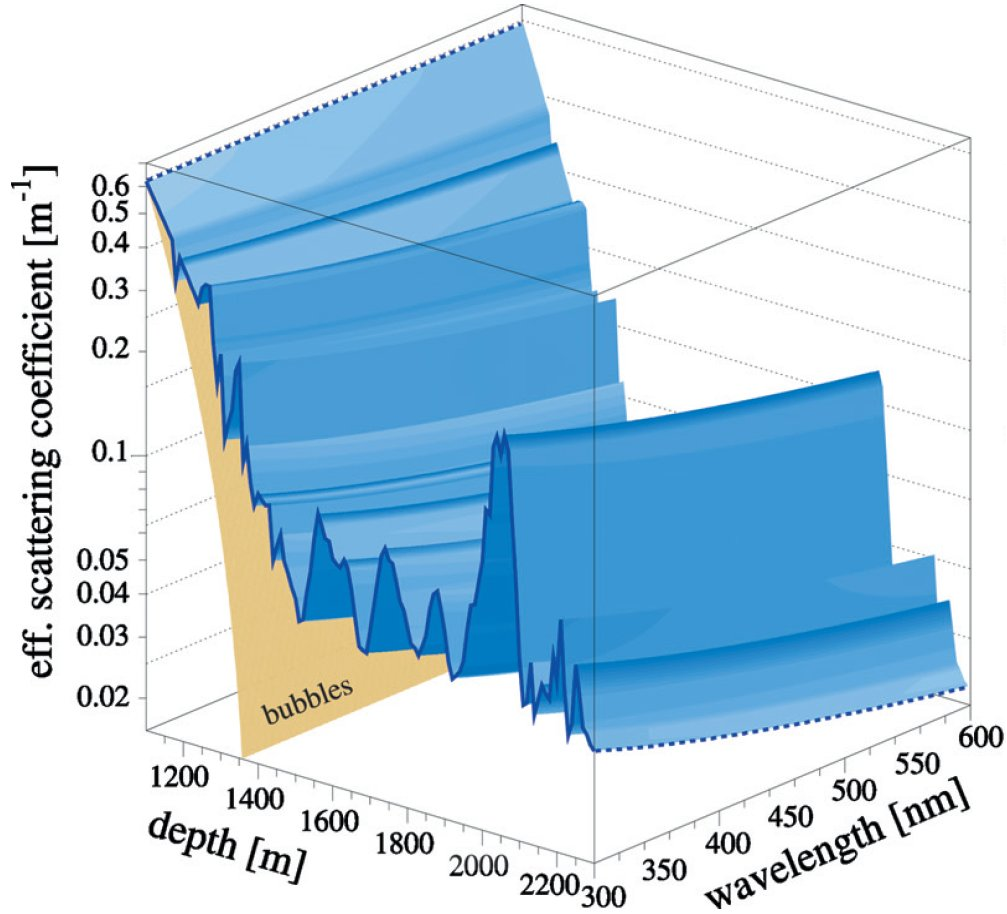}
\hspace{1cm}
\caption{   %\texttt{
Absorption coefficient (left) and scattering coefficient (right) in the South
Polar ice as functions of depth and wavelength  
\cite{Amanda-ice-2006}.
\label{SP-Ice}
}
\end{figure}

In Fig.~\ref{SP-Ice}, absorption and scattering coefficients are shown as
functions of depth and wavelength \cite{Amanda-ice-2006}. The variations with
depth are due to bubbles at shallow depth leading to very strong scattering and,
at depths greater than 1400\,m, to dust and other material transported to Antarctica during
varying climate epochs. The quality of the ice improves substantially below a
major dust layer at a depth of about 2000--2100\,m, with a scattering
length about twice as large as for the region above 2000\,m. The depth
dependence of the optical properties complicates the analysis of the
experimental data. Furthermore, the large delays in photon propagation due to
the strong scattering cause worse angular resolution of deep-ice detectors
compared to water. On the other hand, the large absorption length,
with a cut-off below 300\,nm instead of 350--400\,nm in water, results in better
photon collection.

The relatively short distance between optical modules and surface electronics
allowed for transporting the analogue signals of the photomultipliers to surface
over 2\,km of cable instead of digitizing them in situ. This requires a large
output signal of the photomultiplier, a specification met by the 8-inch R5912-2
from Hamamatsu with 14 dynodes and a gain of $10^9$. The first ten strings used
copper cables for both high-voltage supply and signal transmission; for the last
9 strings the anode signal was used to drive a LED, and the light signal was
transmitted to surface via optical fibers. The feasibility of transmitting 
analog signals via 2\,km optical fibers was all but obvious at the beginning.  
The idea is due to Albrecht Karle, at that time postdoc at DESY. It was later
also used in the MAGIC gamma telescope at the Canary Islands
(although via shorter distances). 
The time resolution achieved was better than
5\,ns; given the strong smearing of photon arrival times due to light
scattering in ice, this jitter appeared to be acceptable. For optical fiber
transmission the time resolution was slightly better than for copper cable,
as was the dynamic range.
%An event was defined
%by a majority trigger formed in the surface counting house, requesting $\ge8$
%hits within a sliding window of $2\mu$s.

A big advantage of deep ice compared to deep water is the small optical
noise rate, about 0.5\,kHz in an 8-inch tube, compared to 20--60\,kHz due to
K$^{40}$ decays and bio-luminescence in lakes and oceans. The contamination of
hit patterns from particle interactions with noise hits is thus small and makes
hit selection much easier than in water and allows identifying burst-like
low-energy events from Supernovae (see section \ref{sec-icecube}).

The angular resolution of AMANDA for muon tracks was $2^\circ$--$2.5^\circ$,
with a lower energy threshold around 50\,GeV. Although better than for Lake
Baikal ($3^\circ$--$4^\circ$), this was much worse than for ANTARES ($<0.5^\circ$,
see below). This is the result of the strong light scattering which deteriorates
the original information contained in the Cherenkov cone. 
%The effect is even
%worse for cascades, where the angular resolution achieved with present
%algorithms is only $25^\circ$ (compared to $5^\circ--8^\circ$ in water
%\cite{Antares-casc-2006}).

\begin{figure}[ht]
%\sidecaption
\includegraphics[width=10cm]{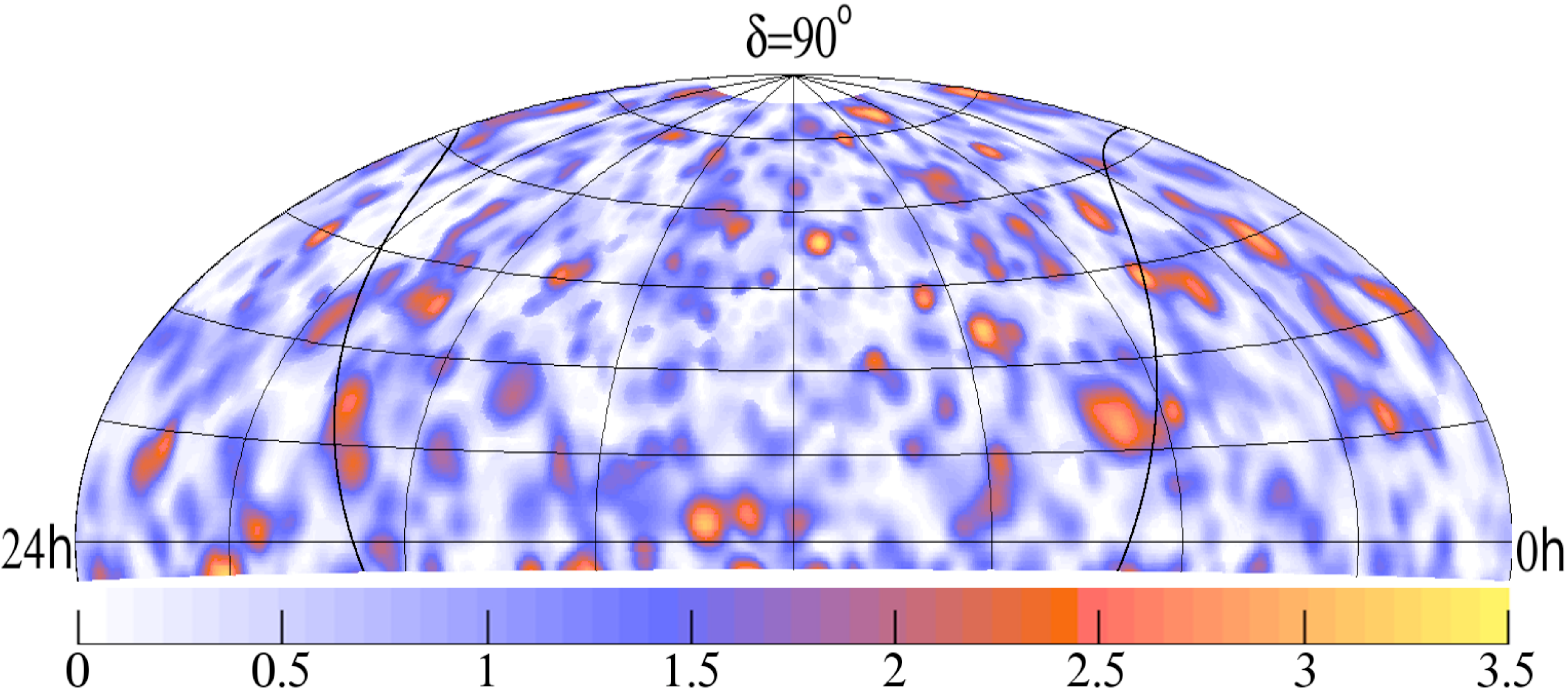}
\caption{   %\texttt{
7-year significance map of the Northern hemisphere derived with AMANDA
\cite{Amanda-7-years}.
  %}
\label{AMANDA-skyplot}
}
\end{figure}

AMANDA was switched off in April 2009, after more than 9 years of data
taking in its full configuration.  Figure \ref{AMANDA-skyplot} shows the
skyplot derived from of 6959 neutrinos taken in the years 2000-2007.
None of the spots is statistically significant, therefore only upper limits
could be derived.

AMANDA provided record limits on fluxes for cosmic neutrinos, 
be it for diffuse fluxes, for point sources or for transient sources like Gamma Ray Bursts.
These limits ruled out the first models on neutrino production in cosmic
sources.

The AMANDA limits on the diffuse neutrino flux have to be seen in the context of
the theoretical models and bounds existing at that time.
The commonly accepted "benchmark" bound was been derived from charged cosmic
ray fluxes ("Waxman-Bahcall
bound'' \cite{Waxman-Bahcall-1999}) and normalised to the cosmic ray flux at
about $10^{19}$\,eV. Assuming a generic $E^{-2}$ spectrum for all extragalactic
sources, the authors obtained a limit of $E^2 \cdot dN/dE^2 =
1-5\times10^{-8}$ GeV\,cm$^{-2}$\,s$^{-1}$\,sr$^{-1}$, with the uncertainty given by different cosmic
evolution models. This estimate
assumed that the sources are sufficiently dilute, so that neutrons can escape,
decay and provide the observed cosmic rays to which the estimate is normalised. 
If the sources are opaque even for neutrons, the only remaining estimator is
electromagnetic radiation. The decay of $\pi^0$s (co-produced with charged pions
producing neutrinos) yields gamma rays. Those develop electromagnetic cascades
leading to gamma rays of lower energies. 
Measurements of gamma rays in the MeV-GeV energy range from a satellite 
had set a bound to the neutrino flux, 
which hardly could be circumvented by more sophisticated
assumptions on the character of the sources. This 
"gamma limit" (first introduced in \cite{Berezinsky-1975}) was close to a second 
bound derived from cosmic ray spectra: Contrary to Waxman and Bahcall, 
the authors Mannheim, Protheroe and Rachen (MPR) 
\cite{Mannheim-Protheroe-Rachen-2001} assumed that a significant part of
the observed cosmic ray spectrum below $10^{19}$ eV was due to
extragalactic rather than Galactic sources. Interpreting the cosmic ray spectrum
between $10^{16}$ eV and $10^{19}$ eV essentially as a superposition of spectra
from many extragalactic source classes, each with a different cut-off, the
neutrino bound considerably weakened to  $E^2\cdot dN/dE \approx
5\times10^{-7}$ GeV\,cm$^{-2}$\,s$^{-1}$\,sr$^{-1}$ at a few $10^{14}$ eV.
AMANDA could exclude the MPR limit, but the exclusion of the WB limit 
was only achieved with IceCube (see section \ref{where-do-we-stand}).

\begin{figure}[ht]
\sidecaption
\includegraphics[width=7cm]{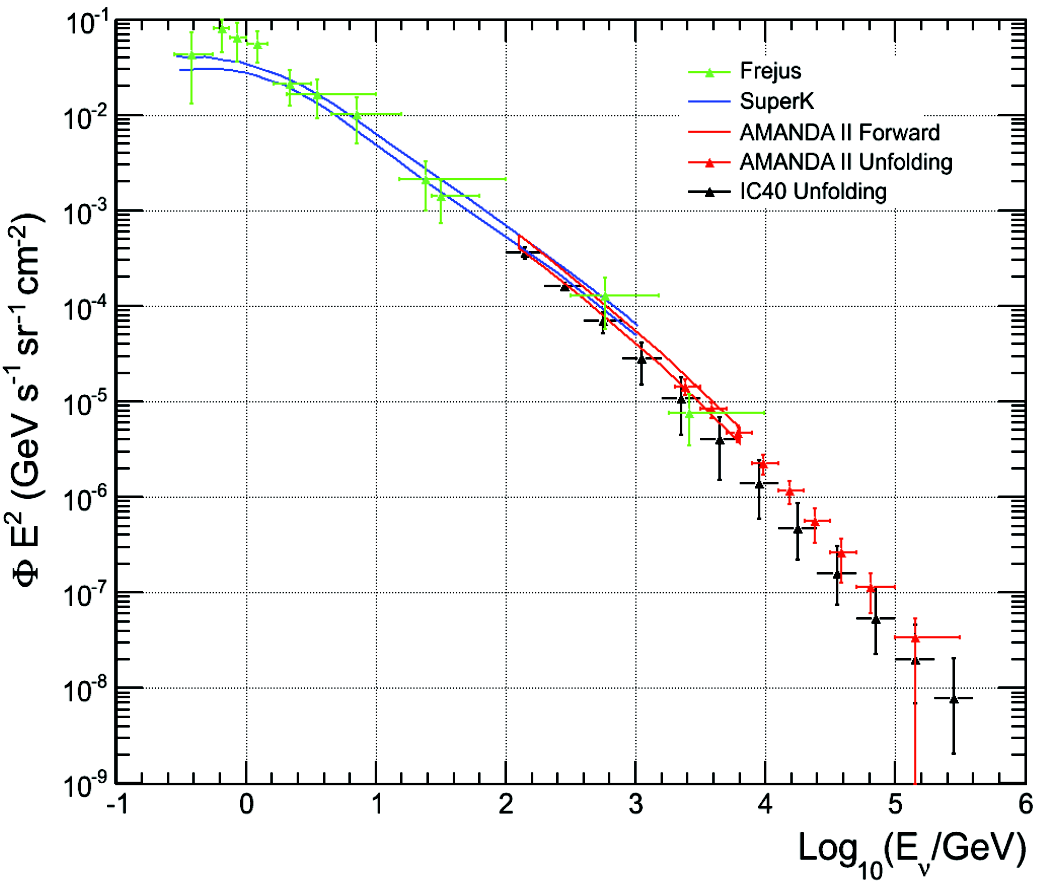}
\caption{
Energy spectrum of atmospheric neutrinos \cite{IceCube-atm}.
%Green triangles: Frejus
%\cite{Daum-1995}; blue band: Super-Kamiokande \cite{SK-atmnu}; red band:
%AMANDA forward folding analysis \cite{icecube-2009d}; red triangles: AMANDA
%unfolding analysis \cite{icecube-2010d}; black triangles: IceCube-40 unfolding
%analysis \cite{icecube-2011c}.
}
\label{atm-energy}
\end{figure}

AMANDA extended the measured spectrum of atmospheric neutrinos by nearly
two orders of magnitude, from a few TeV to 200\,TeV (Fig.\ref{atm-energy}).
It also established record limits on indirect  dark matter search, on the
flux of magnetic monopoles, and on effects violating Lorentz invariance
(see for a summary of results and references \cite{Katz-Spiering-2011}).
It would have detected neutrinos from a supernova burst in our Galaxy
(if such a burst would have appeared!), and it provided results on
the spectrum and composition of cosmic rays.

\begin{figure}[ht]
\sidecaption
\includegraphics[width=8cm]{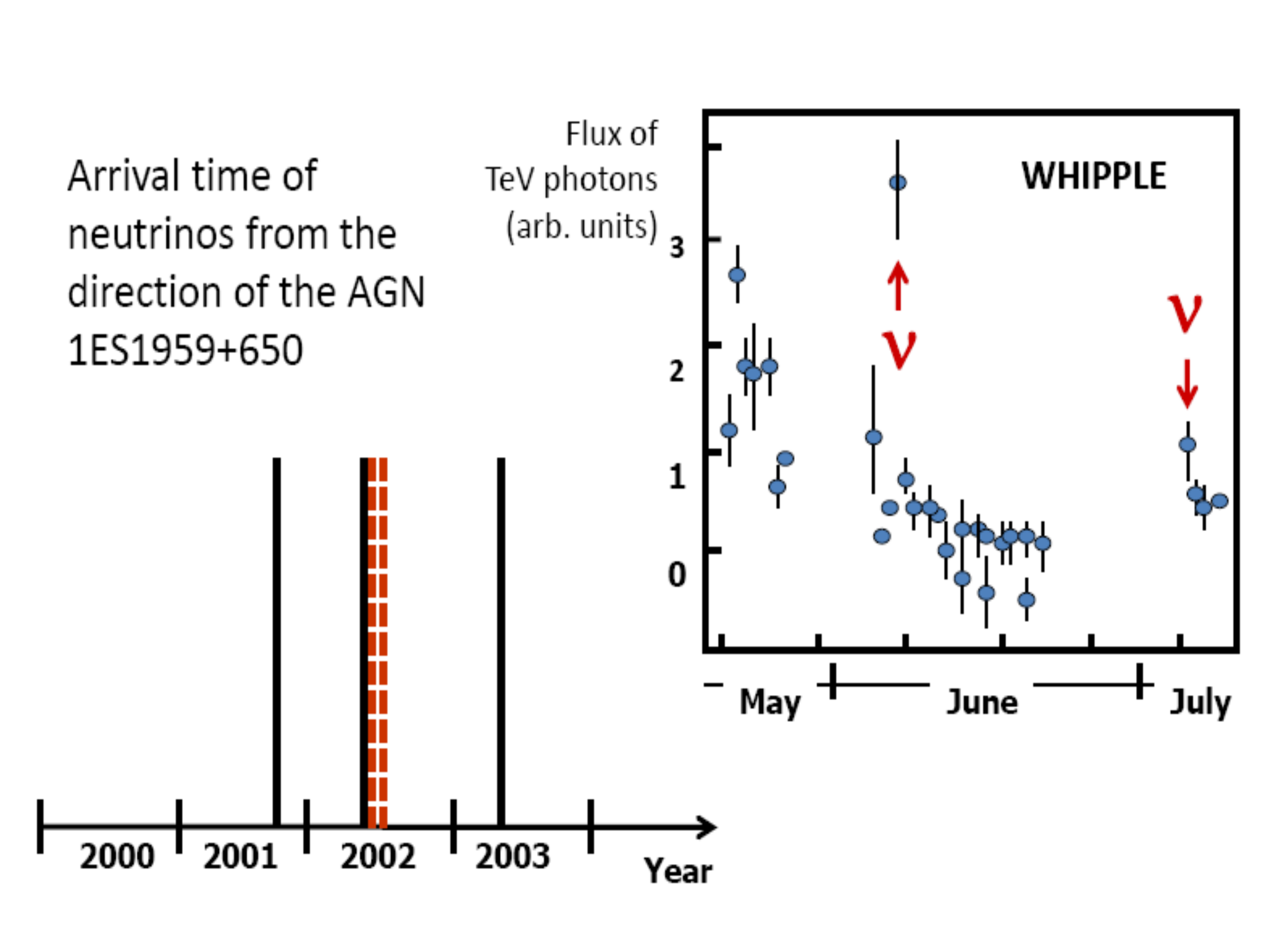}
\caption{   %\texttt{
The "curious" coincidence of neutrino events from the direction of
an AGN with gamma flares from the same source. The second and third
of the three events recorded in 2002 (dashed) coincide within about one
day with peaks seen by WHIPPLE. 
  %}
\label{ES}
}
\end{figure}

What was hoped for in optimistic dreams -- the discovery of an extraterrestrial
source of neutrinos -- did not happen. But there was one moment when the 
adrenaline level of some of us
went up and we thought we were close to a discovery. While analyzing
in 2005 the data taken from 2000-2003, five events where identified from the direction of
the Active Galaxy 1ES1959+650. Interestingly, three of these came within 66~days 
in 2002 \cite{Markus-thesis}. Two of
the three neutrinos were coinciding within about a day with gamma-ray flares
observed by the gamma-ray telescopes HEGRA and Whipple
 -- see Fig\ref{ES}. Excitingly, one of these two flares was not
accompanied by an X-ray flare, a so-called "orphan flare", which one would
expect for a hadron flare where the X-ray flux accompanying electron flare is
absent. This result was quickly followed by two theoretical papers, one claiming
that the corresponding neutrino flux would not fit any reasonable assumption on
the energetics of the source \cite{Reimer-2005}, the other claiming that
scenarios yielding such fluxes were conceivable \cite{Halzen-Hooper-2005}. Since
the analysis was not a fully blind analysis, it turned out to be impossible to
determine chance probabilities for this event, and actually the result was never
published in a journal. However, it initiated considerations to send alerts to
gamma-ray telescopes in case time-clustered events from a certain direction would
appear. Such a "Target-of-Opportunity'' alert is currently operating
between IceCube and the gamma-ray telescopes MAGIC (La Palma) and
VERITAS (Arizona).

\section{Mediterranean Sea: NESTOR, ANTARES, NEMO }
\label{sec-medi}

Whereas  AMANDA and Baikal developed coherently, 
the Mediterranean community split in
three branches which are related to three different locations and
host countries.

The NESTOR collaboration presented the full concept of a "tower" in 1993
\cite{nestor-1994}. 
After a long phase of tests and developments, 
a cable was installed to a site at 4\,km depth.
In 2004, a single prototype floor was deployed, connected and operated for
about one month \cite{nestor-test}. 
Then, its operation had to be terminated due to a failure of the cable to shore.
However, the data taken with this prototype demonstrated the detector functionality 
and provided a measurement of the atmospheric muon flux \cite{nestor-atmuons}.

Comparing the original plan to deploy an array of towers within a decade
(see Fig.\ref{Nestor-towers}) with the reality of a single floor 
operated over just a month, demonstrates the enormous challenges which
these projects face. The deep sea medium is hostile and unforgiving, 
the iterative approach is not what funding agencies like.
A professional management and a coherent collaboration are necessary
for any large projects, and if problems from this corner add to the
inherent problems of deep-underwater projects, delays or even
failure are inevitable.

Currently NESTOR is part of the KM3NeT framework which is directed
towards a multi-cubic kilometer detector in the Mediterranean Sea.

\vspace{0.3cm}

French collaborators who temporarily had been members of NESTOR,
pursued an independent strategy from the mid-nineties.  Together with
collaborators from Italy and the Netherlands they presented a full proposal for a 
12-string detector in 1999 \cite{antares-proposal}. In 2001 also
a German group at the University Erlangen joined the experiment.
ANTARES stands for Astronomy with a Neutrino Telescope and Abyss environmental
RESearch) \cite{antares-web}.  
This proposal was based on the operation of a demonstrator string
\cite{antares-demo1,antares-demo2} as well as on the results of extensive site
exploration campaigns in the region off Toulon at the French Mediterranean
coast, indicating that the optical background \cite{antares-light} as well as
sedimentation and biofouling \cite{antares-sedi} are acceptable at that site. 
However, taken all together (depth, optical clarity, optical background and
sedimentation), the site is inferior to the Greek and Italian sites. 

The construction of ANTARES started in 2002 with the deployment of a shore cable
and a junction box, the central element connecting the shore
cable to the detector. In 2002/2003, a preproduction string was deployed and
operated for a few months. Several technical problems were identified that
required further studies, design modifications and the operation of a mechanical
test string \cite{antares-line0}. The detector in its final 12-string configuration was
installed in 2006--2008 and has been operational since then, with a
break of a few months in 2009 due to a failure of the main cable that required
repair.

\begin{figure}[ht]
%sidecaption
\includegraphics[width=11cm]{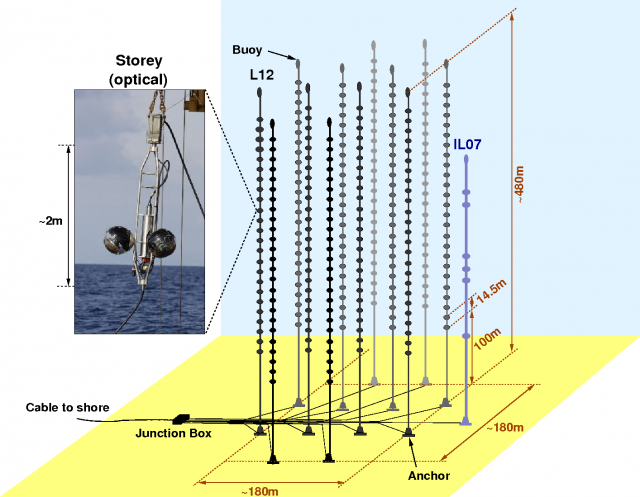}
\caption{
Schematic of the ANTARES detector. Indicated are the 12 strings and the
instrumentation line in its 2007 configuration (IL07). Shown as an inset is the
photograph of a storey carrying 3 photomultipliers.}
\label{ANTARES}
\end{figure}

ANTARES consists of 12 strings, each carrying 25 "storeys'' equipped
with three optical modules, an electronics container and calibration devices. 
The optical module %\cite{antares-om} 
consists of a 17-inch glass sphere
housing a hemispherical 10-inch photomultiplier (Hamamatsu R7081-20).
A further string, the "instrumentation line'', carries devices
for environmental monitoring. 
%The inter-string distances are about 70\,m, the
%vertical distance between adjacent storeys is 14.5\,m. 
The depth at the
ANTARES site is 2475\,m. The schematic setup is shown in Fig.~\ref{ANTARES}, a
detailed technical description can be found in \cite{antares-detector}.

\begin{figure}[ht]
\includegraphics[width=6.2cm]{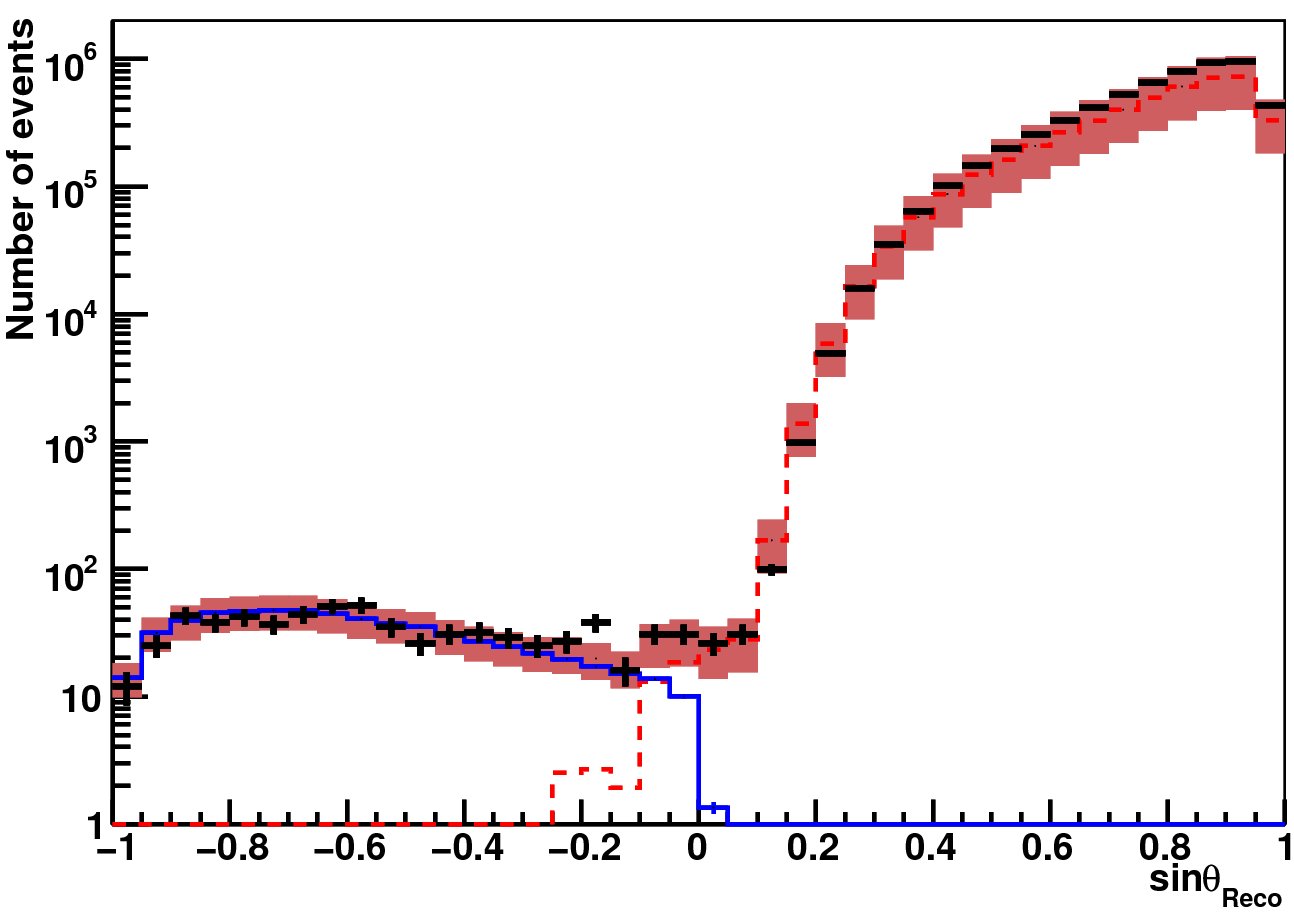}
\hspace{0.2cm}
\includegraphics[width=6.2cm]{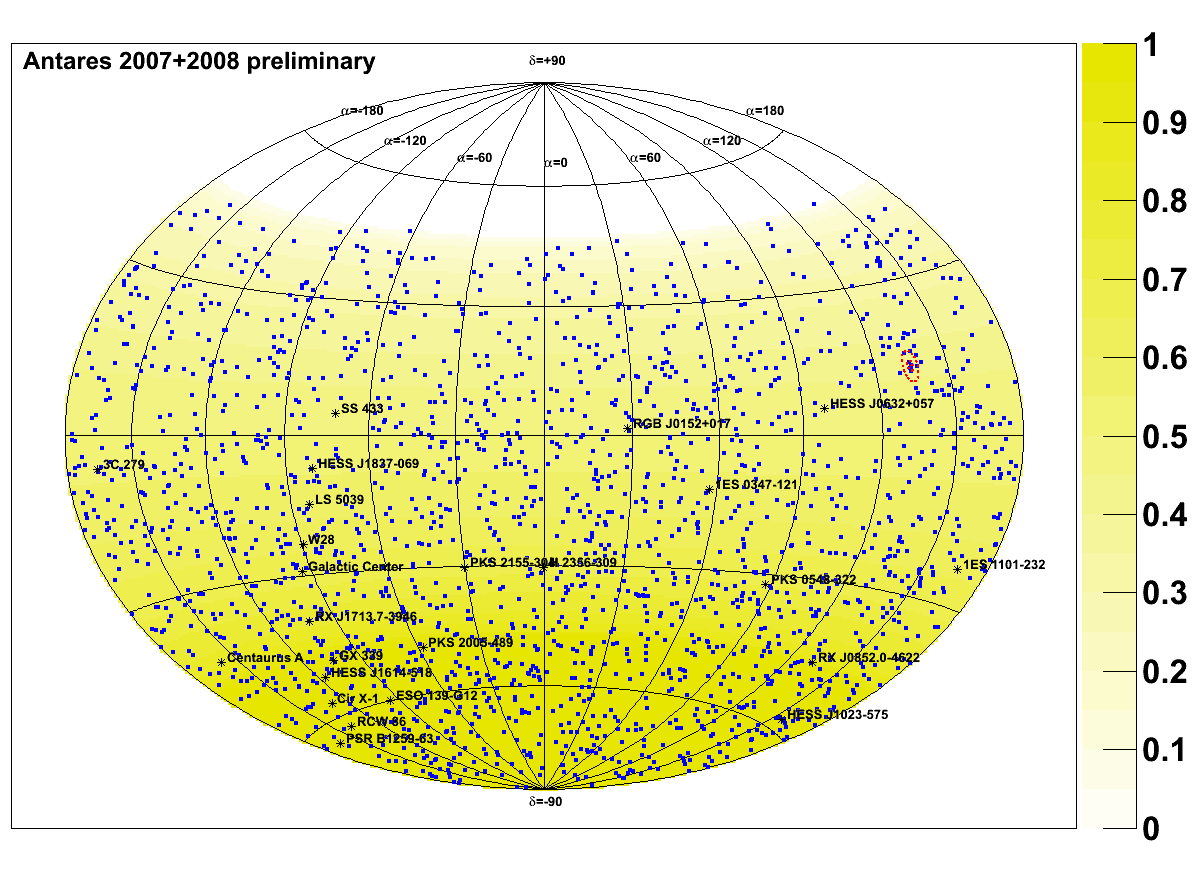}
\caption{
Left: Number of reconstructed muons in the 2008 ANTARES data, as a function of the
reconstructed zenith angle (black error bars). Also
indicated are the simulation results for atmospheric muons (red dashed), and
muons induced by atmospheric neutrinos (blue). The shaded band indicates the
systematic uncertainties. Figure taken from \cite{Antares-atmnuplot}.
Right: Equatorial skymap of neutrino-induced muon events from 295 days of ANTARES
data in 2007/2008. The background color scale indicates the sky visibility in
percent of the time. The most significant accumulation of events, marked with a
red circle, is fully compatible with the background expectation \cite{Antares-pointlims}.
}
\label{atm-ang}
\end{figure}

An almost background-free separation of neutrino-induced upward-going muons from
the huge background of downward-going muons is the central requirement for an
underwater or under-ice telescope. Baikal and AMANDA, followed by ANTARES,
have quickly mastered
this challenge, even more so IceCube. Figure~\ref{atm-ang}(left)  shows the rate of
muons as a function of the zenith angle $\theta$ as measured with ANTARES. Below
the horizon ($\theta<0$) the rate is well described by the expectation for
atmospheric neutrinos, above the horizon by that for atmospheric muons.
The right side of the figure shows the 1-year skymap of neutrino candidates --
as to be expected without a clear source signal.

\vspace{0.3cm}

The newest Mediterranean project is NEMO 
(NEutrino Mediterranean Observatory) \cite{nemo-web}.
It was launched in 1998 after Italian groups left NESTOR.
%hey had hoped for a faster realization of their plans. 
After it had turned
out that the original goal  to deploy a tower before the end of 1997
was not realistic, they had counted on 
a test set-up in the shallow water of the Navarino Bay not far from the NESTOR shore station.
But further delays 
%and managerial problems 
led to the termination of Italian NESTOR participation in 1998.  

In the same year the Italians started to discuss a possible 
continuation of Cherenkov underwater 
neutrino astronomy without the frictions they felt the years before.
Antonino Capone, the chair of the Rome group, was contacted by Emilio Migneco from 
Lab. Nazionale del Sud (LNS) in Catania, Sicily. As a result, Italians
started activities for the characterization of deep-sea sites close to the Italian Mediterranean coasts. 
Migneco presented the NEMO R\&D
project to the corresponding INFN Board and the project started,
with groups from Bari, Catania, Frascati, Messina and Roma.
Still in 1998, first sea campaigns followed and studied various
sites close to Italy.
 
From the beginning, the objective of this project was to study the
feasibility of a cubic kilometer detector, to develop corresponding
technologies and to identify and
explore a suitable site, in this case close to Sicily, rather than to build an 
separate detector of medium size. 
Therefore it is no surprise that in 2000 the Italian NEMO
members joined ANTARES, in parallel to their own
activities directed to the cubic kilometer scale.

\begin{figure}[ht]
\sidecaption
\includegraphics[width=2cm]{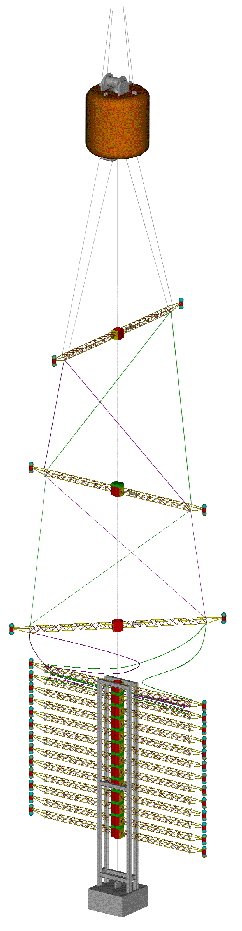}
\caption{
Sketch of the principle of a flexible NEMO tower during the unfurling process. Note
that exact configuration and the packaging of the bars does not correspond to
the current design framework.}
\label{nemo-tower}
\end{figure}

The basic unit of NEMO are towers composed by a sequence of 
floors. Different to NESTOR, floors consist of horizontal "bars", 
originally foreseen to be 15 m long and
each equipped with four 10-inch PMs. The floors are tilted against each 
other and form a three-dimensional structure.
\cite{nemo-status-2009}
(see Fig.~\ref{nemo-tower}). 
A tower can be folded together and deployed to the sea floor as a compact object
that is subsequently unfurled. Contrary to single strings and similarly
to the NESTOR concept, the 3-dimensional arrangement of photomultipliers per tower 
allows for local reconstruction of muon directions.

The most suitable site at a depth of
3.5\,km, about 100\,km off Capo Passero on the South-Eastern coast of Sicily has
been identified and investigated during various campaigns.
During the first prototyping phase, a cable to a test site
near Catania at a depth of 2\,km was installed and equipped with a junction box. 
In 2007, a "mini-tower'' with 4 bars was deployed, connected and operated for
several weeks. Although the data taking period was limited to a few months due
to technical problems, the mini-tower provided the proof of concept for the
technologies and most of the components employed. 
The flux of atmospheric muons
was determined in good agreement with the expectations from simulation
\cite{nemo-atmuons}.

The setup of a second phase \cite{nemo-latest} includes shore infrastructure at Capo
Passero and a 100\,km long cable to the site at 3.5\,km depth; both are
currently in place. A remotely operated vehicle (ROV) is available for the
deep-sea operations. A mechanical test tower of limited size was successfully
deployed and unfurled in early 2010. The plans to deploy a full-size prototype
tower will be pursued in the KM3NeT framework.

\section{IceCube}
\label{sec-icecube}

%-------------------------------------------------------------------------------
%         From DeepIce to IceTop
%-------------------------------------------------------------------------------
%\subsection{From DeepIce to IceCube}

\begin{figure}[ht]
\sidecaption
\includegraphics[width=6.5cm] {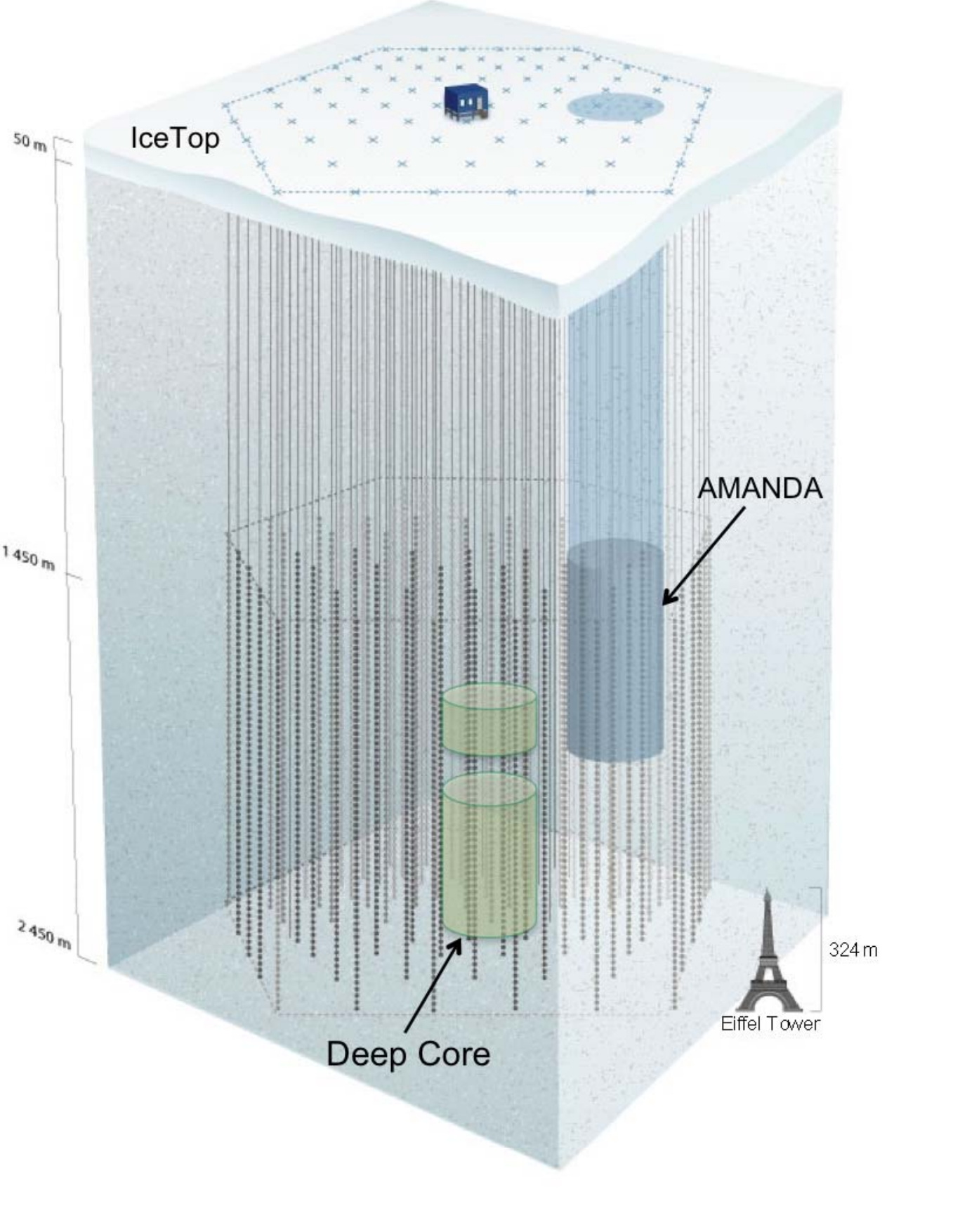}
\caption{
Schematic view of the IceCube neutrino observatory. AMANDA was replaced by
DeepCore, a nested low-threshold array. At the surface, the air shower array
IceTop and the IceCube counting house are indicated.}
\label{IceCube}
\end{figure}

IceCube \cite{icecube-web} is located at the South Pole
and actually incorporates its predecessor AMANDA as a sub-array
(see Fig.\,\ref{IceCube}). With this detector, the idea of 
a cubic-kilometer detector was finally 
realized. However, the path towards the first installation was all but smooth.

Actually, the first initiative beyond AMANDA was a concept called {\it DeepIce},
a proposal for multidisciplinary investigations, including neutrino and
cosmic ray astrophysics, glaciology, glacial biology, seismology and climate research.
%limited compared to Ocean detectors, but nevertheless manifold. Light
%propagation in deep ice is affected by remnant air bubbles at shallow depths
%\cite{Amanda-shallow-1995} and by 
As an example, we note the relation between the layered impurities from dust 
and climatic effects or volcano eruptions \cite{Amanda-ice-2006}. 
%-- see Fig.~\ref{AMANDA} in Sect.~\ref{sec-amanda}. 
%Motivated by these findings, a
%project called {\it 
DeepIce was proposed in 1999 to NSF, but was not funded: Several of the referees 
concluded
that a neutrino detector was sold under the flag of multi-disciplinary
research, (mis)using the NSF funding model for multidisciplinary centers.
The advice was to go ahead with a dedicated project for a neutrino
telescope. 

As a consequence, already in November of the same year 
a first 67-page IceCube proposal  was submitted to NSF
\cite{IceCube-Proposal-1999}. 
It had been worked out essentially by the collaborators 
of the old AMANDA collaboration.
i.e. the 
US Universities in Atlanta, Baton-Rouge, Berkeley, Delaware, Irvine
(Univ. of California), Kansas, 
Madison, Philadelphia (Pennsylvania Univ.) and IAS Princeton and LBNL Berkeley. 
Germany participated with DESY and the Universities Mainz and Wuppertal,  
Sweden and Belgium with the Universities Brussels, Stockholm and Uppsala.
%This was essentially the old AMANDA collaboration.
Soon, a number of additional institutions became interested and a new collaboration
was formed, the IceCube collaboration, which meanwhile has grown to 
nearly 40 institutions. 
%The IceCube principal investigator is
%Francis Halzen. Spokespersons have been Per-Olof Hulth (U. Stockholm,
%2001-2005), Christian Spiering (DESY, 2005-2007), 
%Tom Gaisser (U. Delaware, 2007-2011)
%and Greg Sullivan (U., since 2011).
%from six countries: USA (13), Germany (7), Belgium (4) Sweden (2), 
%Barbados (1), New-Zealand (1), Japan (1), Switzerland (1), UK (1). 
Paradoxically, the two collaborations co-existed until 2005
%(with Steve Barwick, UCI, and Christian Spiering being Amanda
%spokespersons from 1997 on), 
then joining
to one collaboration, IceCube.

Naturally, the mixture of AMANDA pioneers 
%(many with a a cosmic ray rather than
%an accelerator physics background) 
and the new members 
%(mostly coming from accelerator particle physics) 
did not go without some tensions. 
Some AMANDA pioneers feared the newcomers would ignore the merits of AMANDA, 
%and jump into a "prepared bed",  and 
some new IceCube members suspected that AMANDA colleagues would look too much back and not forward.
These tensions were exemplified by a hot debate on the best
concept for data transmission. A group headed by DESY favored analog transmission
via optical fibers, another group headed by LBNL preferred digital transmission.
The first solution was proven to be feasible in AMANDA but suffered from the vulnerability 
of optical fibers and connectors and a rather difficult calibration procedure, the second solution
was standard in accelerator experiments, 
appeared to be more modern, but there were doubts whether the long-term reliability
and the timing accuracy could be as good as promised. 
The digital concept itself did not come as a total novelty: the 18th string
of AMANDA had been equipped, in addition to the analog readout,
with a digital readout which incorporated all basic elements
of what was later used in IceCube \cite{string18}.
In 2001, the controversy was decided
following the recommendation of an external expert panel led by 
Barry Barish, which prioritized the digital transmission.
In the mean time we definitely know that this was the right decision.
With its fantastic reliability and stable 2\,ns timing accuracy, the DOM concept
(DOM for Digital Optical Module) has exceeded all expectations.

The IceCube Technical Office is located at the University of Wisconsin,
with project director Jim Yeck. Under the control
of the Technical Office, IceCube components were developed
in various labs of the collaboration. The DOM was developed
in LBNL, the communication card to the DOMs from surface
in DESY. The integrated
design, the HV system and a "LED flasher board" were
developed in Wisconsin. Assembly and testing of the DOMs was
performed at three locations, in Wisconsin, in Stockholm/Uppsala
and in Zeuthen.

%IceCube sensitivity paper 

%PDD in 2001 \cite{IceCube-PDD-2001}

%IceCube technical office in UW. 

% Module production at 3 places. 

%Approval in US, Germany, Belgium, Sweden
%\vspace{0.5cm}

For IceCube construction, the thermal power of the hot-water drill factory was
upgraded to 5\,MW, compared to 2\,MW for AMANDA.  
This reduced the average
time to drill a 2450\, m deep hole 
with a diameter of 60\,cm 
to $35$\,hours. 
%The subsequent installation of a string required typically 12 hours. 
The commissioning of the drill during the first deployment season 2004/05
turned out to be extremely challenging, but eventually a first, single string was
deployed in January 2005: The first step was made! The following seasons resulted
in 8, 13, 18, 19, 20 and 7 strings, respectively. 
The last of 86 strings was deployed at Dec.\,18, 2010.

%\subsection{The IceCube detector}

IceCube consists of 5160 digital optical modules (DOMs) installed on 86 strings at
depths of 1450 to 2450\,m.
A string carries 60~DOMs with 10-inch photomultipliers Hamamatsu 
R7081-02 housed in a 13-inch glass sphere. 
Signals  are digitized in the DOM and sent to the surface via copper cables.
% \cite{IceCube-PDD-2001}. 
320~further DOMs are installed in IceTop, an array of detector stations on the
ice surface directly above the strings (see Fig.~\ref{IceCube}). AMANDA,
initially running as a low-energy sub-detector of IceCube, was decommissioned in
2009 and replaced by DeepCore, a high-density sub-array of eight strings at large
depths (i.e.\,in the best ice layer) at the center of IceCube. 
DeepCore collects photons with about six times the efficiency of full
IceCube, due to its smaller spacing, the better ice quality and the
higher quantum efficiency of new PMTs.
Together with the veto provided by IceCube, this results in an expected
threshold of about 10\,GeV. This opens a new window for oscillation physics
and indirect dark matter search.

The muon angular resolution achieved by present reconstruction algorithms is
about $1^\circ$ for 1\,TeV muons and below $0.5^\circ$ for energies above 10 TeV. 
%The very clear ice below a depth of 2100\,m has a
%particular potential for improved resolution. This will be even more important
%for the angular reconstruction of cascades. The presently achieved angular
%resolution for cascades is only $30^\circ$, much worse than for water, mainly
%due to the strong light scattering in ice.
%-------------------------------------------------------------------------------
Unlike underwater detectors with their environment of high optical noise,
IceCube can be operated in a mode that is only possible in
ice: The detection of burst neutrinos from supernovae. The low dark-count rate
of the PMTs  allows for detection of the
feeble increase of the summed count rates of all PMTs during several
seconds, which would be produced by millions of interactions of few-MeV
neutrinos from a supernova burst \cite{icecube-sn-2011}. IceCube
records the counting rate of all PMTs in millisecond steps. A
supernova in the center of the Galaxy would be detected with extremely high
significance and the onset of the pulse could be measured in unprecedented
detail. Even a SN\,1987A-type supernova in the Large Magellanic Cloud would
provide a recognizable signal and be sufficient to provide a trigger to the
SuperNova Early Warning System, SNEWS \cite{SNEWS-2004}.

\section{Where do we stand?}
\label{where-do-we-stand}

With IceCube, the sensitivity to point sources and to diffuse fluxes
has been improved by nearly a factor of thousand when compared to
the situation of the mid nineties. But alas -- no indication for
extraterrestrial sources has found yet, but only ever 
tightening upper limits on fluxes have been established.

\begin{figure}[ht]
\center{
\includegraphics[width=11cm]{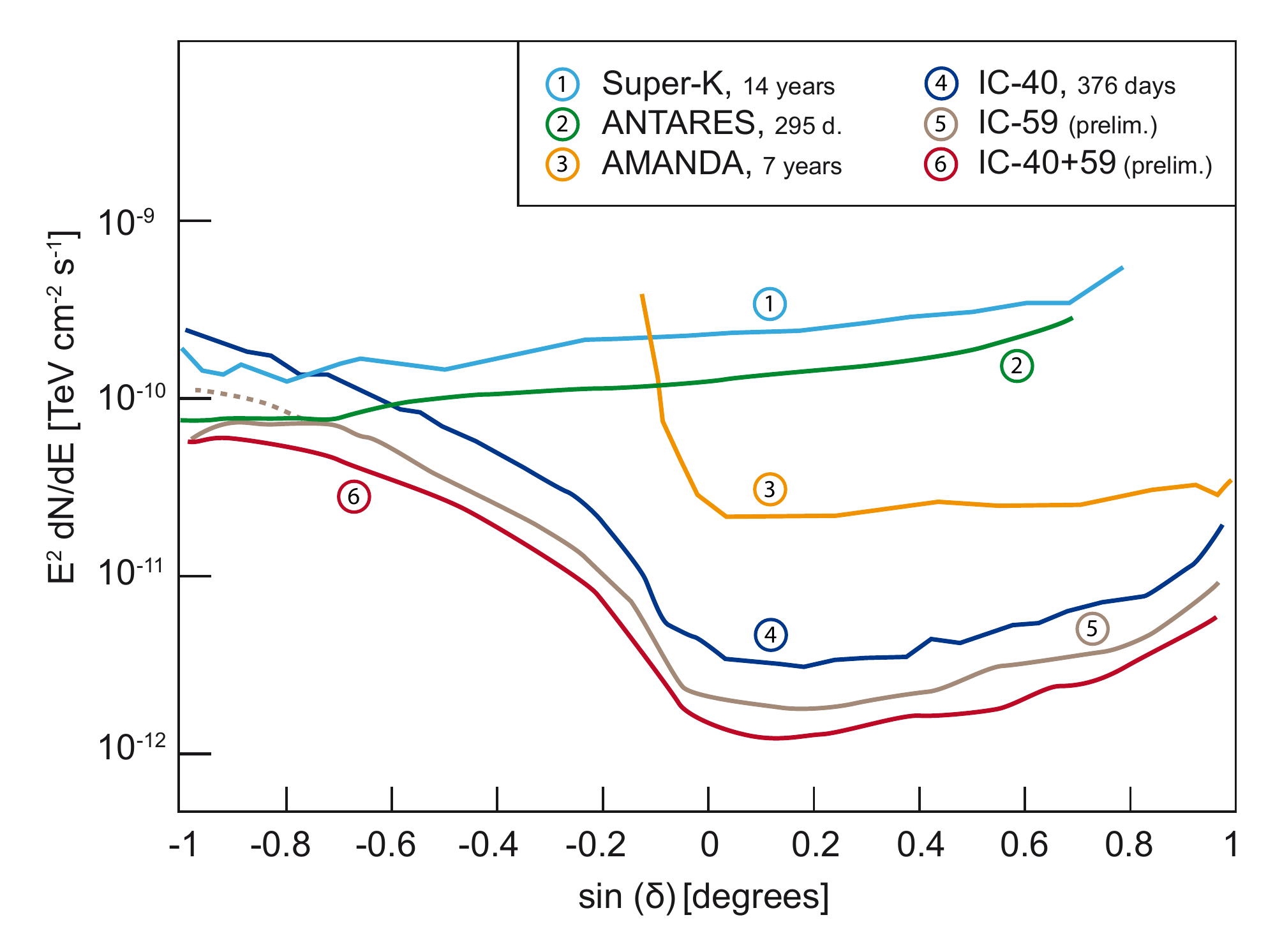}}
\caption
{Point source limits from various experiments.
%Super-Kamiokande
%\cite{SuperK-poiNT2009}, AMANDA \cite{icecube-2009a}, IceCube-40
%\cite{IC-40-point}, IceCube-59, the sum of IceCube-40 and IceCube-59
%(preliminary results) and ANTARES \cite{Antares-pointlims}. 
See \cite{Katz-Spiering-2011} for references. 
%\cite{km3net-tdr}.
}
\label{point-limits}
\end{figure}

Figure~\ref{point-limits} compiles the limits from previous experiments, from
the different IceCube stages and from ANTARES.
Note that the combined data of IceCube-40 and IceCube-59 surpass the
mark of 1\,km$^3\times1\,$year and thus exceed 1 year worth of data from the full
IceCube detector. When this article is printed, a factor of 1000 improvement of the 
sensitivity to point sources will have been reached when compared to the 
very first AMANDA point source paper from the year 2000 \cite{AMANDAB10-2000}.
Searching for coincidences of neutrinos with Gamma Ray Bursts reported by satellites,
IceCube also challenges the hypothesis that Gamma Ray Bursts are the sole origin of cosmic rays
of highest energies \cite{grb-2012}.

%\begin{figure}[ht]
%\sidecaption
%\includegraphics[width=7cm]{figs/skymap-IceCube.pdf}
%\caption{
%Equatorial skymap of pre-trial significances of the all-sky point source search
%with IceCube-40 \cite{IC-40-point}. Each dot represents one neutrino event, the
%colour scale indicates the significance of event accumulations. The Galactic
%plane is shown as black curve.}
%\label{skymap-IceCube}
%\end{figure}

\begin{figure}[ht]
\sidecaption
\includegraphics[width=7cm]{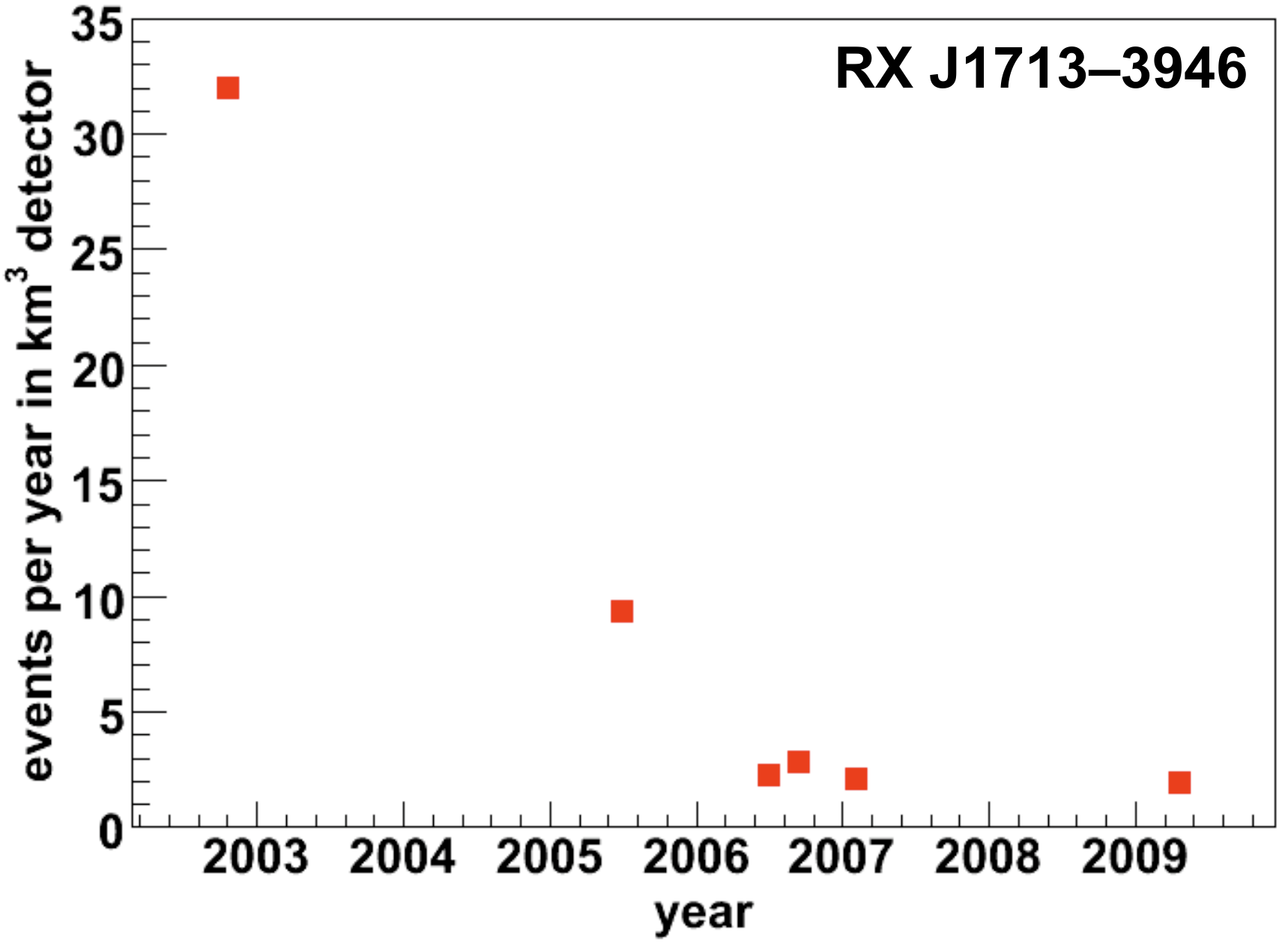}
\caption{
Number of events per year in a cubic kilometer detector
expected from the supernova remnant RX J1713-3946
Predictions are shown for 
\cite{Alvarez-Muniz and Halzen 2002}, \cite{Costantini and Vissani 2005},
\cite{Distefano 2006,Kistler and Beacom 2006},
\cite{Kappes etal 2007} and \cite{Morlino etal 2009}.
Courtesy of Alexander Kappes. 
}
\label{RX-time}
\end{figure}

At the same time, predictions for neutrino fluxes from extraterrestrial sources 
went lower and lower. This is best demonstrated by Figure \ref{RX-time}.
It shows the number of events per year in a cubic kilometer detector
expected from the supernova remnant RX J1713-3946, the most
preferred galactic candidate source.

The drop from the 2002 paper of Alvarez-Muniz and Halzen 
(shortly before the construction of IceCube
started) to the
paper of Costantini and Vissani from 2005 is due to that fact
that in the first paper oscillations have not been taken into account
\footnote{Oscillations turn the original ratio 
$\nu_e : \nu_{\mu} : \nu_{\tau} = 1 : 2 : 0$ to
$\nu_e : \nu_{\mu} : \nu_{\tau} = 1 : 1 : 1$, thereby reducing
the number of muon neutrinos as the only neutrino suitable for 
point source searches by a factor of 2. See also the paper of Learned and 
Pakvasa \cite{Pakvasa-1995} where the authors propose to search
for "double-bang" events for tau-neutrino identification.}, and that better gamma-ray
data on the source were available. The following reduction is
due to further improved data and more realistic calculations.
After 2005 the predictions stabilize, however on a disappointingly
low level: with 2-3 events per year on a much larger background
of atmospheric neutrinos, a significant detection would need more
than ten years of data taking. Note that even this is an optimistic
assumption, since all the calculations assume the the gamma-rays
observed from this source are due to the decay of $\pi^0$s 
from hadronic interactions (with charged pions decaying
to neutrinos as counterparts), and not
to inverse Compton scattering 
$e^- + \gamma_{\mbox{low energy}} \rightarrow e^- +  \gamma_{\mbox{high energy}}$.

Point-source searches use the directional and energy information to reduce the
background from atmospheric neutrinos. 
Cosmic neutrinos from a given source would cluster around the source
direction. 

\begin{figure}[ht]
\center{
\includegraphics[width=12cm]{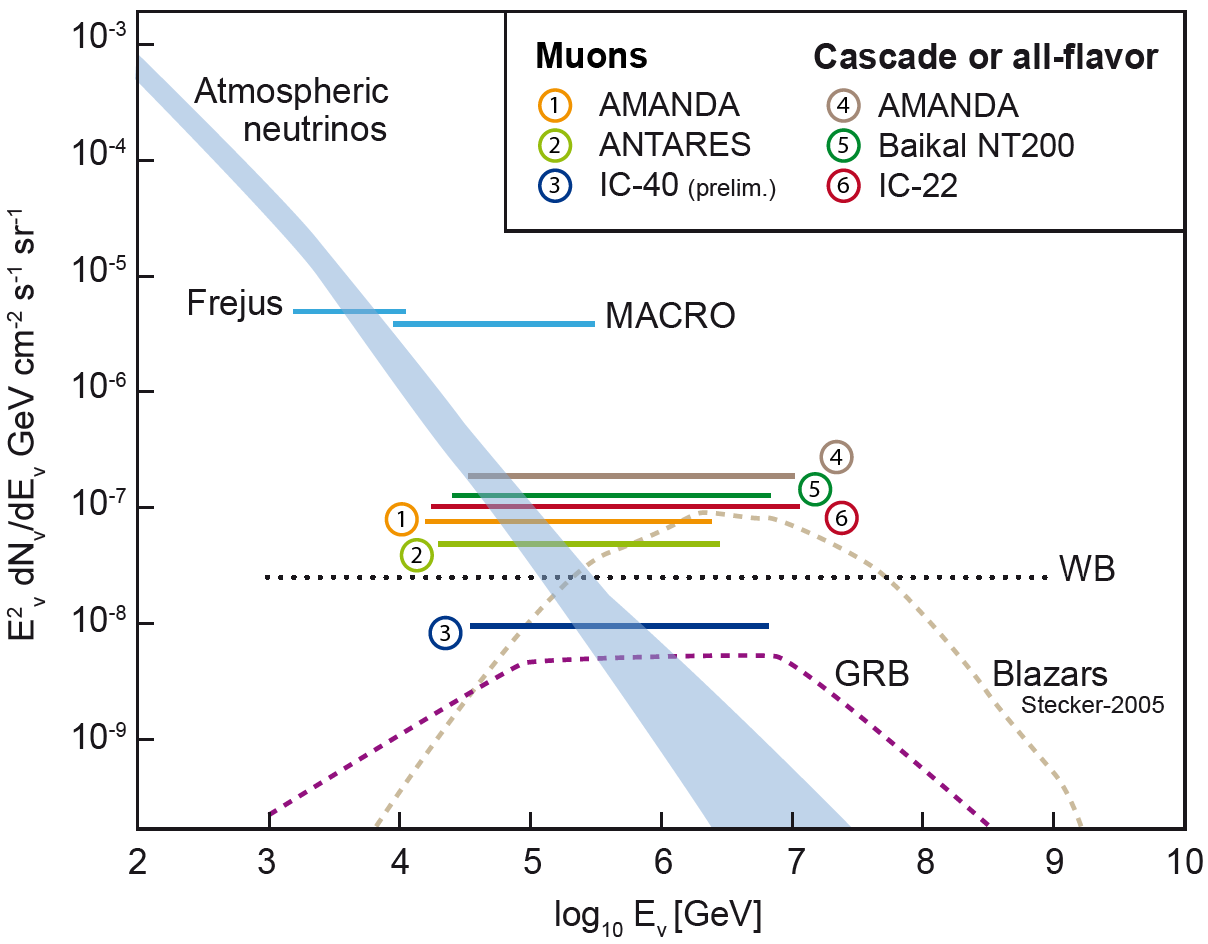}}
\caption{
90\% C.L.\ upper limits on the diffuse flux of extraterrestrial
neutrinos. The horizontal lines extend over the energy range which would cover
90\% of the detected events from an $E^{-2}$ source (5\% would be below and 5\%
above the range). All model predictions have been normalized to one flavor,
i.e.\ all of the all-flavor limits have been divided by 3. The colored band
indicates the measured flux of atmospheric neutrinos (see also
Fig.~\ref{atm-energy}), the broadening at higher energies reflects the
uncertainties for prompt neutrinos. The limits on muon neutrinos are from 807
days AMANDA 
%\cite{icecube-2007b}, 
334 days ANTARES 
%\cite{Antares-diffuse}, 
and 375 days IceCube-40. Cascade/all flavor limits are from 807 days
AMANDA 
%\cite{icecube-2011b}, 
1038 days Baikal-NT200
%\cite{Baikal-diff-Aynutdinov-2006,Baikal-diff-2009}, and 
257 days IceCube-22. See \cite{Katz-Spiering-2011} for references.
%\cite{IC-22-cascades}. The Frejus and MACRO limits have been published in
%\cite{Rhode-1996} and \cite{MACRO-diffuse}, respectively. 
Also indicated is the Waxman-Bahcall (WB) bound \cite{Waxman-Bahcall-1999}}
\label{UHE-diffuse}
\end{figure}

If the extraterrestrial signal is not concentrated to 
individual strong sources but distributed over all the sky, the
signal has to be identified in diffuse fluxes.
Searches for diffuse fluxes can only use the measured energy as criterion for
separating cosmic and atmospheric neutrinos, searching for
an excess at high energies. These studies are not restricted
to muon neutrino interactions with a muon
in the final state. Since the directional information
is not of prime importance, one can also study events without a
long track but only a particle cascade in the final state. Such events
emerge from electron- and tau-neutrino interactions and from all
neutral current interactions. Due to neutrino oscillations, two
third of the extraterrestrial neutrinos arrive as $\nu_e$ or
$\nu_{\tau}$.
No significant excesses over atmospheric neutrinos or other kinds of background
has been observed so far, resulting in upper limits on the diffuse flux of
extraterrestrial high energy neutrinos. Figure~\ref{UHE-diffuse} summarizes the
limits obtained in the TeV--PeV region. For each experiment and each
method only the best limit is shown. Remarkably, from the first limit derived
from the underground experiment Frejus (1996) to the 2010 IceCube-40 limit, a
factor of 500 improvement has been achieved. Several models 
such as e.g.\ the blazar
model of Stecker \cite{Stecker-2005} shown in the figure can be excluded,
and the Waxman-Bahcall bound (see section \ref{sec-amanda})
was eventually passed.
A further factor of 10 improvement is expected over the next 2--3 years, using the
full IceCube detector and combining muon and cascade information. The expected
sensitivity is more than an order of magnitude below the Waxman-Bahcall bound,
and prompt atmospheric neutrinos will be detectable for all but the lowest
predictions \cite{Kowalski-2005}.

Five decades after the first conceptual ideas, and three decades after first practical
attempts to build high-energy neutrino telescopes, we may be close to a turning
point. IceCube, 
%the first cubic kilometer neutrino telescope, 
has started data taking in its full cubic-kilometer configuration, provided 
results equivalent to 1 km$^2 \times$year, but has not
yet detected an extraterrestrial neutrino signal. 
%Meanwhile,  
%counterparts on the Northern hemisphere are being prepared.

The strong case for high-energy neutrino astronomy has remained unchanged over
time, but the requirements on the necessary sensitivity have tightened
continuously. Whereas underground detectors on the kiloton mass scale
(or on the $10^3$\,m$^2$ muon area scale) seemed sufficient in the sixties,
predictions from the seventies and eighties already favored scales of $10^5-10^6$\,m$^2$. 
Actually, DUMAND was conceived as a cubic kilometer configuration in 1978.
On the other hand, underground detectors like MACRO or Super-Kamiokande were 
still given a certain 
potential for high energy neutrino astronomy. Therefore it is no surprise that, in spite
of their declared goal of the kilometer scale, also
the underwater/ice community did hope 
for early discoveries with NT200, AMANDA, NESTOR and ANTARES.  
This hope turned out
to be elusive. Neither did the observations of GeV and TeV gamma rays
in the last two decades support higher flux expectations nor has any
of these detectors seen a signal indication with more than 3$\sigma$ significance.
Therefore the  detection of first extraterrestrial high-energy neutrinos sources 
%from outer space 
lies still ahead. With some optimism, we may expect it within the next few
years. Galactic "Pevatrons'' such as those observed in gamma rays by the
Milagro detector are within reach after a
few years of IceCube data taking if the corresponding predictions are correct. 
Models assigning the most energetic cosmic rays to Gamma Ray Bursts are challenged
by recent IceCube data \cite{grb-2012} and will be more strongly scrutinized within 
a couple of years. However, clear detections are all but
guaranteed.

\section{Alternative technologies for highest energies -- a parallel story}
\label{new-methods}

To detect the feeble fluxes at 100\,PeV -100\,EeV (see Fig.\,\ref{all-nu})
one needs detectors one to three orders larger than IceCube.
The region is dominated by "cosmogenic" neutrinos, a guaranteed source
(although with a flux uncertain by 1-2 orders of magnitude) and
a possible admixture of neutrinos emerging from jets of AGN.
In order to reach such volumes one has to use emissions
which propagate with kilometer-scale attenuation and allow a
sparse instrumentation. There are three methods to reach such volumes.

\begin{itemize}
\item Acoustic detection 
\item Radio detection 
\item Detection via air showers
\end{itemize}

Although the roots of all three methods lay in the seventies and
eighties, it was only in the last decade that a broader community got engaged.
The actual status of these very dynamic fields is nicely reflected
in the series of ARENA Workshops 
\cite{Arena-Zeuthen,Arena-Scotland,Arena-Rome,Arena-Nantes}. 
ARENA stands for Acoustic and Radio EeV Neutrino Detection Activities.
Table 4 compiles the corresponding projects, together
with the optical instruments underwater and in ice.

\begin{table}[h]
\begin{center}
\begin{tabular}{|l|l|l|l|} 
\hline
{\bf Experiment} & {\bf Location} & {\bf Technique} & {\bf Status}  \\
\hline
\hline
DUMAND & Hawaii & Water Cherenkov & turned down 1996 \\ % & \cite{DUMAND87} \\
\hline
NT200+ & Lake Baikal& Water Cherenkov &  operating   \\ %&\cite{Baikalweb,Baikal1} \\
\hline
GVD & Lake Baikal & Water Cherenkov  & design phase \\ %& \cite{GVD} \\
\hline
AMANDA & South Pole & Water Cherenkov  &  terminated 2009 \\ % & \cite{Amanda-1} \\
\hline
IceCube & South Pole & Water Cherenkov & operating \\ %& \cite{IceCubeweb,IceCube-0} \\
\hline
ANTARES & Mediterranean & Water Cherenkov & operating \\ %& \cite{Antaresweb,Antares-1a} \\
\hline
NESTOR & Mediterranean & Water Cherenkov &  R\&D for KM3NeT \\ %& \cite{Nestorweb,Nestor-1} \\
\hline
NEMO & Mediterranean & Water Cherenkov &  R\&D for KM3NeT \\ %& \cite{Nemoweb,Nemo-1} \\
\hline
KM3NeT & Mediterranean & Water Cherenkov & design phase \\ %& \cite{KM3NeTweb,CDR}  \\
\hline
HIRES & USA & Air shower & terminated 2009 \\ %& \cite{HIRES} \\
\hline
Auger & Argentina & Air shower & operating \\ %& \cite{Augerweb,Auger-nu} \\
\hline
TA & USA & Air shower &  operating \\ %& \cite{TA} \\
\hline
JEM-EUSO & Satellite & Air shower & construction \\
\hline
ASHRA & Hawaii & air shower & partial operation \\ %& \cite{mountain} \\
\hline
CRTNT & China & air shower & planned \\ %& \cite{mountain} \\
\hline
ANITA & Antarctica (balloon) & Radio (ice) & flights continuing \\ % & \cite{ANITA} \\
\hline
RICE & South Pole & Radio (ice) & terminated \\ % & \cite{RICE} \\
\hline
ARA & South Pole & Radio (ice) &  construction stage 1 \\ %& \cite{AURA} \\
\hline
ARIANNA & Antarctic shelf & Radio (ice)  & construction stage 1 \\ %& \cite{ARIANNA} \\
\hline
SALSA & open & Radio (salt mine) & conceptual phase \\ %& \cite{SALSA} \\
\hline
SAUND & Caribbean Sea & Acoustic & terminated  \\ % & \cite{acoustics-3} \\
\hline
SPATS & South Pole & Acoustic & test array operating \\ %& \cite{SPATS,speed} \\
\hline
AMADEUS & Mediterranean Sea& Acoustic & test array operating  \\ %& 
%\cite{ARENA-Zeuthen,ARENA-Scotland,ARENA-Rome}  \\
\hline
ON$\nu$DE & Mediterranean Sea & Acoustic &  test array finished  \\ %& 
%\cite{ARENA-Zeuthen,ARENA-Scotland,ARENA-Rome} \\
\hline
Baikal & Lake Baikal & Acoustic &  R\&D  \\ %& 
%\cite{ARENA-Zeuthen,ARENA-Scotland,ARENA-Rome} \\
\hline
GLUE & USA & Radio (moon) & terminated \\ %&  \cite{GLUE} \\
\hline
NUMOON & Netherlands & Radio (moon)  & operating \\% &  \cite{NUMOON} \\
\hline
Kalyzhin & Russia & Radio (moon) & operating \\
\hline
LORD & Satellite & Radio (moon) & planned \\
\hline
FORTE & Satellite & Radio (Earth) & terminated \\ %&  \cite{FORTE} \\
\hline
\end{tabular}
\caption{Instruments and R\&D projects
for high energy astrophysical neutrino detection}
\end{center}
\end{table}

\subsection{Acoustic detection}
\label{subsec-acoustic}

The principle of acoustic particle detection was explained in
\ref{subsec-acoustic-DUMAND}.
An historical review of acoustic particle detection can be
found in \cite{acoustic-history} and a snapshot of the
present situation in \cite{Nahnhauer-2012}. Here, I sketch some of the
most important milestones.

Acoustic detection has first been intensively discussed in the context
of DUMAND (see section \ref{subsec-acoustic-DUMAND}). Since
after 1980 DUMAND focused all forces to the optical method, acoustic
detection went dormant until it had a certain revival within the
NESTOR project. But also the NESTOR collaboration had its hands full
with developing optical equipment, therefore acoustic activities 
did not really move ahead. 

Given the limited resources within the big neutrino underwater
projects, in the early nineties Igor Zheleznykh from INR Moscow proposed to use
military acoustic arrays in the Black Sea and at the Russian
Pacific coast to dig for neutrino signatures. The idea was first
realized in the last decade, however not in Russia but 
by using part of a very large, sparsely
instrumented hydrophone array of the US Navy, close to the Bahamas
\cite{acoustic-Lehtinen-2002}. 
The array covered an area of 250 km$^2$, 
could trigger on events above 100\, EeV with a tolerable
background rate and derived upper limits on the flux
of ultra-high energy neutrinos \cite{acoustic-SAUND-2010}.

Presently, R\&D on acoustic detection is pursued in the Mediterranean
Sea close to Sicily (initially running under the name O$\nu$DE) and 
to Toulon (project AMADEUS,
linked to the ANTARES detector),  in Lake Baikal, at the Scottish Coast and 
at the South Pole (see for references \cite{Nahnhauer-2012} and \cite{Katz-Spiering-2011}). 
For ice, the signal
itself is expected to be higher and ambient noise to be lower than in sea water. 
A test array, SPATS (South Pole Acoustic Test Setup), has been deployed at the
South Pole in order to determine the depth dependence of the speed of sound,
\cite{acoustic-SPATS-2009},
the attenuation length of acoustic signals 
\cite{acoustic-SPATS-2010} 
and the ambient noise 
\cite{acoustic-SPATS-2011}. 
The results for the attenuation length  
%of the latter two measurements are 
is slightly discouraging as the
attenuation length turns out to be about 300\,m, an order of magnitude smaller
than predicted by Buford Price in the nineties \cite{Price-1996}.
However,  the noise level seems to be smaller than in the deep sea at a
calm sea state.
 
As a variation of the ice approach, even the use of permafrost
as medium has been discussed \cite{acoustic-Nahnhauer-2008}.

\subsection{Radio Detection}
\label{subsec-radio}

In 1983, G.\,Gusev and I.\,Zheleznykh proposed the Radio Antartic Muon
And Neutrino Detector (RAMAND) with radio antennas "listening" to
the Antarctic ice massive \cite{Zheleznykh-1983}.  
The proposal was based on an effect
predicted  in 1962 by Gurgen Askaryan \cite{radio-Askaryan-1962}.
Electromagnetic cascades emit coherent Cherenkov radiation at radio frequencies. 
It is due to electrons from the material traversed which
are swept into the developing shower,
which thus acquires an electric net charge. This charge propagates like a
relativistic pancake of about 1\,cm thickness and 10\,cm diameter. For
wavelengths exceeding the cascade diameter, coherent emission of electromagnetic
radiation occurs. The signal amplitude increases with the square of the
net charge in the cascade, i.e. it is proportional to $E_{\nu}^2$, thus making
the method particularly attractive for high-energy cascades. The resulting
bipolar pulse is in the radio frequency band and has a width of 1-2\,ns. 
In 2001, the effect was confirmed by
measurements at accelerators
\cite{radio-Saltzberg-2001,radio-Gorham-2007}. 

In ice, attenuation lengths of several hundred meter to more than a kilometer 
have been reported for radio
signals, depending on the frequency band and the ice temperature. This implies that
for energies above hundred PeV radio detection may become competitive or
superior to optical detection (with its attenuation length of the order 100\,m)
\cite{Price-1996}.

The RAMAND detector of Gusev and Zheleznykh was planned to be deployed
at the Soviet Antarctic station Vostok, but with the decay of the Soviet Union
the chances for realization faded away. About ten years later, RICE, the
Radio Ice Cherenkov Experiment,  was installed at the American
Amundsen-Scott Station at the South
Pole, with 20 receivers and emitters buried at depths between 120 and 300\,m. 
From the non-observation of very large pulses, limits on the diffuse flux of
neutrinos with $E>100$ PeV and on the flux of relativistic magnetic monopoles
have been derived \cite{radio-RICE-2006}.

Much larger volumes could be observed with the 
"Antarctic Impulsive Transient Array" (ANITA) 
\cite{radio-ANITA-2006}. 
This is an array of radio antennas which has been flown at a balloon on an Antarctic
circumpolar path in 2006 and 2008/09. 
%(see Fig.~\ref{ANITA}, left). 
From 35\,km
altitude it searched for radio pulses from neutrino interactions in the thick
ice cover and monitored, with a threshold in the range of several EeV, a volume of the
order of $10^6$ Gigatons. The resulting neutrino flux limits 
\cite{Anita-2009} 
are  presented in Fig.~\ref{EHE-diffuse}.

Future plans for radio detection in ice foresee large arrays of antennas on the
surface of the antarctic ice shelf (project ARIANNA) \cite{radio-ARIANNA-2007} 
or in the South Polar ice close to surface (project ARA)
\cite{ARA-2011}. Both projects envisage a first stage on the
scale of hundred cubic kilometers. 
%Dependent on the results one may
%guess that they join in the next step which would be on the
%thousand cubic kilometer scale.

Most exotic is the search for radio emission from extremely-high energy
cascades induced by neutrinos or cosmic rays skimming the moon surface. 
It was first proposed by I.\,Zheleznykh in 1988 \cite{Zheleznykh-1988}
under the name RAMHAND (Radio Astronomical Method
of Hadron And Neutrino Detection).
After first attempts at the Parks Radio Observatory in the
mid nineties, the experiment was firstly realized by the 
"Goldstone Ultra-high Energy Neutrino Experiment" (GLUE) which
used two NASA antennas and reached a maximum sensitivity at several
thousand EeV=1000
\cite{radio-GLUE-2004}. 
With the same method, the NuMoon experiment at the Westerbork Radio Telescope 
was searching for extremely
energetic neutrinos 
\cite{radio-NUMOON-2008}.  Presently, the
Kalyzin 65-m radio telescope in Russia is pursuing similar observations.
Corresponding activities are also under
preparation in the context of the LOFAR experiment \cite{LOFAR-2010}.
An overview of RAMHAND-type activities is given in \cite{RAMHAND}.

\subsection{Detection via air showers}
\label{subsec-air}

\begin{figure}[ht]
\begin{center}
\includegraphics[width=11cm]{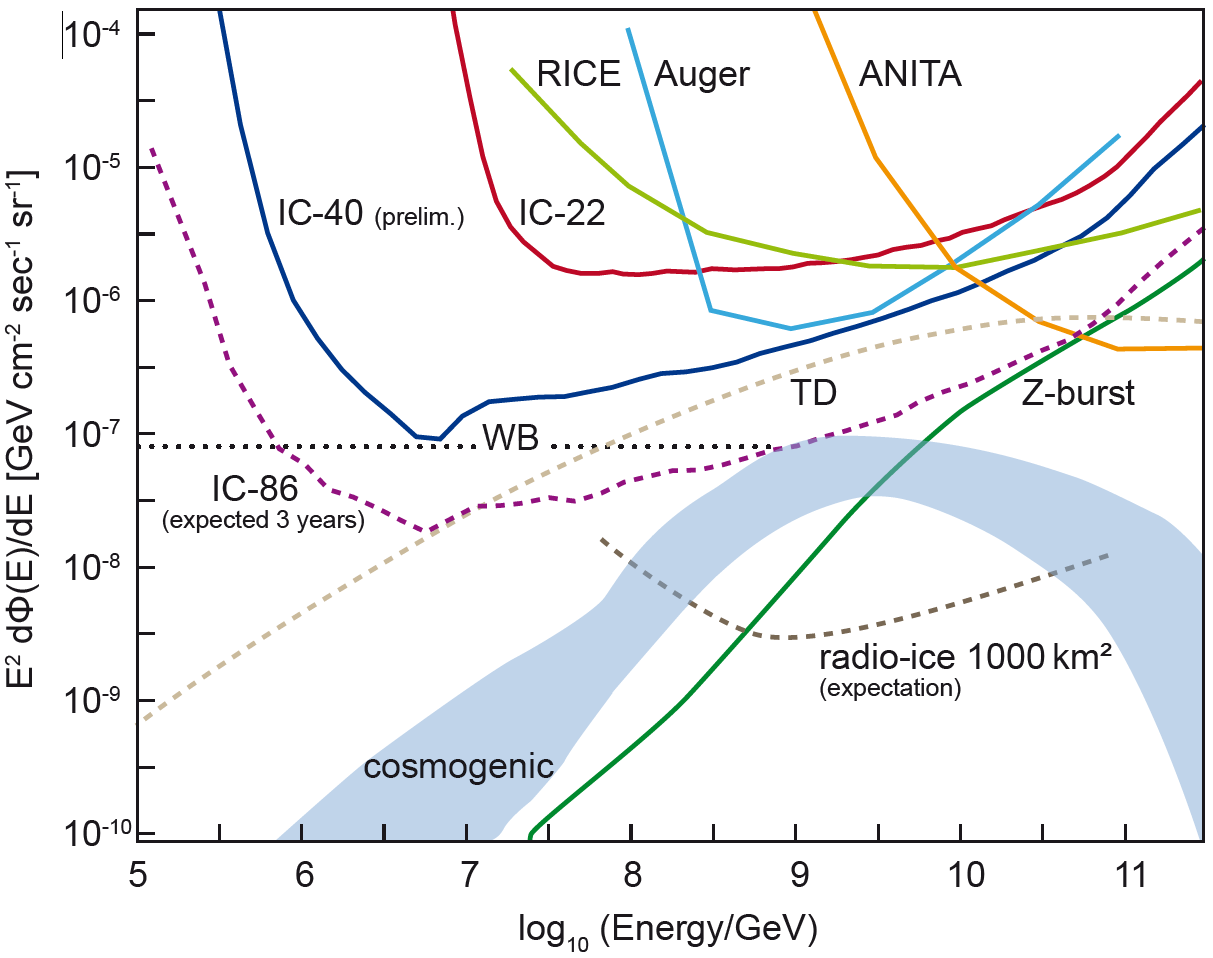}
\caption{
All-flavour 90\% C.L.\ differential upper limits on the flux of extraterrestrial
neutrinos in the PeV--EeV region. Limits are from the under-ice radio
array RICE, 
%\cite{RICE-2006}, 
the air shower detector Auger, 
%\cite{Auger-2010},
the radio balloon experiment ANITA, 
%\cite{ANITA-2009}, 
IceCube-22,
%\cite{icecube-2010d}, 
and IceCube-40. See for references \cite{Katz-Spiering-2011}.
%\cite{icecube-EHE2011}. 
Also given are the expectations for 3 years of operation of the full IceCube
detector and for a 1000\,km$^2$ shallow radio detector at the South Pole. The
colored band corresponds to different predictions for cosmogenic neutrinos 
processes. References for the these scenarios, the Z-burst prediction and the
SUSY top-down scenario (TD) can be found in \cite{icecube-EHE2011}.}
\label{EHE-diffuse}
\end{center}
\end{figure}

At energies above $10^{17}$\,eV, large air shower arrays like the Pierre Auger
Observatory in Argentina \cite{Auger-Web} or the Telescope Array in Utah,
USA \cite{TA-2003} are searching for horizontal air showers induced by neutrino
interactions deep in the atmosphere (showers caused by charged cosmic ray
interactions start much higher up in the atmosphere). 
The optimum
sensitivity window for this method is at 1-100\,EeV.
% the effective target
%mass is up to 20\,Gigatons. 
An even better sensitivity might be obtained for tau
neutrinos, $\nu_{\tau}$, scratching the Earth and interacting close to the array
\cite{Fargion-2002,Bertou-2002,Fargion-2004}. The charged $\tau$ lepton produced in
charged-current interaction can escape the rock around the array (in contrast to
electrons) and mostly decays into hadrons (branching ratio ca.\,65\%) after a
short path length (in contrast to muons). If this decay happens in the field of
view of the fluorescence telescopes of the Pierre Auger Observatory
or the Telescope Array, the decay cascade can be recorded. 
The limits derived by Auger are included in Fig. \ref{EHE-diffuse}.

%Provided the experimental pattern allows for a clear identification, the acceptance for
%this kind of signals can be large. For the optimal energy scale of $1\eev$, the
%present $\nu_\tau$ limit for an $E^{-2}$ tau neutrino flux is about
%$E^2\phi<10^{-7}\flunit$ 
%\cite{Auger-nutau}.

Space-based observation of extended air showers is an approach to even further
increase the target mass for highest-energy cosmic-ray and neutrino detection,
at energies beyond $10^{19}$\,EeV. This the rationale for
the Extreme Universe Space Observatory
(EUSO) which has been proposed around the year 2000. 
With a wide-field camera, EUSO will  observe the
atmosphere from an orbit at several 100\,km height and register the
fluorescence light from extended air showers and, if circumstances allow, also
the reflection at Earth surface of the Cherenkov light emitted in shower
direction. After several years of technical development and a
long phase of uncertainty concerning the space carrier, plans are now to install
the device -- meanwhile renamed to JEM-EUSO \cite{JEM-EUSO-Web} -- on the
Japanese Experiment Module (JEM) of the International Space Station, with the
launch expected around 2015. JEM-EUSO will observe an atmospheric target volume
with a mass of more than one Tera-ton and will thus exceed the Auger sensitivity by
two orders of magnitude for energies above some $10^{19}$\,eV
(see for the actual status of the mission \cite{JEM-EUSO-2012}). 
%Similarly as for PAO, neutrino-induced air showers can be
%separated from those from hadrons or gammas by the depth of the interaction
%point in the atmosphere. The JEM-EUSO physics opportunities in the neutrino
%channel have been studied in general \cite{Fargion-2002b,JEM-EUSO-2010b} and
%specifically for cosmogenic neutrinos \cite{Kotera-2010} and GRB neutrinos
%\cite{JEM-EUSO-2009}. It is not obvious that JEM-EUSO will detect neutrinos from
%known astrophysical or cosmogenic sources, but its measurements will explore
%possible neutrino fluxes in a hitherto inaccessible energy region.

%\vspace{1cm}

Figure \ref{EHE-diffuse} summarizes the present limits in the PeE-EeV region.
One sees that IceCube competes well up to several hundred PeV.
At higher energies, ANITA takes over. A realistic chance to
measure more than just a handful of cosmogenic neutrinos will stay
the privilege of detectors of the 100-1000 cubic kilometer scale.

\newpage
\section{What next?}

Whereas the identification of first extraterrestrial neutrinos IceCube
has not yet been achieved, projects of similar or greater size on the Northern hemisphere
are under preparation. 
%In recognition of the fact that at least a cubic-kilometre sized detector will
%be necessary to really observe abundant astrophysical high-energy neutrino
%sources, the High Energy Neutrino Astronomy Panel (HENAP) of the
In 2002, an expert committee 
%PaNAGIC (the Particle and Nuclear Astrophysics and Gravitation International
%Committee) Committee of IUPAP
installed by the International Union of Pure and Applied Physics (IUPAP) concluded 
\cite{henap-2002} that {\it "a
km$^3$-scale detector in the Northern hemisphere should be built to complement
the IceCube detector being constructed at the South Pole''}.
The main physics rationale of this recommendation was that
the field of view of a Northern detector includes the central
parts of the Galaxy, with several potential neutrino sources. 
IceCube can see only the outer parts of the galactic plane --
with the exception of low energies (DeepCore detecting
contained events) and very high
energies (where the background from 
downward moving atmospheric muons becomes small). 
Following this recommendation, the Mediterranean neutrino telescope groups 
have formed the
KM3NeT collaboration to prepare, construct and operate such a device. 
KM3NeT has been included in the priority project list of the European Strategy.
Forum on Research Infrastructures, ESFRI. 

\begin{figure}[ht]
%\sidecaption
\center{
\includegraphics[width=10cm] {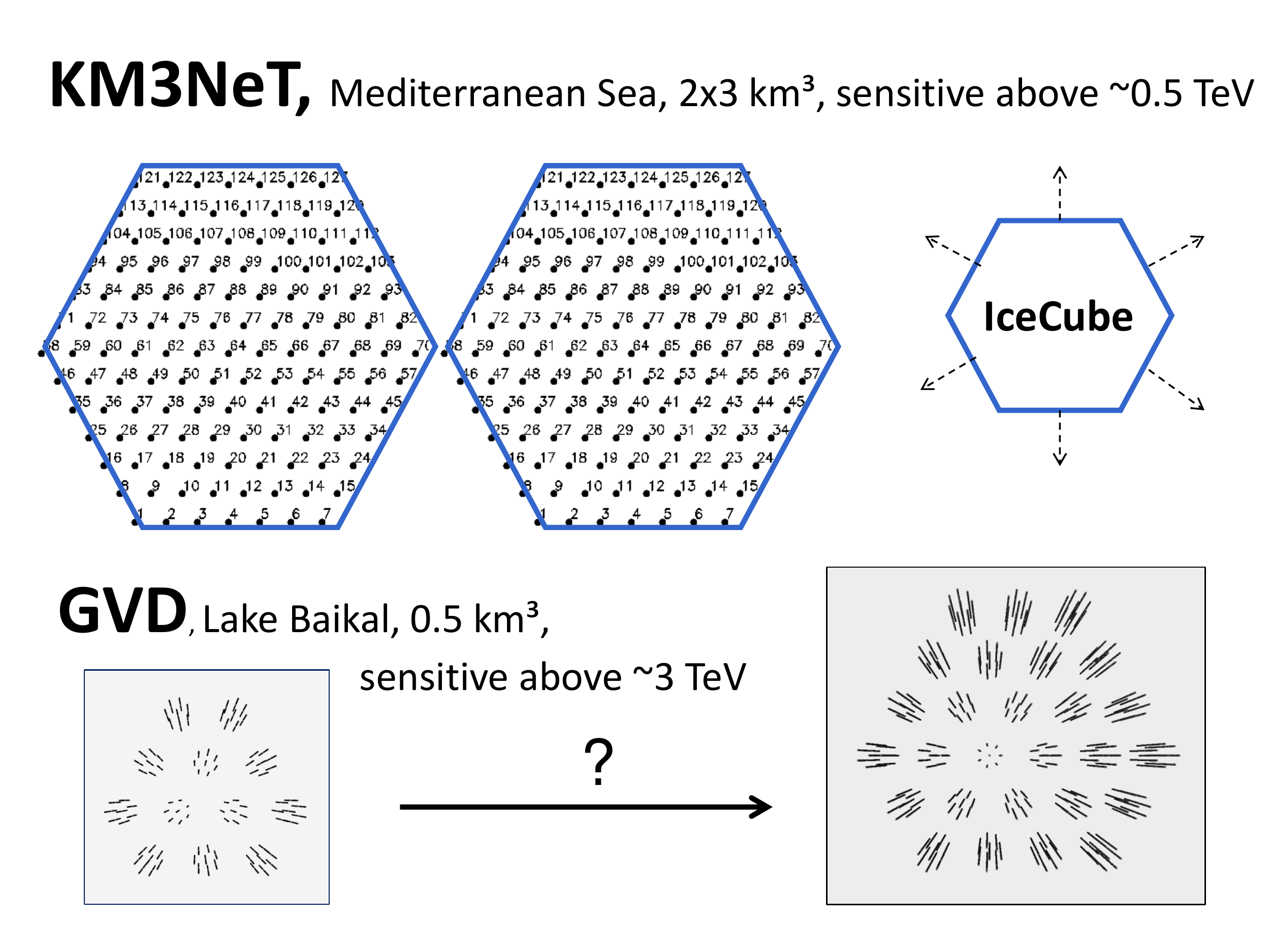}}
\caption{
Top views of planned new detectors at the Northern hemisphere
(KM3NeT and GVD). The are compared to the top view of IceCube.
Arrows symbolize the possibility of an IceCube extension
in case of discovery of extraterrestrial neutrinos.
}
\label{Beyond-cube}
\end{figure}

A design study from 2006 to 2009 resulted
in a Conceptual Design Report (CDR) \cite{km3net-cdr}
and a Technical Design Report (TDR) \cite{km3net-tdr}. 
At present, the project is in a Preparatory Phase and envisages to install
a detector with 6 km$^3$ volume from 2014 on. The total investment cost
is estimated to be around 225 MEuro.  A top view of a possible detector configuration 
consisting of two blocks, each 3\,km$^3$, is sketched in
Fig.\,\ref{Beyond-cube}.

After initial funding in the Netherlands some years ago, substantial funding for
building engineering array(s) has been recently assigned in Italy and in
France. At present, the partners are preparing construction of this/these
demonstrator(s). 

In Russia, the Baikal Collaboration plans the stepwise installation of a 
kilometer-scale array in Lake Baikal, the Gigaton Volume Detector, GVD
\cite{Baikal-GVD-Aynutdinov-2007}.  In the years 2008-2010 
the basics elements -- new optical modules, readout with Flash-ADCs.
underwater communication and trigger system -- have been tested with
stationary prototype strings and resulted in a Conceptual Design Report
\cite{gvd-cdr}. In April 2012 an engineering array with a first full-scale string 
and two half-strings have been deployed for a long-term test.
Realizing that the originally
planned size of half a cubic kilometer is no longer enough, a four times
larger array is presently being studied, as sketched bottom-right of
Fig.\,\ref{Beyond-cube}. Note that due to the shallower depth
of Lake Baikal, the height of GVD will be smaller than that of
KM3NeT and IceCube.

The realization of these projects depends on several factors.
First of all, IceCube results will play a strong role. 
Secondly, future gamma-ray data must provide 
stronger indications that the observed gamma-rays
are pion-decay counterparts of neutrinos and
not only the result of inverse Compton scattering. And
last but not least, the considerable funding must be found.

Missing or marginal evidence for sources from IceCube may have various
consequences. If one is going to continue the venue of detectors which explore
the energy range most characteristic for GRBs and AGNs, one has to envisage 
an order-of-magnitude step in sensitivity, i.e. beyond
what is presently scheduled by KM3NeT and GVD.
% This requires configurations with
% significantly larger spacing than the present KM3NeT design, resulting in a
% penalty at energies below a few tens of TeV, and essentially in sacrificing
% most Galactic sources.

The second option would be an even larger leap in size. 
It would address energies above 100 PeV with the help of new technologies like
radio or acoustic detection and envisage 100--1000 cubic kilometers of
instrumented volume. This option might still have sensitivity to neutrinos from
AGN jets but would also cover well the energy range of neutrinos from
cosmic ray interactions with the 3-Kelvin microwave background. 
In contrast to optical detectors, new-technology detectors are still in the R\&D
phase and also have no natural calibration source like atmospheric neutrinos for
optical detectors. 

The third option would define, at least for the time being,
an end to the search for neutrinos from cosmic accelerators. It would focus on
optical detection with small spacing optimized to investigate oscillations with
accelerator neutrinos (Mediterranean Sea) and atmospheric neutrinos, or, even
more pretentious, to study Supernova bursts beyond our own Galaxy or even proton decay.

Taken all together, we may be close to a turning point. 
%For the moment, however,  there is no reason to be pessimistic. 
We have made a
factor-of-thousand step in sensitivity compared to a dozen years ago. This is
far more than the traditional factor of ten which so often led to the discovery
of new phenomena \cite{Harwit}. For instance, looking across 
our own field, the prospects for
discovery had not been estimated too highly before launching the first X-ray
rocket in 1962, or before detecting the Crab Nebula in TeV gamma rays in 1989. 
%To quote F.\,Halzen:  "Nothing is guaranteed, but history is on our side" 
History has told another story, as we know today.
The same may be the case for high energy neutrino astronomy.
The journey is not yet finished!

\section*{Acknowledgments}

I want to thank Albrecht Karle, John Learned, Rolf Nahnhauer,
Per-Olof Hulth, Alan Watson and Ralf Wischnewski for carefully reading the
manuscript and for many extremely helpful comments. 
I also acknowledge valuable information and comments from
Mark Bowen, Grigorij Domogatsky, Zhan-Arys Dzhilkibaev, 
Stavros Katsanevas, Ulrich Katz,
Christopher Wiebusch and Igor Zhelesnykh.

\end{document}